\documentclass[conference]{IEEEtran}
\IEEEoverridecommandlockouts
\usepackage{amsmath,amssymb,amsfonts}
\usepackage{algorithm}
\usepackage{algpseudocode} %
\usepackage{graphicx}
\usepackage{textcomp}
\usepackage{xcolor}
\def\BibTeX{{\rm B\kern-.05em{\sc i\kern-.025em b}\kern-.08em
    T\kern-.1667em\lower.7ex\hbox{E}\kern-.125emX}}

\usepackage{nicefrac}
\usepackage{fancyref}
\usepackage{bm}		%
\usepackage{caption}
\usepackage{hyperref}

\usepackage[normalem]{ulem} %
\usepackage{multirow}

\usepackage{mathtools}
\mathtoolsset{showonlyrefs}

\usepackage{siunitx}
\makeatletter
\@ifpackagelater{siunitx}{2021/05/17} %
{
	\sisetup{per-mode=fraction,
		inter-unit-product=\ensuremath{{\cdot}},
		exponent-product=\ensuremath{\cdot},
		fraction-function=\nicefrac}
}
{	
	\sisetup{per=frac,
		fraction=nice,
		decimalsymbol=comma,
		binary-units=true,
		loctolang={DE:ngerman,UK:english}}
}%
\makeatother

\newcommand{\ttt}{\! \left( t \right)}

\newcommand{\tft}{\! \left( t_\text{f} \right)}
\newcommand{\tftt}{\! \left( t_\text{f,t} \right)}
\newcommand{\tzt}{\! \left( 0 \right)}
\newcommand{\tat}{\! \left( a_\text{max} \right)}

\newcommand{\xt}{\bm{x} \ttt}
\newcommand{\ut}{\bm{u} \ttt}

\newcommand{\xtutt}{\! \left( \xt \! , \, \ut \! , \, t\right)}

\newcommand{\tct}{\! \left( \cdot \right)}

\newcommand{\wrt}{w.r.t. }%

\newcommand{\ocpS}{\mathtt{OCP}\text{-}\mathtt{{S}}}
\newcommand{\ocpJ}{\mathtt{OCP}\text{-}\mathtt{{J}}}
\newcommand{\ZV}{\mathtt{ZV}}
\newcommand{\ZVD}{\mathtt{ZVD}}

\newcommand{\SCurve}{\mathtt{S}\text{-}\mathtt{{Curve}}}
\newcommand{\EI}{\mathtt{EI}}

\newcommand{\Fir}{\mathtt{FIR}}
\newcommand{\FirImp}{\mathtt{FIR}_\mathtt{Imp}}

\newcommand{\colorZtft}{black}
\newcommand{\colorTOne}{black}
\newcommand{\colorTTwo}{black}

\begin{document}
\title{On trajectory design from motion primitives for near time-optimal transitions for systems with oscillating internal dynamics \\ %
\thanks{This work has been submitted to the IEEE for possible publication. Copyright may be transferred without notice, after which this version may no longer be accessible.}
}

\author{\IEEEauthorblockN{1\textsuperscript{st} Thomas Auer}
\IEEEauthorblockA{\textit{IACE - UMIT TIROL}\\
Hall in Tirol, Austria \\
thomas.auer@umit-tirol.at}
\and
\IEEEauthorblockN{2\textsuperscript{nd} Frank Woittennek}
\IEEEauthorblockA{\textit{IACE - UMIT TIROL}\\
Hall in Tirol, Austria \\
frank.woittennek@umit-tirol.at}
}

\maketitle

\begin{abstract}
	An efficient approach to compute near time-optimal trajectories for linear kinematic systems with oscillatory internal dynamics is presented. Thereby, kinematic constraints with respect to velocity, acceleration and jerk are taken into account. The trajectories are composed of several motion primitives, the most crucial of which is termed jerk segment. Within this contribution, the focus is put on the composition of the overall trajectories, assuming the required motion primitives to be readily available. Since the scheme considered is not time-optimal, even decreasing particular constraints can reduce the overall transition time, which is analysed in detail. This observation implies that replanning of the underlying jerk segments is required as an integral part of the motion planning scheme, further insight into which has been analysed in a complementary contribution. Although the proposed scheme is not time-optimal, it allows for significantly shorter transition times than established methods, such as zero-vibration shaping, while requiring significantly lower computational power than a fully time-optimal scheme. \\

\end{abstract}

\begin{IEEEkeywords}
	Motion and Path Planning, Optimization and Optimal Control, Motion Control, Semiconductor Manufacturing
\end{IEEEkeywords}

\section{Introduction}
Pick-and-place processes in the electrical industry demand accuracy and speed. Trends in the electronics industry point to miniaturisation and further integration of circuits from year to year \cite{LindaS_Wilson2023,Lau2025,Ikegami2024,Haneda2024}. High accuracy is required for part placement to ensure electrical connections work as intended. In order to decrease cost, short transitions are desirable. Forces due to motion acting on the machine frame in combination with finite stiffness lead to oscillation and inaccuracy in component placement, as well as increased machine wear \cite{Esau2022,Bilal2023}. The oscillation amplitudes of the machine frame can be bigger, than the allowed accuracy requirements, which are fractions of $\si{\micro\meter}$'s \cite{Lau2025,Ikegami2024,Haneda2024}. Specifically planned motion profiles, also called trajectories, are used to achieve rest-to-rest transitions for the system.

\subsection{Previous research in this topic}
The so called $\SCurve$ \cite{ClaudioMelchiorri2008} is the fastest time optimal trajectory, that is limited in jerk, acceleration and velocity for rest-to-rest transitions of ideally stiff systems. Since production machines exhibit flexibility originating from the finite stiffness of the machine frame, it is usually not feasible to use the $\SCurve$ directly, since it leads to oscillation. A method to plan rest-to-rest trajectories for systems exhibiting flexibility is the application of shaping techniques to those trajectories. This is commonly referred to as trajectory shaping. A common shaper, that is used for this purpose is the zero-vibration $\left(\ZV\right)$-shaper \cite{Singhose1990,Singhose1994}. A key advantage of trajectory shapers lies in its ease of implementation and the robustness \wrt to parameter uncertainty \cite{UH2001}. The sensitivity to parameter uncertainty of the $\ZV$-shaper can be further reduced by increasing the system damping \cite{Singhose1994}. Further and more substantial improvements can be achieved by utilizing $\ZVD$-shapers \cite{Pao1998,Kasprowiak2022,Kang2019} or $\EI$-shapers \cite{Singhose1997_EI_shaper,Singer1999, Vaughan2008}. A very detailed overview concerning the shaping methods can be found in \cite{Singhose2009}. Decreasing the sensitivity to parameter uncertainty has been studied extensively, because it is an important topic. However, if the parameters of the system are known with a certain accuracy, methods offering shorter transition times can be of greater interest.
A method to decrease transition times, when using trajectory shapers is to employ negative-impulse shapers \cite{SinghoseMarch1996}. However, even if the original trajectory does not violate any kinematic constraints, due to the nature of negative-impulse shapers, violations of kinematic constraints can occur for the shaped trajectory. While there are strategies to avoid violation of kinematic constraints \cite{Sorensen2008} and negative-impulse shapers allow for faster transitions, they are limiting the shape of the underlying trajectory and might still lack time-optimality in order to avoid violation of the kinematic constraints.
Methods to adjust the underlying $\SCurve$ trajectories in order to remove as much oscillation as possible have been studied in \cite{Meckl1998,Bearee2014,Kim2018}. By adjusting the maximal jerk and the duration, considerable amounts of oscillation can be removed without utilizing trajectory shapers. %
The general idea has also been extended ($\Fir$-methods) and developed for full trajectories \cite{Biagiotti2015,Besset2017,Biagiotti2019,Biagiotti2021,Biagiotti2020,Yalamanchili2024}, leading to a potential reduction of transition time at the expense of sensitivity to parameter uncertainty. A drawback of these approaches is, that the damping of the system cannot be taken into account, even if it is known. This leads to residual oscillation, even for the nominal model (no parameter uncertainty). It is possible to account for system damping according to \cite{Bearee2014}. However, this has not been applied to the expanded full trajectories.

Employing feedback control concepts is a further possibility to improve the oscillation behaviour \cite{Bandopadhya2006, Pilbauer2018, Singh2020, She2021}. This allows to use fast shapers or trajectories planned without shapers and the insensitivity to parameter uncertainty can be ensured by the active control concept \cite{Dharne2007}. Approaches with feedback still benefit from fast trajectories that offer rest-to-rest transitions for the nominal model \cite{Auer2024Case}. In this case, the controller reduces the oscillation at the end of a transition. A disadvantage is, that additional measurement equipment might be required. Using trajectories planned to achieve oscillation free rest-to-rest transitions is still beneficial. %
Trajectories planned with optimal-control approaches guarantee the fastest transitions while respecting the kinematic constraints. Different planning methods including two optimal-control concepts were compared in \cite{Auer2023IFAC}. Calculation through optimization of truly time-optimal trajectories can cause issues due to the discontinuous nature of the input signals required \cite{Silva2010}. Implementation of the required numeric algorithms on programmable logic controllers (PLC) can be especially problematic due their limited computing power. Another issue that prevents such approaches from being used in actual pick-and-place machines consists in the lack of guarantees concerning the convergence of the underlying algorithms. This is particularly problematic in cases where it is not possible to pre-calculate and store trajectories, for example, when system parameters are constantly updated by measurements.
A method to plan the motion out of trajectory segments, where each segment has to satisfy a number of constraints has been proposed in \cite{Dijkstra2007}. This method offers a time-advantage over a $\ZV$-shaped trajectory at the expense of more complicated calculation, however the calculation effort required can be prohibitive, when it comes to using it on industrial machinery.

\subsection{Contribution of this paper to the field}
This publication introduces the trajectory planning method called $\ocpJ$. Here, the full trajectories are assembled from individual motion primitives (which are themselves computed in a time-optimal manner, as explained in \cite{tau_ocpJ_assembly_part2}). Crucially, in order to ensure minimal transition times of the complete trajectories, the proposed algorithm allows overlapping of individual motion primitives while avoiding a violation of the kinematic constraints. Moreover, this publication shows that lowering the acceleration limit for the proposed approach can lead to faster overall rest-to-rest transitions. It allows for faster transitions than the existing $\ZV$-shaper and offers better behaviour \wrt sensitivity to parameter uncertainty than $\Fir$-methods. The analyses presented compare the time advantages with other trajectory planning approaches and shows the sensitivity to parameter uncertainty. An extensive measurement study performed on a laboratory system (with all of the trajectories calculated on the laboratory system's PLC) demonstrates and validates the $\ocpJ$ approach and highlights its performance characteristics. Furthermore, because the calculation is not based on an iterative numerical optimization, but on direct calculation, the computation time is known, constant and sufficiently small to allow trajectories to be calculated as they are needed. This allows the system parameters to be continuously updated to take parameter drift into account.

\bigskip
\subsection{Organisation of paper}
The mathematical model, the parameters and the most important mathematical background are presented in \autoref{sec:model_and_paramters}. Then the calculation of the full trajectory with the three different cases is explained in \autoref{sec:calc_full_trajectory}. The formulation of an overlying optimization problem to ensure minimum transition times with the $\ocpJ$ approach is described in \autoref{sec:optim_full_trajectory} and the algorithms required are also summarised there. This section shows why the quick calculation of the jerk segments may be an advantage in some cases (cf. \cite{tau_ocpJ_assembly_part2} for the algorithm). A comparison with other existing approaches (commonly used in the field and newly developed) in terms of transition time and parameter sensitivity is given in \autoref{sec:comparison_to_regular}. This comparison is extended to different parameter sets for a more detailed analysis. Eventually, measurement results from a laboratory system are used to validate the approach in \autoref{sec:measurements_laboratory_system}. The results are then discussed in \autoref{sec:summary_and_comparison} and points for relevant follow-up research are provided in \autoref{sec:outlook_and_conclusion}.

\section{Preliminaries}\label{sec:model_and_paramters}
This section presents a mathematical model of a system to which the presented trajectory planning method can be applied. As the equations used for calculation in this publication and in \cite{tau_ocpJ_assembly_part2} (calculation of motion primitives) show, the trajectory planning approach is solely influenced by the kinematic constraints of the actuator and the system's oscillatory behaviour. The kinematic constraints are the maximum velocity $v_\text{lim}$, acceleration $a_\text{lim}$ and jerk $j_\text{lim}$, while the systems behaviour is characterised by the eigenfrequency $\omega_{\text{d}}$ and damping factor $\delta$, which is allowed to be zero. The laboratory system used to validate the approach via measurements corresponds to the model presented. The use of this model as a basis allows for better presentation, while also maintaining the possibility of its application to other systems.

\subsection{Mathematical model}
\begin{figure}[!ht]
	\centering
	\def\svgwidth{0.95\linewidth}
	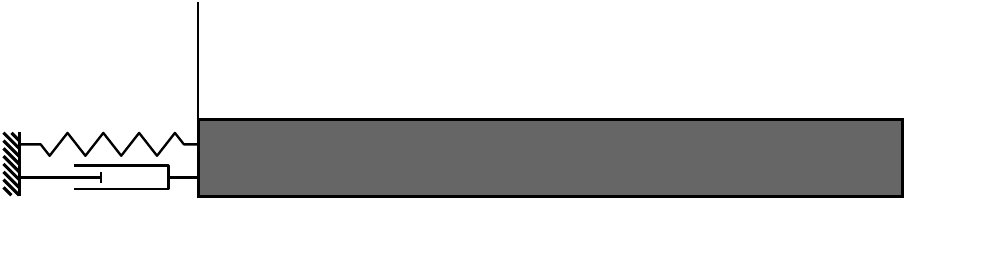
	\caption{Symbolic representation of one axis of the pick-and-place machine.}
	\label{fig:model_of_the_system}
\end{figure}

A pick-and-place machine exhibiting flexibility could be modelled using a continuum mechanic description, if every effect of the oscillation should be captured in detail. However, since the essential oscillation of the machine frame can be captured by the first mode, a simplified system description with a lumped model has been chosen. This model consists of two masses, a spring and a viscous damping element and is shown schematically in \autoref{fig:model_of_the_system}. While this is a rather simple system, it is sufficient in modelling the motion and the focus of the contribution can be put on the effective method of calculation of the trajectory. The spring $k$ represents the stiffness of the machine frame and the viscous damping element $d$ the natural damping of the machine frame. The masses are the mass of the baseframe $m_\text{b}$ and the mass of the slider $m_\text{s}$. In an actual production machine this slider would mount a gripper for component handling and it is commonly referred to as endeffector.
The equations governing the motion of the system \cite[cf.]{Auer2023IFAC} can be derived via momentum balance and are given by
\begin{subequations}\label{eq:dynamical_equations_of_system}
	\begin{align}
		m_\text{b}  \ddot{x}\ttt + d \, \dot{x}\ttt + k \, x\ttt &= -F\ttt \, \text{,} \label{eq:movement_equation_for_base} \\
		m_\text{s} \! \left(\ddot{x}\ttt + \ddot{z}\ttt\right) &= F\ttt \label{eq:force_acting_in_system} \, \text{,}
	\end{align}
\end{subequations}
whereas any effects arising from friction are included in the force $F\ttt$. Using a position controller for the slider position $z\ttt$ on the actual production machines removes the influence of those forces by basically using $\ddot{z}\ttt$ as input. The parameters used for the simulation studies in this contribution are listed in \autoref{tab:table_machine_parameters_ex_Pap}.
\begin{table}[!ht]
	\captionsetup{width=\linewidth}
	\caption{Parameters and resulting eigenfrequencies}
	\vspace{-1em}
	\renewcommand{\arraystretch}{1.25}
	\centering
	\begin{tabular}{|l|l|l|l|}
		\hline
		$m_\text{s} = \SI{25}{\kilo\gram}$ & $m_\text{b} = \SI{500}{\kilo\gram}$ &
		$k = \SI{15e6}{\newton\per\metre}$ & $d = \SI{5e3}{\kilo\gram\per\second}$ \\ \hline
		\multicolumn{2}{|l|}{$f_\text{0} = \SI{26.9}{\hertz}$} & \multicolumn{2}{l|}{$f_\text{d} = \SI{26.8914}{\hertz}$} \\ \hline
	\end{tabular}
	\label{tab:table_machine_parameters_ex_Pap}
\end{table}
The kinematic constraints listed in \autoref{tab:table_listing_kincont_ex_Pap} correspond to an exemplary production machine. All limits are imposed by the physical limits of the drives while, additionally, the jerk constraint has an influence on  wear and tracking accuracy \cite{Kyriakopoulos1988,Piazzi2000,Macfarlane2003,Porawagama2014,Bilal2023}. The kinematic constraints result from values provided by an industrial partner.
\begin{table}[!ht]
	\captionsetup{width=\linewidth}
	\caption{Kinematic constraints of slider.}
	\vspace{-1em}
	\renewcommand{\arraystretch}{1.3}
	\centering
	\begin{tabular}{|l|l|l|}
		\hline
		$\left|\dot{z}\ttt\right| \le v_\text{lim}$ & $\left|\ddot{z}\ttt\right| \le a_\text{lim}$ &
		$\big|z^{\left(3\right)}\ttt\big| \le j_\text{lim}$ \\ \hline
		$v_\text{lim} = \SI{1.5}{\meter\per\second}$ & $a_\text{lim} = \SI{20}{\meter\per\square\second}$ &
		$j_\text{lim} = \SI{800}{\meter\per\second\cubed}$ \\ \hline
	\end{tabular}
	\label{tab:table_listing_kincont_ex_Pap}
\end{table}
It is advantageous to use a dynamic extension with $z^{\left(3\right)}\ttt$ as input for the calculations required later on. Adding \eqref{eq:movement_equation_for_base} and \eqref{eq:force_acting_in_system} together to remove $F\ttt$ and using ${m_\text{g} = m_\text{s} + m_\text{b}}$ leads to the state-space representation
\begin{align}\label{eq:opt:sys}
	\bm{\dot{x}} &= %
	\begin{bmatrix}
		0 & 1 & 0& 0& 0 \\
		-k^\star & -d^\star &0&0&-m^\star \\
		0&0&0&1&0 \\
		0&0&0&0&1 \\
		0&0&0&0&0
	\end{bmatrix} \cdot \begin{bmatrix} x \\ \dot{x} \\ z \\ \dot{z}  \\ \ddot{z} \end{bmatrix} + 
	\begin{bmatrix} 0 \\ 0 \\ 0 \\ 0  \\ 1 \end{bmatrix} \! z^{\left(3\right)} \\
	k^\star &= \nicefrac{k}{m_\text{g}} \qquad d^\star = \nicefrac{d}{m_\text{g}} \qquad m^\star = \nicefrac{m_\text{s}}{m_\text{g}}
\end{align}
with the state $\bm{x}$ and input $u$ given by
\begin{align}
	\bm{x} &= \left[x, \dot{x}, z, \dot{z}, \ddot{z}\right]^T \, \text{,} & u &= z^{\left(3\right)} \, \text{.} \label{eq:sys_state_for_Axbu}
\end{align}
If $z$ is understood as the output of the system it can be readily observed that the system possesses relative degree 3 and is given in Byrnes-Isidori form. The internal dynamics, which obviously is not observable over this output corresponds to a damped harmonic oscillator with undamped eigenfrequency $\omega_0=\sqrt{k^\star}$ and damping ratio $\frac{d^\star}{2\omega_0}$, which is excited by the slider acceleration $\ddot{z}$. 

\subsection{Solution of the slider dynamics}\label{Subsec:shape_of_trajectories}
Twice continuously differentiable trajectories $t\mapsto z\ttt$ are considered, which are piecewise constant in the input $u\ttt = z^{\left(3\right)}\ttt$. As a consequence, the jerk $z^{\left(3\right)}\ttt$ is a sum of $n$ steps of amplitudes $a_1,\dots,a_n$ occurring at $t_1<\dots<t_n$: 
\begin{equation}\label{eq:trajectory:ansatz}
	z^{\left(3\right)}\ttt = \sum\limits_{i=1}^{n} a_i H\!\left(t-t_i\right) \, \text{.}
\end{equation}
Therein, $H\tct$ represents the Heaviside step function. The terminal time of the trajectory motion is denoted by $t_\text{f,t}$. Integration of \eqref{eq:trajectory:ansatz} leads to
\begin{subequations}\label{eq:calc_z_to_zppp_from_snap}
	\begin{align}
		\ddot{z}\ttt &= \ddot{z}(0) + \sum\limits_{i=1}^{n} a_i H\left(t\! - \!t_i\right)\cdot \left(t\! - \!t_i\right) \, \text{,} \label{eq:calc_zpp_from_imp} \\
		\dot{z}\ttt &= \dot{z}(0) + \ddot{z}(0) \, t + \frac{1}{2} \, \sum\limits_{i=1}^{n} a_iH\left(t\! - \!t_i\right) \cdot \left(t\! - \!t_i\right)^2 \, \text{,} \label{eq:calc_zp_from_imp} \\
		z\ttt &= z(0)\! +\! \dot{z}(0) \, t \!+\! \ddot{z}(0) \, \frac{t^2}{2} + \frac{1}{6} \, \sum\limits_{i=1}^{n} a_i H\left(t\! -\! t_i\right)\cdot \left(t\! - \!t_i\right)^3\,. \label{eq:calc_z_from_imp}
	\end{align}\noeqref{eq:calc_zpp_from_imp}\noeqref{eq:calc_zp_from_imp}\noeqref{eq:calc_z_from_imp}
\end{subequations}
As the considered trajectories should correspond to rest-to-rest transitions of the slider, they must satisfy\footnote{Formally, the condition   $z^{(3)}(t_{\text{f,t}})=0$ is not really required in the formulation of the corresponding optimal control problem. However, it can always be satisfied by choosing $t_n=t_\text{f,t}$ with an appropriate choice of $a_n$, leaving $t\to z^{(i)}\ttt$, $i=0,1,2$ unchanged on $[0,t_\text{f,t}]$. In fact, it ensures $\ddot{z}(t)=\dot{z}(t)=0$ not only at $t=t_\text{f,t}$ but for all $t\ge t_\text{f,t}$, i.e., the system remains in the desired equilibrium.
Moreover, the formulation chosen essentially simplifies the algebraic computations.}
\begin{subequations}\label{eq:boundary_constraints_zp_zpp_zppp_zero}\noeqref{eq:boundary_constraints_zp_zpp_zppp_zero}
	\begin{align}
		z^{\left(3\right)}\tzt &= 0 \, \text{,} & \ddot{z}\tzt &= 0 \, \text{,}  & \dot{z}\tzt &= 0 \, \text{,}\label{eq:boundary_constraints_zp_zpp_zppp_zero:0}\noeqref{eq:boundary_constraints_zp_zpp_zppp_zero:0} \\
		z^{\left(3\right)}\tftt &= 0 \, \text{,} & \ddot{z}\tftt &= 0 \, \text{,} & \dot{z}\tftt &= 0 \, \text{.}\label{eq:boundary_constraints_zp_zpp_zppp_zero:f}\noeqref{eq:boundary_constraints_zp_zpp_zppp_zero:f}
	\end{align}
\end{subequations}
Evaluating the initial and final conditions \eqref{eq:boundary_constraints_zp_zpp_zppp_zero} in view of \eqref{eq:trajectory:ansatz} and \eqref{eq:calc_z_to_zppp_from_snap} yields the following relations for the coefficients in \eqref{eq:trajectory:ansatz}:
\begin{align}\label{eq:conditions_on_coefficients}
	\sum_{i=1}^n a_i &= 0 \, \text{,} & \sum_{i=1}^n a_i \, t_i &= 0 \, \text{,} & \sum_{i=1}^n a_i \, t_i^2 &= 0 \, \text{.}
\end{align}
An example for such a trajectory is the so-called $\SCurve$ \cite{ClaudioMelchiorri2008} which is also used as reference trajectory in $\ZV$ and $\ZVD$-shaping as shown in \autoref{fig:example_optimal_control_sCurve_ZV}.

\subsection{Optimal control formulation}\label{SubSec:OcpS_formulation}
For a general time-invariant non-linear system with state-space description
\begin{equation}\label{eq:sys:nonlin}
	\bm{\dot{x}} = \bm{f} \! \left(\xt,u\ttt\right)\text{,} \qquad \xt \in \mathbb{R}^n \text{,} \, \bm{u}\ttt \in \mathbb{R}^m %
\end{equation}
an optimal control problem aims to compute the minimum 
\begin{subequations}
	\begin{align}
          \min_{\bm{x}, \bm{u}, t_\text{f,t}} &J \!\left(\bm{x}, \bm{u}, t_\text{f,t}\right) \label{Eq_OPC_allg_1} \, \text{,}\\
\intertext{of a cost functional}          
		J \tct = V& \! \left( \bm{x} \! \left( t_\text{f,t} \right)\! , \, t_\text{f,t} \right) + \int_{0}^{t_\text{f,t}} \ell \xtutt  \text{dt},\label{eq:OPC_allg_2}
	\end{align}
\end{subequations}
involving the Lagrangian $\ell$ and the Mayer term $V$ (cf.\ \cite{Athans2007,Hocking1991} for details).
Thereby, for $t \in \left[0, \, t_\text{f,t}\right]$ the variables $\bm{x}$ and $\bm{u}$ are constrained by the state-space equation \eqref{eq:sys:nonlin} and
\begin{align*}
	\ut &\in   \mathcal{U}, & %
	\xt &\in  \mathcal{X}, & \bm{x}&\tzt \in \mathcal{X}_0\, ,&  \bm{x} \! \left(t_\text{f,t} \right) &\in \mathcal{X}_f\, \text{.}
\end{align*}
Start- and terminal-conditions of the state are taken into account by $\mathcal{X}_0\subset\mathbb{R}^n$ and $\mathcal{X}_f\subset\mathbb{R}^n$ respectively. The state is constrained to the subset $\mathcal{X}\subset\mathbb{R}^n$ and the input to $\mathcal{U}\subset\mathbb{R}^m$.

\begin{figure}[!ht]
	\centering
	\def\svgwidth{0.95\linewidth}
	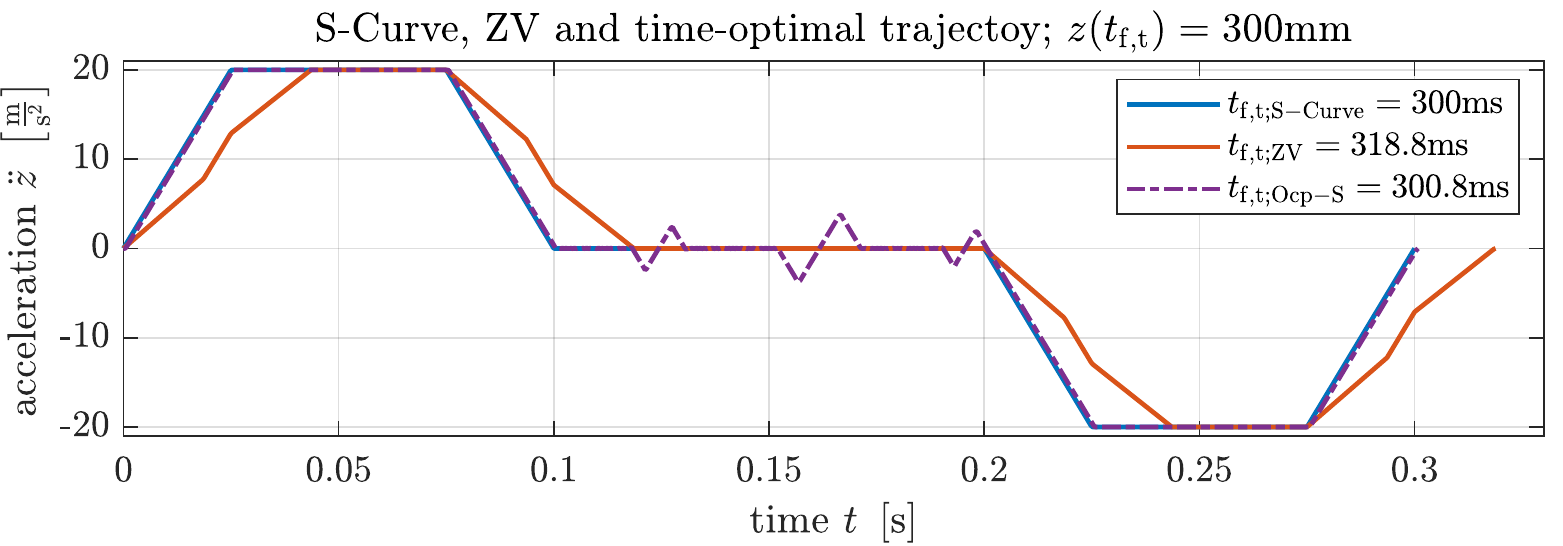 %
	\caption{Trajectories: $\SCurve$, $\ZV$ and $\ocpS$ for a comparison of transition times for the motion distance $z\tftt = \SI{300}{\milli\meter}$.}
	\label{fig:example_optimal_control_sCurve_ZV}
\end{figure}

The main objective of contribution consists in computing time-optimal trajectories or at least nearly time-optimal rest-to-rest trajectories for the system \eqref{eq:opt:sys}. The corresponding optimal control problem \cite{Auer2021_7te_IftommDach} is obtained by using the terminal time of the transition $t_\text{f,t}$ directly as cost 
\begin{equation}
	J\tct = t_\text{f,t},
  \end{equation}
which corresponds to the choice $\ell(\bm{x}(t),\bm{u}(t),t)=1$ and $V(\bm{x}(t_\text{f,t}),t_\text{f,t})=0$
in \eqref{eq:OPC_allg_2}. Moreover, \eqref{eq:sys:nonlin} is replaced by \eqref{eq:sys_state_for_Axbu}, and the initial and terminal constraints are given by
\begin{subequations}
	\begin{align}
		\bm{x} \tzt &= \left[ 0, 0, 0, 0, 0 \right]^T \, \text{,} \\
		\bm{x}\tftt &= \left[ 0, 0, z_\text{f}, 0, 0 \right]^T
	\end{align}
\end{subequations}
with $z_\text{f} = z\tftt$ being the target point for the transition $t_\text{f,t}$.
As a consequence, the complete system including the internal dynamics is at rest at the end of the slider transition as long as $u$ is chosen according to $u\ttt=z^{\left(3\right)}\ttt=0$, $t\ge t_\text{f,t}$. The state variables concerning the slider motion $\dot{z}\ttt$ and $\ddot{z}\ttt$ as well as the input $z^{\left(3\right)}\ttt$ are constrained according to the limits given in \autoref{tab:table_listing_kincont_ex_Pap}. The trajectory presented in \autoref{fig:example_optimal_control_sCurve_ZV} has been calculated using a full discretization approach \cite{Papageorgiou2015,Hargraves1987,Betts1998,TSANG1975, Auer2021_7te_IftommDach} to solve this optimization problem. %
If the optimization problem is formulated for the entire transition and solved in one step, the method is called $\ocpS$ in this contribution. An exemplary trajectory is shown in \autoref{fig:example_optimal_control_sCurve_ZV} and it is compared to the $\SCurve$ trajectory \cite{ClaudioMelchiorri2008} and a trajectory calculated with a $\ZV$-shaper \cite{Singhose1990}. The transition distance of $\SI{300}{\milli\meter}$ shown can be considered a large distance in the context of the pick-and-place machines considered.
The $\SCurve$ represents the fastest transition from a start-point to an end-point, while respecting the kinematic constraints from \autoref{tab:table_listing_kincont_ex_Pap}. However, the excitation of the internal dynamics of the system leads to oscillation and as a result to a loss of accuracy. $\ZV$-shaping is a common method to achieve rest-to-rest transitions. The two main drawbacks of $\ocpS$ are the effort for computing the trajectory (calculation time) and the sensitivity to parameter uncertainties \cite{Auer2023IFAC}. For this reason \cite{Auer2023IFAC} included a comparison to a method called $\ocpJ$. In $\ocpJ$, the trajectory for the full transition is assembled from preplanned motion primitives. A comparison to existing trajectory planning methods is shown in \autoref{sec:comparison_to_regular} (\autoref{fig:exampleDist_SC_ZV_ocpS_ocpJ_Fir_exempl_pap_param_distLong} in \autoref{sec:comparison_to_regular} expands the comparison shown in \autoref{fig:example_optimal_control_sCurve_ZV} to also include the $\ocpJ$ and $\FirImp$ trajectory) and in greater detail also in \cite{Auer2023IFAC}. This contribution explains the algorithms and mathematical background used for this second method $\left(\ocpJ\right)$ in detail and highlights the advantages.

\section{Method explained: $\ocpJ$}\label{sec:calc_full_trajectory}
\begin{figure}[!ht]
	\centering
	\def\svgwidth{0.95\linewidth}
	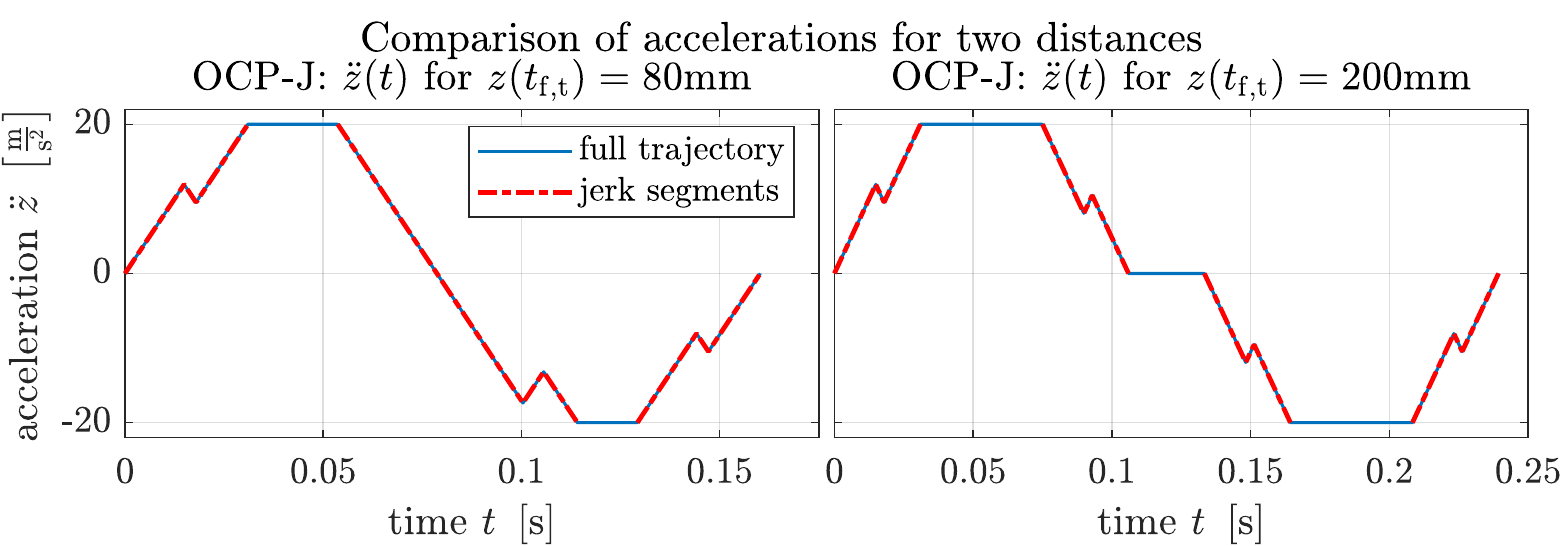 %
	\caption{Two $\ocpJ$ trajectories illustrating the method (picture taken from \cite{Auer2023IFAC}).} %
	\label{fig:two_OcpJ_trajectories_for_demo_of_method}
\end{figure}

This section provides an overview over $\ocpJ$. Instead of solving an optimization problem for the whole trajectory like $\ocpS$, the $\ocpJ$ method assembles trajectories from motion primitives, some of them are the solution of smaller optimization problems. The proposed method is strongly inspired my the standard $\SCurve$ (cf.\ \cite{ClaudioMelchiorri2008}). However, in contrast to the latter, which constitutes the time-optimal trajectory when neglecting the internal dynamics in \eqref{eq:opt:sys}, $\ocpJ$ does not result in an exact solution of the optimal-control problem \eqref{Eq_OPC_allg_1}. For fixed kinematic constraints and given plant parameters, the particular sequence of motion primitives depends on the terminal position $z_\text{f}$. Exemplary trajectories planned with this method are shown in \autoref{fig:two_OcpJ_trajectories_for_demo_of_method}.
The trajectory on the left represents a rather short transition distance, where only the maximum acceleration $a_\text{lim}$ is reached, but not the maximum velocity $v_\text{lim}$. In contrast, the trajectory on the right corresponds to a larger distance, where both $a_\text{lim}$ and $v_\text{lim}$ are reached. In both trajectories the segments coloured in red are crucial. These motion primitives are called jerk segments in this paper and are designed such that the internal dynamics at the beginning and at the end of these segments are at rest. A step-by-step explanation of how these are calculated using a direct method that does not rely on optimization can be found in \cite{tau_ocpJ_assembly_part2}. This results in an optimization problem similar to the one introduced in the previous section.  The initial conditions and terminal constraints at the free end time $t=t_{\text{f}}$ are given by
\begin{subequations}\label{eq:BC_for_zero_endOscillation_xp_xpp_xppp}
  \begin{alignat}{4}
    \ddot{x}(0)&=-k^{\star}x(0) - m^\star\ddot{z}(0)&&=0,\quad&	\dot{x}(0)& = 0\label{eq:BC_for_zero_endOscillation_xp_xpp_xppp:1}\\
    \ddot{x}\tft&=-k^{\star}x\tft-m^\star\ddot{z}\tft&&=0,\quad&	\dot{x}\tft &= 0.\label{eq:BC_for_zero_endOscillation_xp_xpp_xppp:2}
  \end{alignat}
\end{subequations}
\noeqref{eq:BC_for_zero_endOscillation_xp_xpp_xppp:1}\noeqref{eq:BC_for_zero_endOscillation_xp_xpp_xppp:2}%
Moreover, within this contribution the following initial and terminal constraints for the acceleration segments are considered
\begin{equation}\label{eq:BC_slider_zpp_zppp}
	\ddot{z}\tft,\ddot{z}(0)\in\{-a_\text{max},0,a_\text{max}\},\quad \ddot{z}(0)\ne\ddot{z}\tft.%
\end{equation}
Moreover, the terminal velocity and terminal position are unconstrained while initial values $z(0)$ and $\dot z(0)$ may be specified. Above, the maximal acceleration $a_\text{max}$ does not necessarily coincide with\footnote{Note that, due to the particular construction of the trajectories it can be beneficial to choose $a_\text{max}<a_\text{lim}$ where $a_\text{lim}$ is the limit according to the specification. This is emphasized in \autoref{SubSec:adjusting_aMax_to_removing_jerkViolations}.} $a_\text{lim}$. 

\begin{figure}[!ht]
	\centering
	\def\svgwidth{0.95\linewidth}
	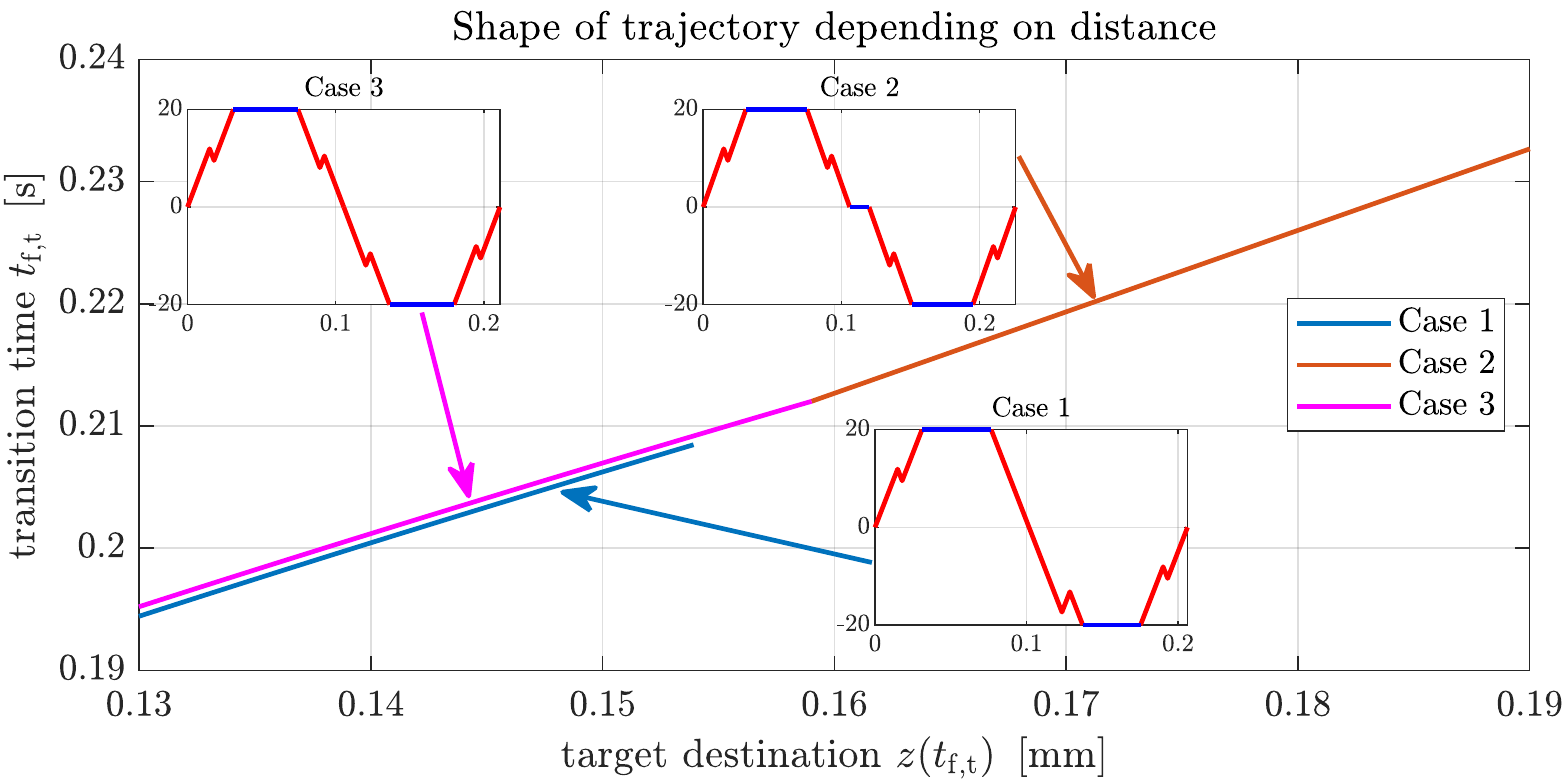 %
	\caption{Trajectories for multiple distances and regions associated with different cases.} %
	\label{fig:OcpJ_trajectories_showing_regions_of_methods}
\end{figure}

By choosing the duration the slider moves at constant acceleration (resp.\ velocity) appropriately, the total transition distance can be adjusted (cf.\ \autoref{fig:two_OcpJ_trajectories_for_demo_of_method}). %
Within this contribution several cases requiring different calculations for trajectory planning are considered. Their dependance on the final position $z_\text{f}$ is illustrated in \autoref{fig:OcpJ_trajectories_showing_regions_of_methods} and characterized as follows:
\begin{itemize}
	\item \underline{Case~1:} A trajectory, where $a_\text{max}$ is reached, but $v_\text{lim}$ is not reached and there is only one jerk segment to get from $a_\text{max}$ to $-a_\text{max}$.
	\item \underline{Case~2:} A trajectory, where $a_\text{max}$ and $v_\text{lim}$ are reached.
	\item \underline{Case~3:} A trajectory, where $a_\text{max}$ is reached and $v_\text{lim}$ is not reached. This case exists for distances to large for Case~1, but to small for Case~2. It is rare and only required for certain distances as shown in \autoref{fig:OcpJ_trajectories_showing_regions_of_methods}. This special case is considered in \autoref{SubSec:Case3_description_of_calculation}.
\end{itemize}

\subsection{Preprocessing of jerk segments}\label{SubSec:Preparation_Values_full_traject}
As \autoref{alg:get_ocp_ind_traject} describes, after calculating the jerk segments for the given $a_\text{max}$, the velocity differences $\left(v_\text{f1, f2, f3}\right)$ and some travel distances $\left(s_\text{f1, f2, f3}\right)$ for the segments themselves have to be calculated. Those values are required when computing the full trajectories. The jerk segments and the corresponding parameters are highlighted in \autoref{fig:showing_ocpJ_dividedSegments_methodOne} as a part of a particular trajectory. Although this trajectory corresponds to Case~1, the parameters computed also apply the remaining cases. %
The jerk $z^{\left(3\right)}\ttt$ of the jerk segments is piece-wise constant, as depicted in \autoref{fig:two_OcpJ_trajectories_for_demo_of_method} and shown in \cite{tau_ocpJ_assembly_part2}. As a consequence, it can be written in the form \eqref{eq:trajectory:ansatz}. Therefore, the corresponding velocity and the position result from \eqref{eq:calc_z_to_zppp_from_snap}. 
Following this procedure, the parameters $s_\text{f1}$ and $v_\text{f1}$, associated with the first jerk segment, are given by
\begin{alignat*}{2}
		s_\text{f1} &= z\!\left(t_\text{f1}\right) \, \text{,}\quad & v_\text{f1} &= \dot{z}\!\left(t_\text{f1}\right),
\end{alignat*}
where $t\mapsto z(t)$ and $t_\text{f1}$ 
can be computed by minimizing $t_\text{f1}$ subject to the given kinematic constraints and the system dynamics \eqref{eq:opt:sys}. The latter is complemented by the initial and final conditions, given by \eqref{eq:BC_for_zero_endOscillation_xp_xpp_xppp} and
\begin{alignat*}{4}
            z(0) &= 0 \, \text{,}\quad & \dot{z}(0) &= 0 \, \text{,} &\quad \ddot{z}(0) &= 0 \,\text{,}\quad& \ddot{z}\left(t_\text{f1}\right) &= a_\text{max} \text{.}
	\end{alignat*}
        Moreover, $s_\text{f2}=z\!\left(t_\text{f2}\right)$, $v_\text{f2}=\dot{z}\!\left(t_\text{f2}\right)$ are obtained in the same way for initial and terminal conditions 
	\begin{alignat*}{4}
            z(0) &= 0 \, \text{,}\quad & \dot{z}(0) &= 0 \, \text{,} &\quad \ddot{z}(0) &= a_\text{max}  \,\text{,}\quad& \ddot{z}(t_\text{f2}) &= -a_\text{max} \text{.}
	\end{alignat*}
Finally, $t_\text{f3}$, $s_\text{f3}=z(t_\text{f3})$, and $v_\text{f3}=-\dot z(t_\text{f3})$ follow from the boundary conditions
\begin{subequations}
	\begin{align}
            z(0) &= 0 \, \text{,} & \dot{z}(0) &= 0 \, \text{,} & \ddot{z}(0) &= -a_\text{max}  \,\text{,}& \ddot{z}\left(t_\text{f2}\right) &= 0 \text{.}
	\end{align}
\end{subequations}

\subsection{Case~1}\label{SubSec:Case1_description_of_calculation}%
\begin{figure}[!ht]
	\centering
	\def\svgwidth{0.95\linewidth}
	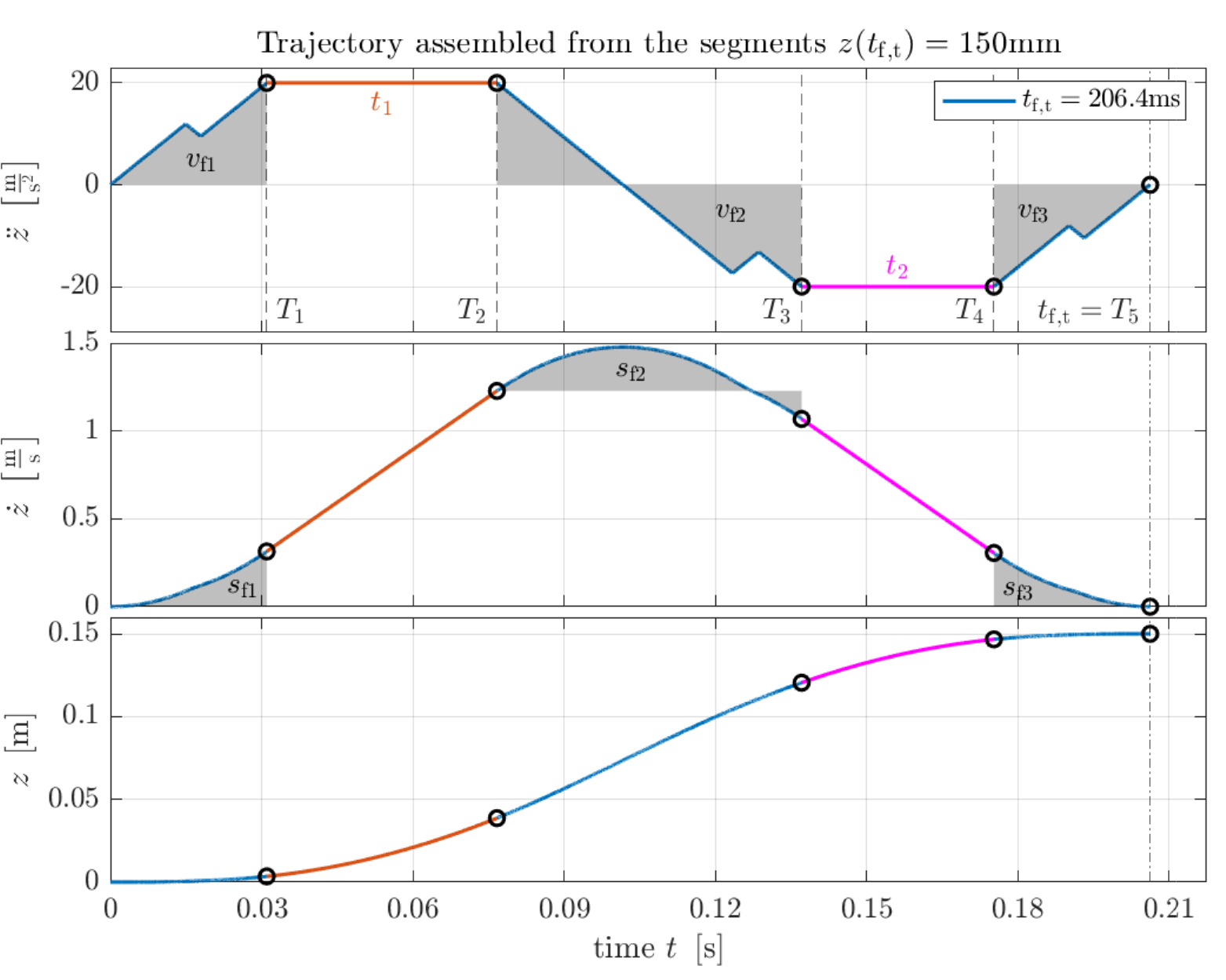 %
	\caption{$\ocpJ$ trajectory for Case~1.}
	\label{fig:showing_ocpJ_dividedSegments_methodOne}
\end{figure}

The general shape of the solution is shown in \autoref{fig:showing_ocpJ_dividedSegments_methodOne}. In order to reach the desired final position with the slider, the jerk segments have to be connected by segments with constant acceleration. In particular, acceleration $\ddot{z}\ttt$ remains at $a_\text{max}$ during the intervals $\left[T_1, \, T_2\right]$ of duration $t_1$ and at $-a_\text{max}$ during the interval $\left[T_3, \, T_4\right]$ of duration $t_2$ as shown in \autoref{fig:showing_ocpJ_dividedSegments_methodOne}. 
To compute the durations $t_1$ and $t_2$, the precomputed values $s_\text{f1, f2, f3}$ and $v_\text{f1, f2, f3}$, as explained in \autoref{SubSec:Preparation_Values_full_traject}, are employed. As mentioned previously, the times $t_\text{f1,f2,f3}$ are the terminal times of the individual jerk segments and they satisfy
\begin{align}
	t_\text{f1} &= T_1 \, \text{,} & t_\text{f2} &= T_3 - T_2 \, \text{,} & t_\text{f3} &= t_\text{f1} = T_5 - T_4 \, \text{.}
\end{align}
The velocities at the concatenation points $T_1, \dots, T_5$ can be derived to be
\begin{subequations}
	\begin{align}
		v \! \left(T_1\right) &= v_\text{f1} \, \text{,} \\
		v \! \left(T_2\right) &= v_\text{f1} + {t_1} \cdot a_\text{max} \, \text{,} \\
		v \! \left(T_3\right) &= v_\text{f1} + {t_1} \cdot a_\text{max} + v_\text{f2} \, \text{,} \\
		v \! \left(T_4\right) &= v_\text{f1} + {t_1} \cdot a_\text{max} + v_\text{f2} - {t_2} \cdot a_\text{max} \, \text{,} \\
		v \! \left(T_5\right) &= v_\text{f1} + {t_1} \cdot a_\text{max} + v_\text{f2} - {t_2} \cdot a_\text{max} - v_\text{f3} \, \text{.} \label{eq:vT5_case1}
	\end{align}
\end{subequations}
Moreover, the respective positions are given by
\begin{subequations}
	\begin{align}
		s \! \left(T_1\right) &= s_\text{f1} \, \text{,} \\
		s \! \left(T_2\right) &= s_\text{f1} + \int\limits_{0}^{{t_1}} v\!\left(\tau\right) \text{d}\tau = s_\text{f1} + {t_1} \cdot v_\text{f1} + \frac{1}{2} \, a_\text{max} \cdot {t_1^2} \, \text{,} \\
		s \! \left(T_3\right) &= s \! \left(T_2\right) + \left(v_\text{f1} + t_1 \cdot a_\text{max} \right) \cdot t_\text{f2} + s_\text{f2} \, \text{,} \\
		s \! \left(T_4\right) &= s \! \left(T_3\right) + \left(v_\text{f1} + t_1 \cdot a_\text{max} + v_\text{f2} \right) \cdot t_\text{2} - \frac{1}{2} \, a_\text{max} \cdot t_2^2 \, \text{,} \\
		s \! \left(T_5\right) &= s \! \left(T_4\right) + s_\text{f3} = z\tftt \, \text{.} 
	\end{align}
\end{subequations}
Therefore, using $s_\text{f} = s_\text{f1} + s_\text{f2} + s_\text{f2}$, the final position is:
\begin{align}
	\textcolor{\colorZtft}{z\tftt} &= s_\text{f} + v_\text{f1} \, t_\text{f2} + \textcolor{\colorTOne}{t_1} \! \left(v_\text{f1} +a_\text{max} \!\left(t_\text{f2} + \textcolor{\colorTTwo}{t_2}\right)\right) +...\\
	& \qquad \textcolor{\colorTOne}{t_1^2}\frac{a_\text{max}}{2} + \textcolor{\colorTTwo}{t_2} \! \left(v_\text{f1}+ v_\text{f2}\right) - \textcolor{\colorTTwo}{t_2^2}\frac{a_\text{max}}{2} \, \text{.} \label{eq:first_eq_to_calc_t1_and_t2}
\end{align}
From \eqref{eq:vT5_case1} and $v \! \left(T_5\right)=0$, it follows, that
\begin{equation}\label{eq:second_eq_to_calc_t1_and_t2}
	v \! \left(T_5\right) = 0 = v_\text{f1} + \textcolor{\colorTOne}{t_1} \cdot a_\text{max} + v_\text{f2} - \textcolor{\colorTTwo}{t_2} \cdot a_\text{max} - v_\text{f3}.
\end{equation}
Solving \eqref{eq:second_eq_to_calc_t1_and_t2} for $t_2$ and inserting into \eqref{eq:first_eq_to_calc_t1_and_t2} leads to
\begin{equation}
	0 = \textcolor{\colorTOne}{t_1^2} + p \cdot \textcolor{\colorTOne}{t_1} + q
\end{equation}
with
\begin{subequations}
	\begin{align}
		p &= \frac{a_\text{max} \, t_\text{f2} + 2 \, v_\text{f1} + v_\text{f2}}{a_\text{max}} \, \text{,} \\
		q &= -\frac{\textcolor{\colorZtft}{z\tftt}}{a_\text{max}} + \frac{s_\text{f1} +s_\text{f2} +s_\text{f3} + v_\text{f1} \, t_\text{f2}}{a_\text{max}} +...\\
		& \qquad \frac{\left(v_\text{f1}+ v_\text{f2}+ v_\text{f3}\right) \cdot \left(v_\text{f1}+ v_\text{f2}- v_\text{f3}\right)}{2 \, a_\text{max}^2} \, \text{.}
	\end{align}
\end{subequations}
This allows to solve for $t_1$ directly. From the two possible solutions for $t_1$, the smaller one, which is always
negative, is ignored.
Therefore, the time $\textcolor{\colorTOne}{t_1}$ is calculated with:
\begin{equation}\label{eq:case1_calculate_t1}
	\textcolor{\colorTOne}{t_1} = -\frac{p}{2} + \sqrt{\frac{p^2}{4} - q}.
\end{equation}
For small transition distances $q$ may be negative which results in negative values for $t_1$. This leads to overlapping jerk segments, which, however, can be a viable solution. A detailed discussion of this case, which is illustrated in \autoref{fig:showing_overlapping_1p5mm} and \autoref{fig:showing_overlapping_200mm}, can be found in \autoref{SubSec:adjusting_aMax_to_removing_jerkViolations}. Having computed $t_1$, the duration $t_2$ follows immediately from \eqref{eq:second_eq_to_calc_t1_and_t2}. The resulting trajectory is checked and it is calculated, if one of the other cases (Case~2 or Case~3) would be the actual solution. This is explained in further detail in \autoref{alg:get_ocp_ind_traject} in \hyperref[App:alg_for_implementation]{Appendix~\ref*{App:alg_for_implementation}}.

\subsection{Case~2}\label{SubSec:Case2_description_of_calculation}%
\begin{figure}[!ht]
	\centering
	\def\svgwidth{0.95\linewidth}
	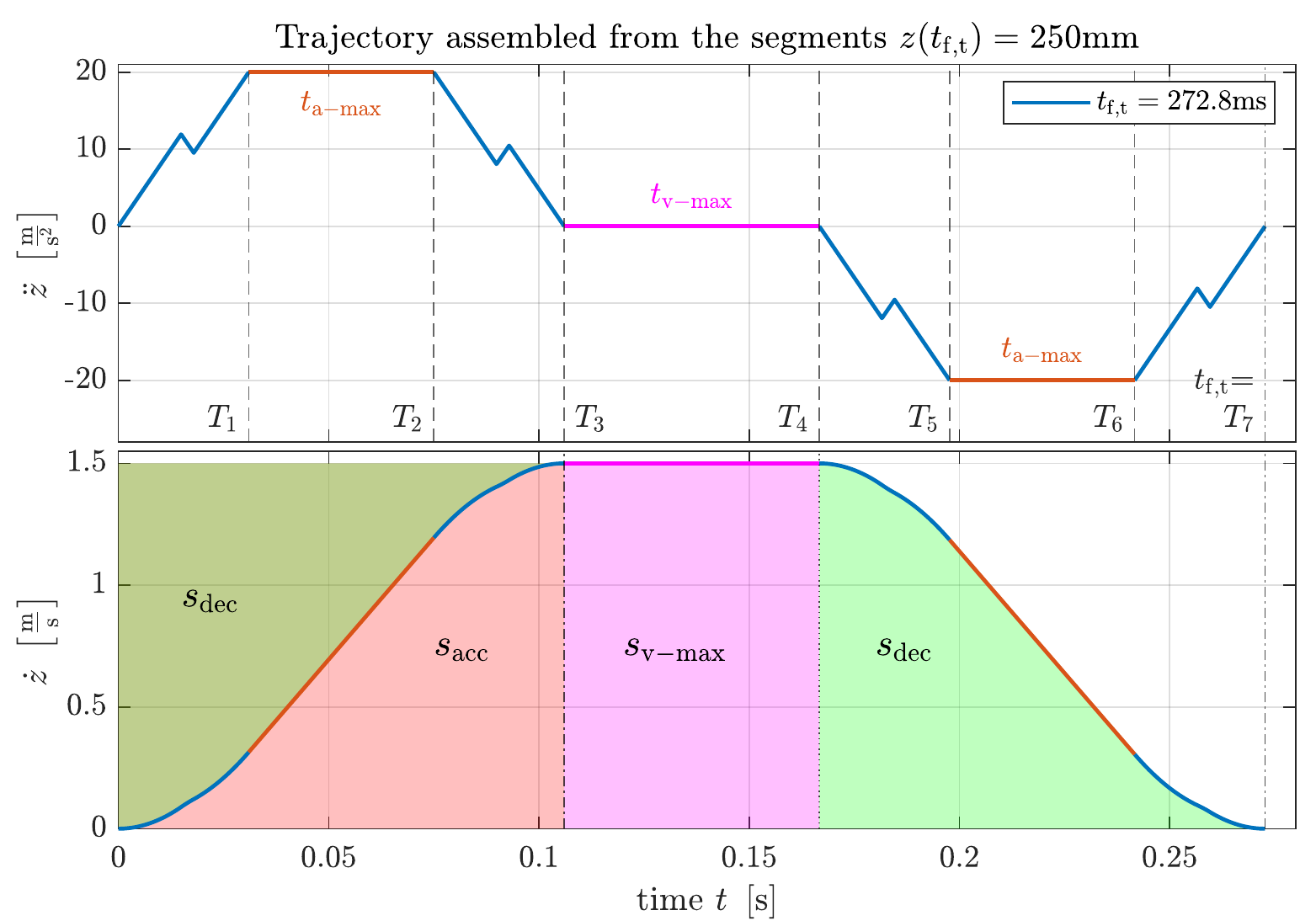 %
	\caption{$\ocpJ$ trajectory for Case~2.}%
	\label{fig:showing_ocpJ_dividedSegments_methodTwo}
\end{figure}

This case has to be considered, if the kinematic velocity constraint would be violated in Case~1 (see \autoref{alg:get_ocp_ind_traject} in \hyperref[App:alg_for_implementation]{Appendix~\ref*{App:alg_for_implementation}}). The general shape of the solution is shown in \autoref{fig:showing_ocpJ_dividedSegments_methodTwo}. It consists of overall seven individual segments, and can be divided into an acceleration phase on the interval $\left[0,T_3\right]$, followed by a constant velocity phase on $\left[T_3,T_4\right]$ with duration $t_\text{v-max}=T_4-T_3$, and, finally, a deceleration phase on  $\left[T_4,T_7\right]$.
Therein the acceleration phase is composed of three segments, a jerk segment on the interval $\left[0,T_1\right]$, a segment of constant acceleration on $\left[T_1,T_2\right]$ and another jerk segment on $\left[T_2,T_3\right]$. Similarly, the deceleration phase consists of a jerk segment on the interval $\left[T_4,T_5\right]$, a segment of constant acceleration on $\left[T_5,T_6\right]$ and another jerk segment on $\left[T_6,T_7\right]$.

First, the time at maximal acceleration $a_\text{max}$ to reach $v_\text{lim}$ results from
\begin{equation}\label{eq:vMax_for_case_2}
	v_\text{lim} = v_\text{f1} +v_\text{f3} + a_\text{max} \, t_\text{a-max},
      \end{equation}      
with the abbreviation $t_\text{a-max}=T_{2}-T_1=T_6-T_5$ and the velocity differences $v_\text{f1}$ and $v_\text{f3}$ according to \autoref{SubSec:Preparation_Values_full_traject}.
This yields
\begin{equation}\label{eq:Case2_t_aMax_calculation}
	t_\text{a-max} = \frac{v_\text{lim} - \left(v_\text{f1} +v_\text{f3}\right)}{a_\text{max}} \, \text{.}
\end{equation}
Note that high acceleration limits $a_\text{max}$ may imply $v_\text{f1} +v_\text{f3} > v_\text{lim}$ which would result in $t_\text{a-max}<0$ and, therefore, intersecting segments, i.e., $T_2<T_1$. This is discussed later in this section (cf.\ also \autoref{fig:showing_overlapping_200mm} in \autoref{SubSec:adjusting_aMax_to_removing_jerkViolations}).
From the considerations in \autoref{SubSec:Preparation_Values_full_traject} the symmetry
\begin{equation}
	\ddot{z}\!\left(t+T_4\right) = -\ddot{z}\ttt \, \text{,} \qquad t\in\left[0,T_3\right]
\end{equation}
can be deduced, which, together with $\dot{z}(T_4)=v_{\text{lim}}$, implies
\begin{subequations}\label{eq:case2:symmetries}
	\begin{align}
		\dot{z}\!\left(t+T_4\right) &= %
                                v_\text{lim} - \dot{z}\ttt\label{eq:case2:vaccel} \\
		z\!\left(t+T_4\right) &= z(T_4)+t v_\text{lim}-z(t),\quad t\in[0,T_3] \text{.}\label{eq:case2:saccel}
	\end{align}\noeqref{eq:case2:saccel}\noeqref{eq:case2:vaccel}%
\end{subequations}%
Denote by $s_\text{acc}=z\!\left(T_3\right)$ and $s_\text{dec}=z\!\left(T_4+T_3\right)-z\!\left(T_4\right)$ the total distances travelled during the acceleration and deceleration phases, respectively. According to \eqref{eq:case2:saccel} these distances satisfy
\begin{equation}
	s_\text{acc} + s_\text{dec} = v_\text{lim} \, T_3=v_\text{lim} \cdot \left(2 \cdot t_\text{f1} + t_\text{a-max}\right) \, \text{.}
      \end{equation}
For the sake illustration, this relation and the underlying symmetries \eqref{eq:case2:symmetries} have been 
visualized in \autoref{fig:showing_ocpJ_dividedSegments_methodTwo}.
As a consequence of the above, the final position $z\tftt$ reached is given by
\begin{equation}
	\textcolor{\colorZtft}{z\tftt} = v_\text{lim} \, t_\text{v-max} + v_\text{lim} \cdot \, \left(2 \cdot t_\text{f1} + t_\text{a-max}\right) \, \text{.}
\end{equation}
Therefore, for a given target position, the time spent at maximum velocity $v_\text{lim}$ reads
\begin{equation}\label{eq:Case2_t_vMax_calculation}
	t_\text{v-max} = \frac{\textcolor{\colorZtft}{z\tftt} - v_\text{lim} \cdot \, \left(2 \cdot t_\text{f1} + t_\text{a-max}\right)}{v_\text{lim}} \, \text{.}
\end{equation} 

\subsection{Case~3}\label{SubSec:Case3_description_of_calculation}%
There are certain situations (travelling distances) where neither Case~1 nor Case~2 applies. For these distances the velocity constraint is violated on the interval $\left[T_2,T_3\right]$ when using Case~1. For the same distances the maximal velocity is not reached when using Case~2 which results in $t_{\text{v-lim}}<0$.
The particular travelling distances (for the parameter set and kinematic constraints used in this publication) corresponding to this situation are shown in \autoref{fig:OcpJ_trajectories_showing_regions_of_methods}.

\begin{figure}[!ht]
	\centering
	\def\svgwidth{0.95\linewidth}
	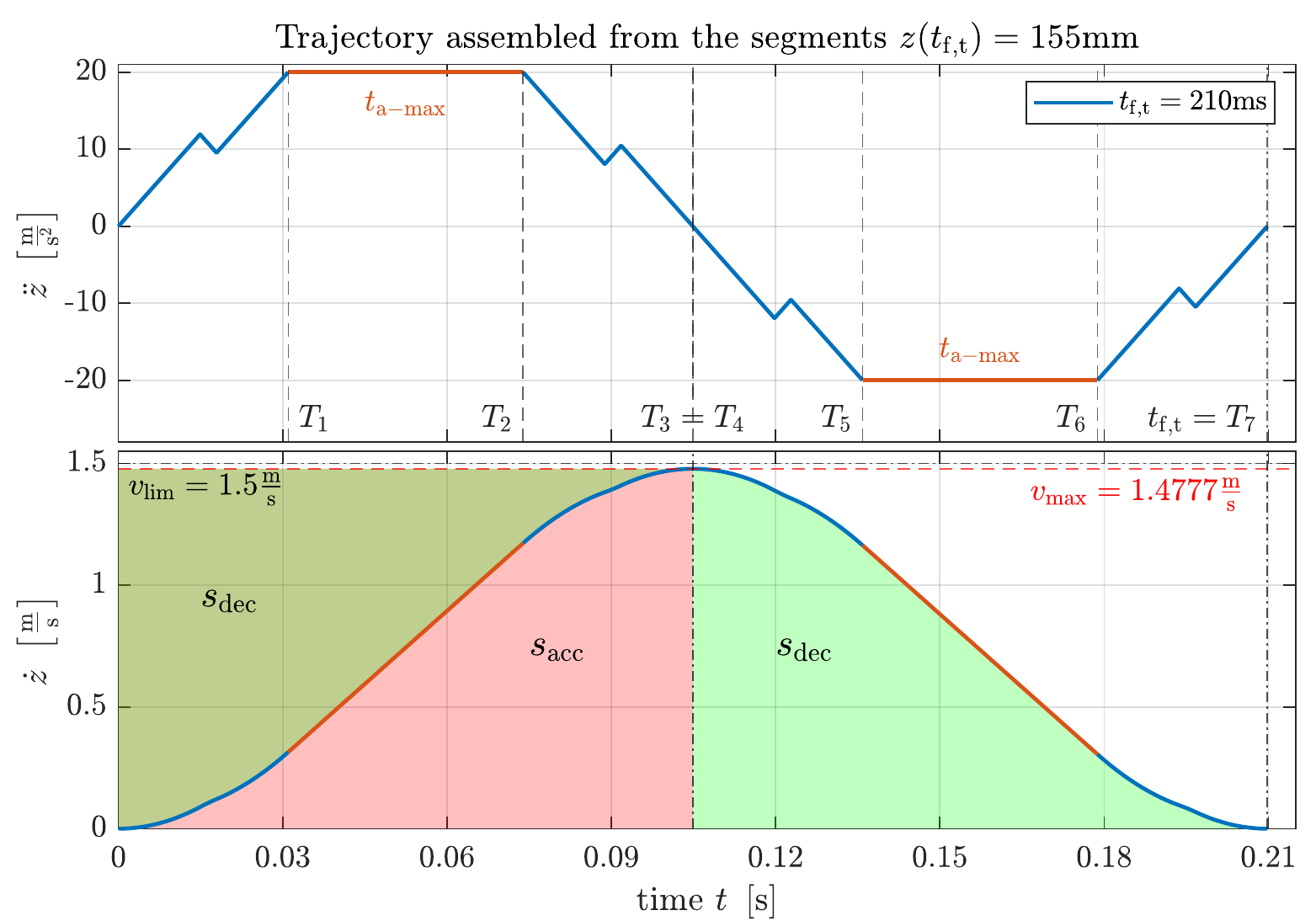 %
	\caption{$\ocpJ$ trajectory for Case~3.} %
	\label{fig:showing_ocpJ_dividedSegments_methodThree}
\end{figure}

In the described situation Case~3 is used for trajectory design. The general shape of the trajectory for this case is shown in \autoref{fig:showing_ocpJ_dividedSegments_methodThree}. It consists of overall six individual segments, and can be divided into an acceleration phase on the interval $\left[0,T_3\right]$, directly followed by the deceleration phase on $\left[T_4,T_7\right]$, whereas $T_3=T_4$. Similarly to Case~2 these phases are composed of two jerk segments connected by a constant-acceleration segment. In contrast to Case~2, maximum velocity $v_{\text{lim}}$ does not need to be reached. Instead, the duration $T_{2}-T_1=T_6-T_5=t_\text{a-max}$ of the constant-acceleration segments are computed such that the final position $z\tft$ is reached (cf.\ \autoref{fig:showing_ocpJ_dividedSegments_methodThree}).
As a prerequisite, the distances travelled during the acceleration and deceleration can be computed by employing the
same symmetries as in Case~2 (cf.\ \autoref{fig:showing_ocpJ_dividedSegments_methodThree}). With $v_{\text{max}}=\dot{z}\!\left(T_3\right)$ the maximum velocity reached, this yields
\begin{subequations}
	\begin{align}
		\textcolor{\colorZtft}{z\tftt} &= v_{\text{max}}\cdot \left(2 \cdot t_\text{f1} + t_\text{a-max}\right)  \\
		 &= \left(v_\text{f1} + v_\text{f3} +a_\text{max} \, t_\text{a-max} \right) \cdot \left(2 \cdot t_\text{f1} + t_\text{a-max}\right) \, \text{.}
	\end{align}
\end{subequations}
Using $v_\text{f} = v_\text{f1}+v_\text{f3}$ allows to rewrite
\begin{equation}
	0 = t_\text{a-max}^2 + t_\text{a-max} \cdot \underbrace{\frac{2 \, a_\text{max} \, t_\text{f1} + v_\text{f}}{a_\text{max}}}_{p} + \underbrace{\frac{2 \, t_\text{f1} \cdot v_\text{f} - \textcolor{\colorZtft}{z\tftt}}{a_\text{max}}}_{q} \, \text{.}
\end{equation}
Determining $t_\text{a-max}$ by choosing the positive branch of the square-root solution yields
\begin{equation}\label{eq:case3_calculate_t_aMax}
	t_\text{a-max} = -\frac{p}{2} + \sqrt{\frac{p^2}{4} - q} \, \text{.}
\end{equation}
As in Case~1, $q>0$ leading to $t_\text{a-max}<0$ is not explicitly excluded.

\subsection{Violation of jerk constraints}\label{SubSec:adjusting_aMax_to_removing_jerkViolations}%
\begin{figure}[!ht]
	\centering
	\def\svgwidth{0.95\linewidth}
	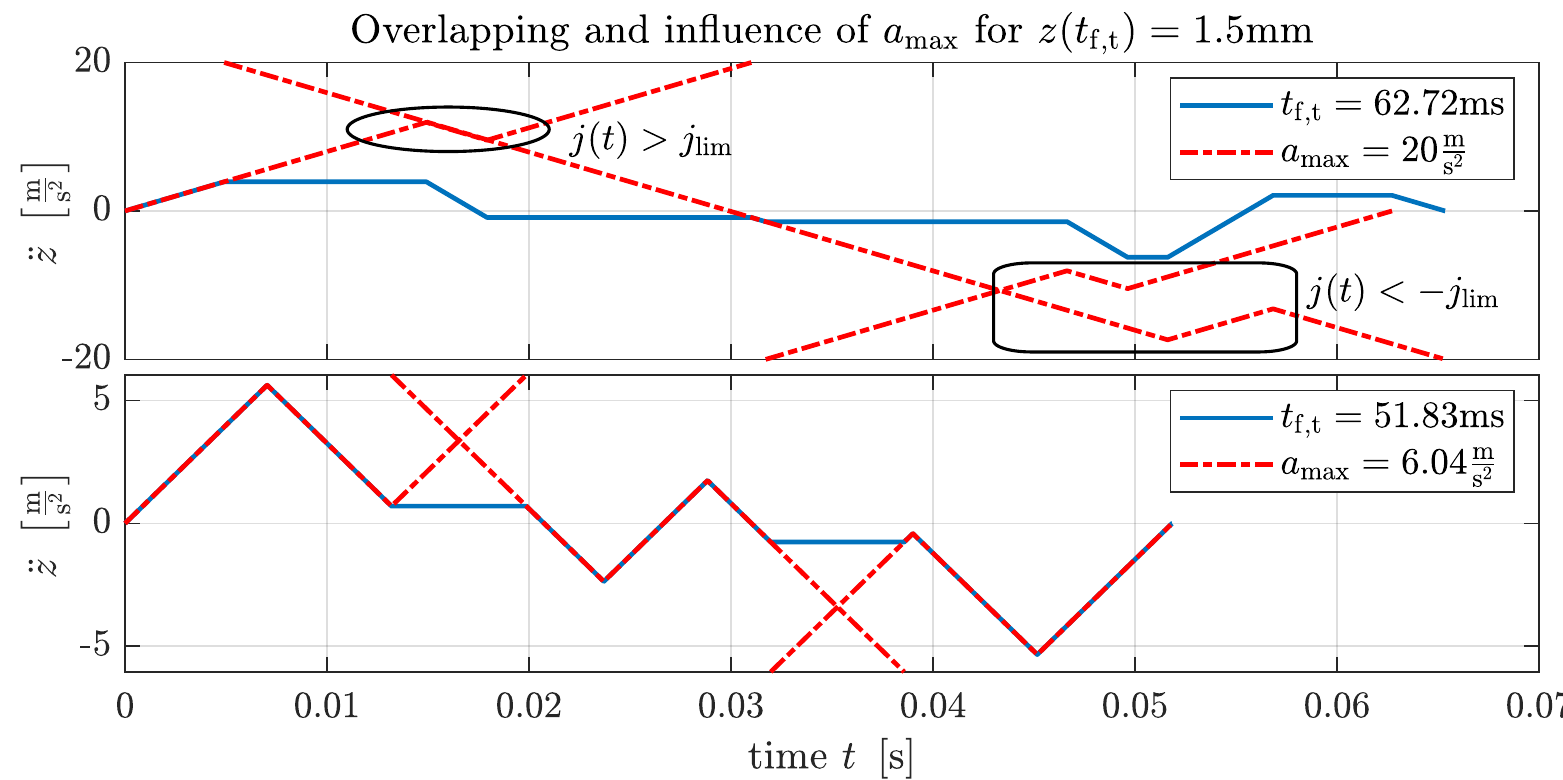 %
	\caption{Superposition of intersecting jerk segments in Case~1. Top: original maximum acceleration causes a violation of the jerk constraint. Bottom: Reduced maximum acceleration avoids violation of the jerk constraint and reduces $t_\text{f,t}$.}
	\label{fig:showing_overlapping_1p5mm}
\end{figure}

\begin{figure}[!ht]
	\centering
	\def\svgwidth{0.95\linewidth}
	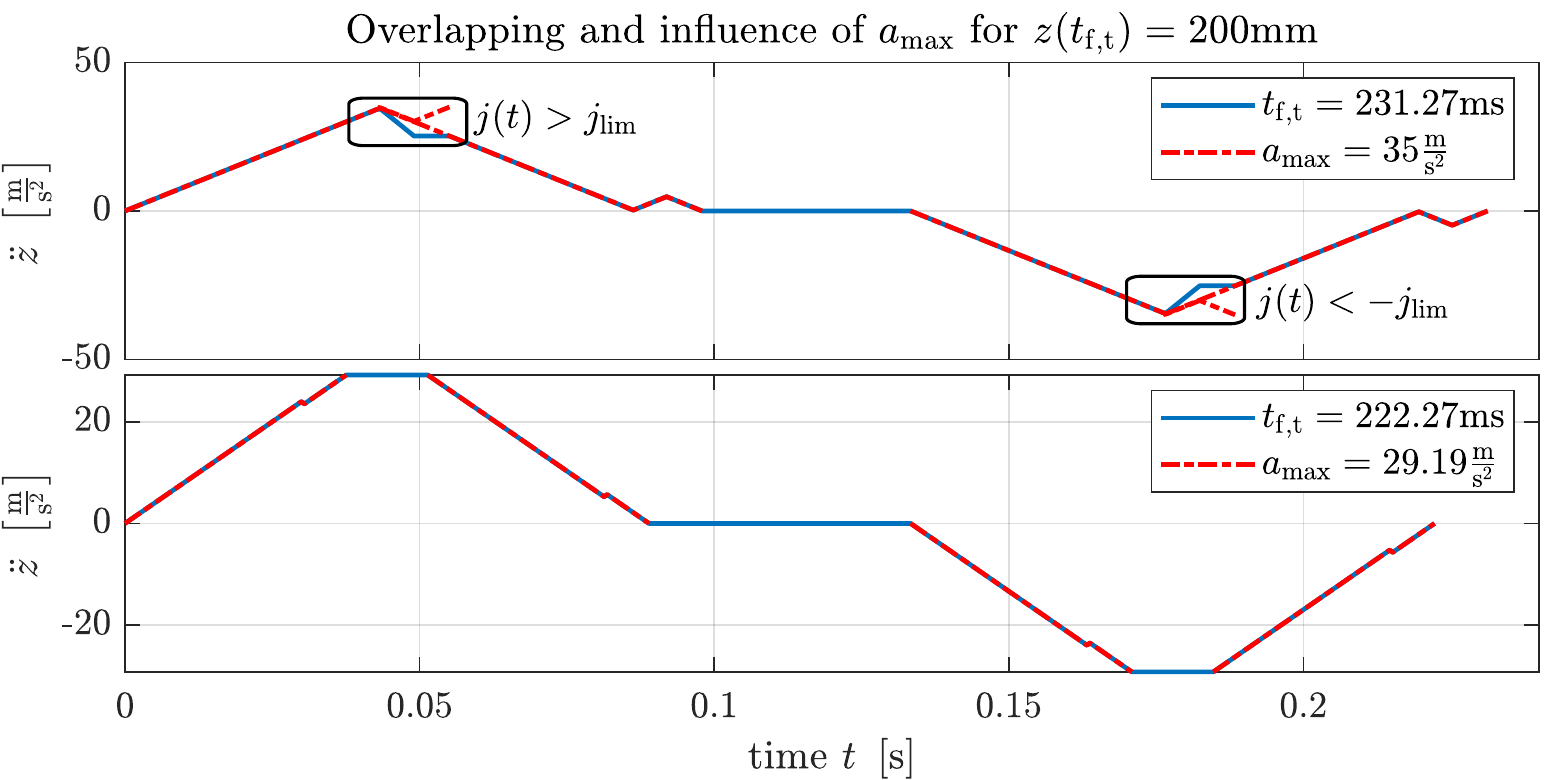 %
	\caption{Superposition of intersecting jerk segments in Case~2. Top: original maximum acceleration causes a violation of the jerk constraint. Bottom: Reduced maximum acceleration avoids violation of the jerk constraint and reduces $t_\text{f,t}$.}
	\label{fig:showing_overlapping_200mm}
\end{figure}

\begin{figure}[!ht]
	\centering
	\def\svgwidth{0.95\linewidth}
	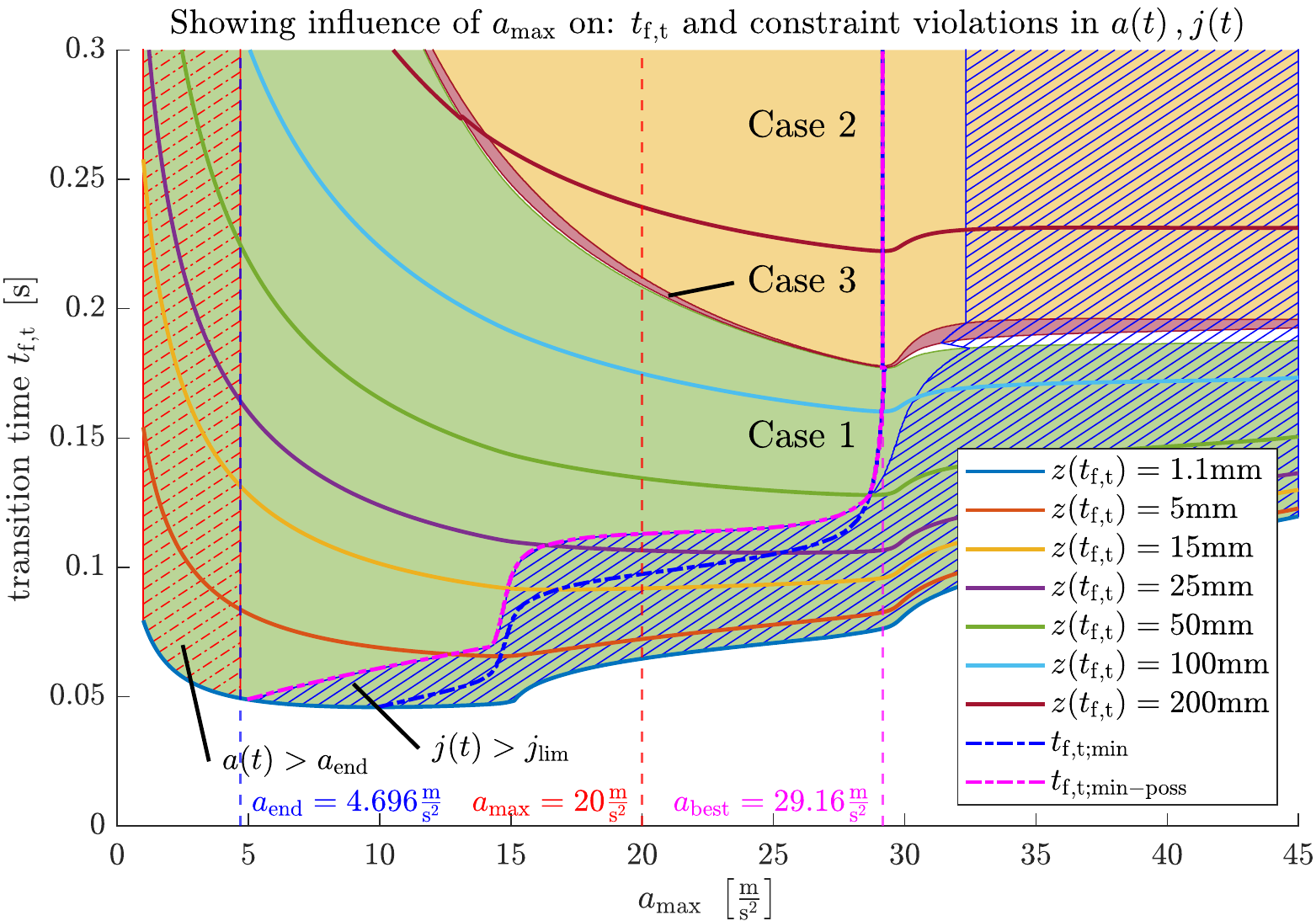 %
	\caption{Influence of $a_\text{max}$ on $t_\text{f,t}$.}
	\label{fig:showing_why_efficient_method_ocpJ_required_amax}
\end{figure}

\begin{figure}[!ht]
	\centering
	\def\svgwidth{0.95\linewidth}
	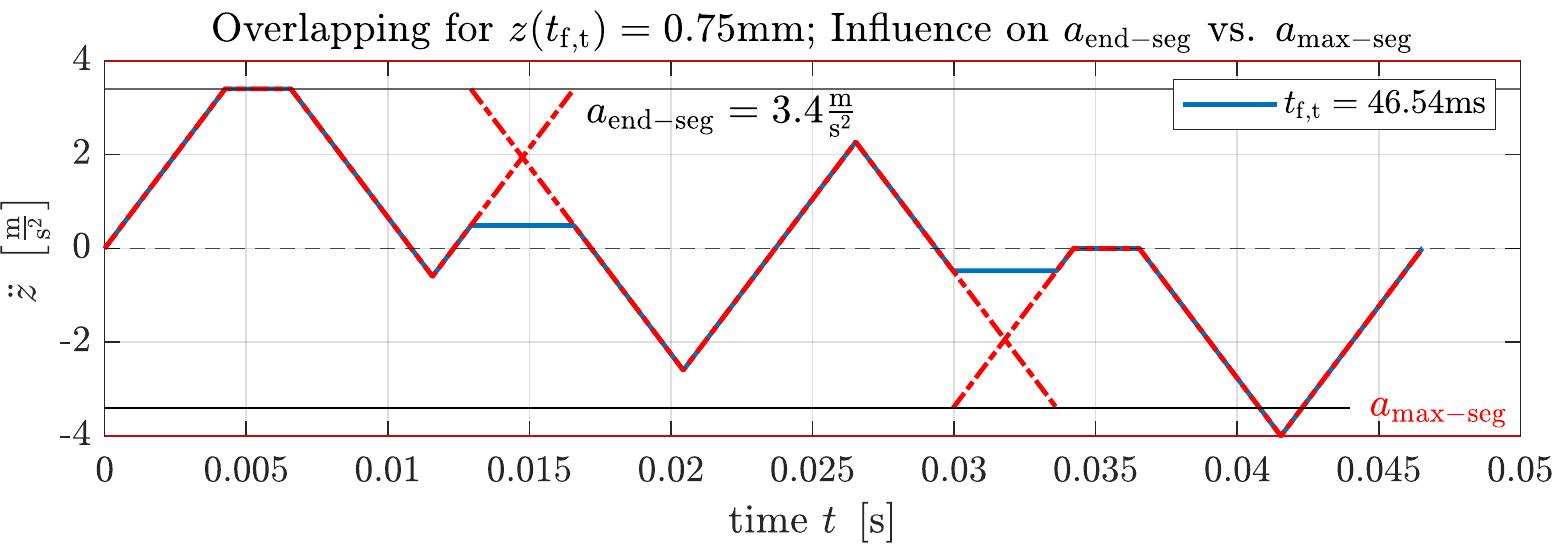 %
	\caption{Trajectory showing, that $a\ttt < a_\text{lim}=\SI{20}{\meter\per\second\squared}$, even though $a\ttt >a_\text{max}$.}
	\label{fig:showing_overlapping_0p75mm}
\end{figure}

As discussed in \autoref{SubSec:Case1_description_of_calculation}, \autoref{SubSec:Case2_description_of_calculation} and \autoref{SubSec:Case3_description_of_calculation} the jerk segments may intersect in certain situations and they are allowed to. However, for particular parameters this overlap may cause a violation of the jerk constraints as shown in \autoref{fig:showing_overlapping_1p5mm} and \autoref{fig:showing_overlapping_200mm}. This subsection analyses these violations in more detail and shows how they can be avoided.
A reduction of $a_\text{max}$ is a possibility to avoid these violations, as shown in \autoref{fig:showing_overlapping_1p5mm} and \autoref{fig:showing_overlapping_200mm}. This is outlined in \autoref{alg:Optimize_aMAx}. For the considered parameter set, a detailed analysis of the influence of the maximum acceleration $a_\text{max}$ on the transition time and the violation of the jerk constraint is shown in \autoref{fig:showing_why_efficient_method_ocpJ_required_amax}. Regions corresponding to a violation of the kinematic constraints are hatched.
This overview suggests, that there is an optimal acceleration for each distance. When using the proposed design method, it is advantageous to use the ideal acceleration instead of the highest possible acceleration. Since the optimal feasible acceleration $a_\text{max}$ depends on the travel distance, the calculation of the jerk segments has to be implemented in an efficient manner. An efficient algorithm for the solution of this optimal control problem is presented in \cite{tau_ocpJ_assembly_part2}. As highlighted by \autoref{fig:showing_why_efficient_method_ocpJ_required_amax}, there is a region where very low accelerations $a_\text{max}$ can lead to $a\ttt > a_\text{max}$ for the finished trajectory as shown in \autoref{fig:showing_overlapping_0p75mm}.
For the parameters presented here, this only occurs for very low accelerations $a_\text{max}$.
In particular, for the example depicted in \autoref{fig:showing_overlapping_0p75mm}, the constraint
$a_\text{lim}$ is satisfied although $\ddot{z}$ exceeds $a_\text{max}$. A method to calculate the ideal acceleration for Case~2: $a_\text{max} = a_\text{best}$ as marked in \autoref{fig:showing_why_efficient_method_ocpJ_required_amax} is outlined in \autoref{sec:optim_full_trajectory}.

\section{Optimizing the full trajectory}\label{sec:optim_full_trajectory}
As mentioned in \autoref{SubSec:adjusting_aMax_to_removing_jerkViolations} and shown in \autoref{fig:showing_why_efficient_method_ocpJ_required_amax}, adjustments of the maximal acceleration $a_\text{max}$ used for planning of the jerk segments can lead to an overall reduction of transition times. This motivates the higher-level optimization problem considered in this section. It is thereby assumed that for given maximal acceleration $a_\text{max}$, the overall trajectory is calculated according to \autoref{sec:calc_full_trajectory}. Based on that, the total transition time $t_\text{f,t}$ is optimized with respect to $a_\text{max}$ as the scalar optimization variable. The calculation is developed in detail for Case~2, while a detailed analysis of the remaining cases is postponed to future work.

\subsection{Ideal acceleration $a_\text{max}$ for Case~2}\label{SubSec:Ideal_aMax_Case2}
\begin{figure}[!ht]
	\centering
	\def\svgwidth{0.95\linewidth}
	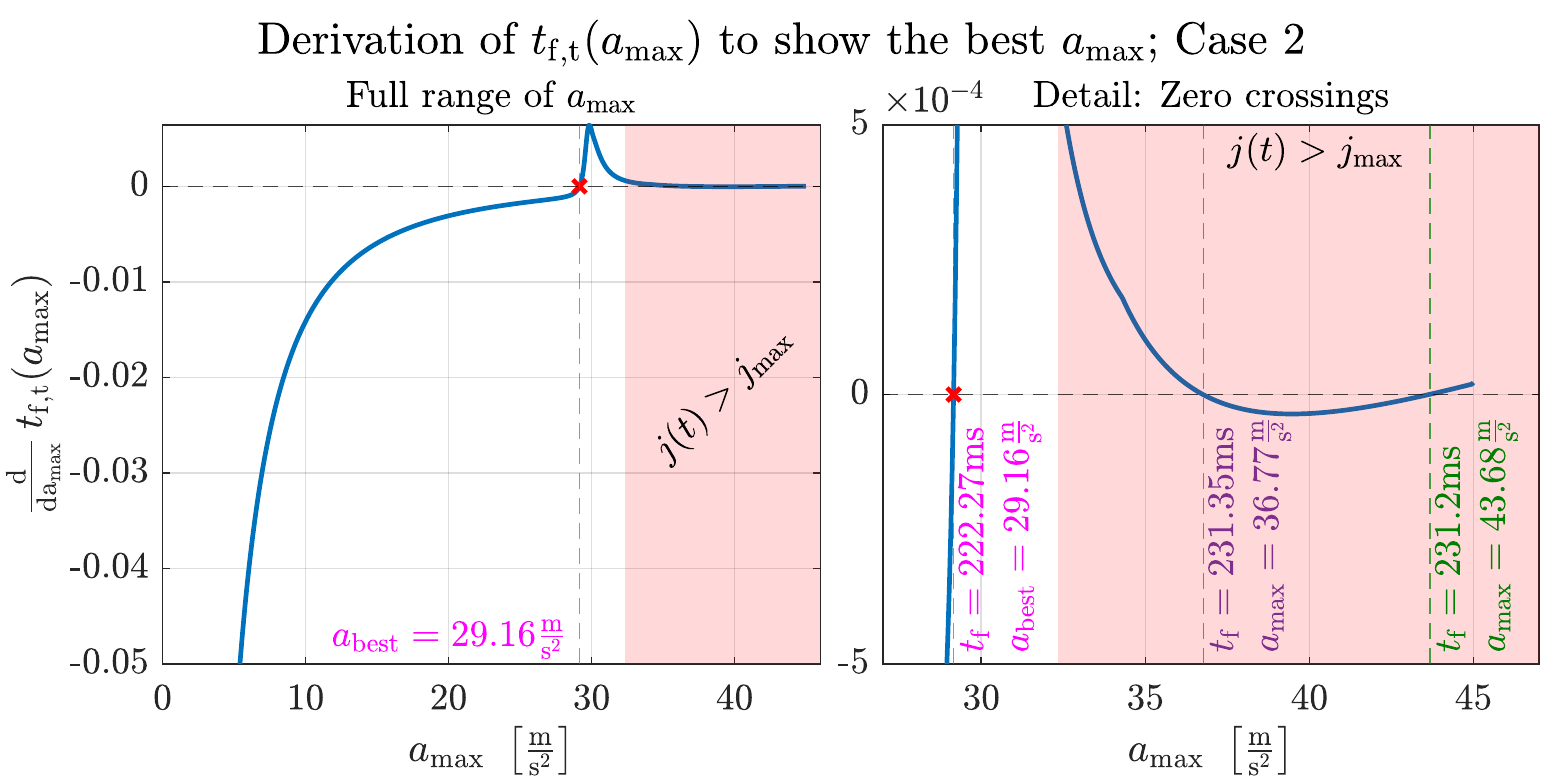 %
	\caption{Best acceleration for this set of parameters (Case~2).}%
	\label{fig:showing_ocpJ_best_aMax_case2}
\end{figure}

The total transition time for the overall trajectory in Case~2 is
\begin{equation}
	t_\text{f,t} = 4\cdot t_\text{f1} + 2 \cdot t_\text{a-max} + t_\text{v-max}
\end{equation}
with $t_\text{f1}$ being the terminal time of a single jerk segment, $t_\text{a-max}$ the time at maximum acceleration and deceleration and $t_\text{v-max}$ the time at maximum slider velocity (see \autoref{fig:showing_ocpJ_dividedSegments_methodTwo}). Using \eqref{eq:Case2_t_aMax_calculation} and \eqref{eq:Case2_t_vMax_calculation} leads to
\begin{equation}
	t_\text{f,t}\tat = \frac{\textcolor{\colorZtft}{z\tftt}}{v_\text{lim}} + \frac{v_\text{lim}}{a_\text{max}} + 2 \, t_\text{f1} \tat - \frac{v_\text{f1}\tat + v_\text{f3}\tat}{a_\text{max}} \, \text{.}
\end{equation}
Searching for the optimal $a_\text{max}$ to minimize $t_\text{f,t}$ is a static optimization problem and the extrema (local minimum or maximum) satisfy
\begin{equation}
	\frac{d}{da_\text{max}} t_\text{f,t}\tat = 0 \, \text{.}
\end{equation}
Differentiating $t_\text{f,t}\tat$ with regards to $a_\text{max}$ in order to minimize $t_\text{f,t}\tat$ gives
\begin{equation}\label{eq:calx_idal_aMax_Case2}
	\begin{aligned}
		\frac{1}{a_\text{max}^2}& \Big[ 2 \, a_\text{max}^2 \frac{d}{d a_\text{max}}  t_\text{f1} \tat + v_\text{f1} \tat + v_\text{f3} \tat -  \\
		& v_\text{lim}-a_\text{max} \cdot \frac{d}{d a_\text{max}} \left(v_\text{f1} \tat + v_\text{f3} \tat \right) \Big] = 0 %
	\end{aligned}
\end{equation}
Solving \eqref{eq:calx_idal_aMax_Case2} for $a_\text{max}$ gives the optimal acceleration $a_\text{best}$ for Case~2. This confirms, that the optimal acceleration for Case~2 is independent of the transition distance as suggested by the results shown in \autoref{fig:showing_why_efficient_method_ocpJ_required_amax}. However, as Case~2 can only be used on a certain interval of distances as shown in \autoref{fig:OcpJ_trajectories_showing_regions_of_methods}, the optimal value for  $a_\text{max}$ may still depend on the transition distance. The result for the derivation of $t_\text{f,t}\tat$ is plotted in \autoref{fig:showing_ocpJ_best_aMax_case2}.
The result shows, that there exists several local optima, at least for the particular parameters used here.

\subsection{Optimal maximum acceleration for Cases~1~and~3}\label{SubSec:IDeal_aMax_Case1_and_3}
\begin{figure}[!ht]
	\centering
	\def\svgwidth{0.95\linewidth}
	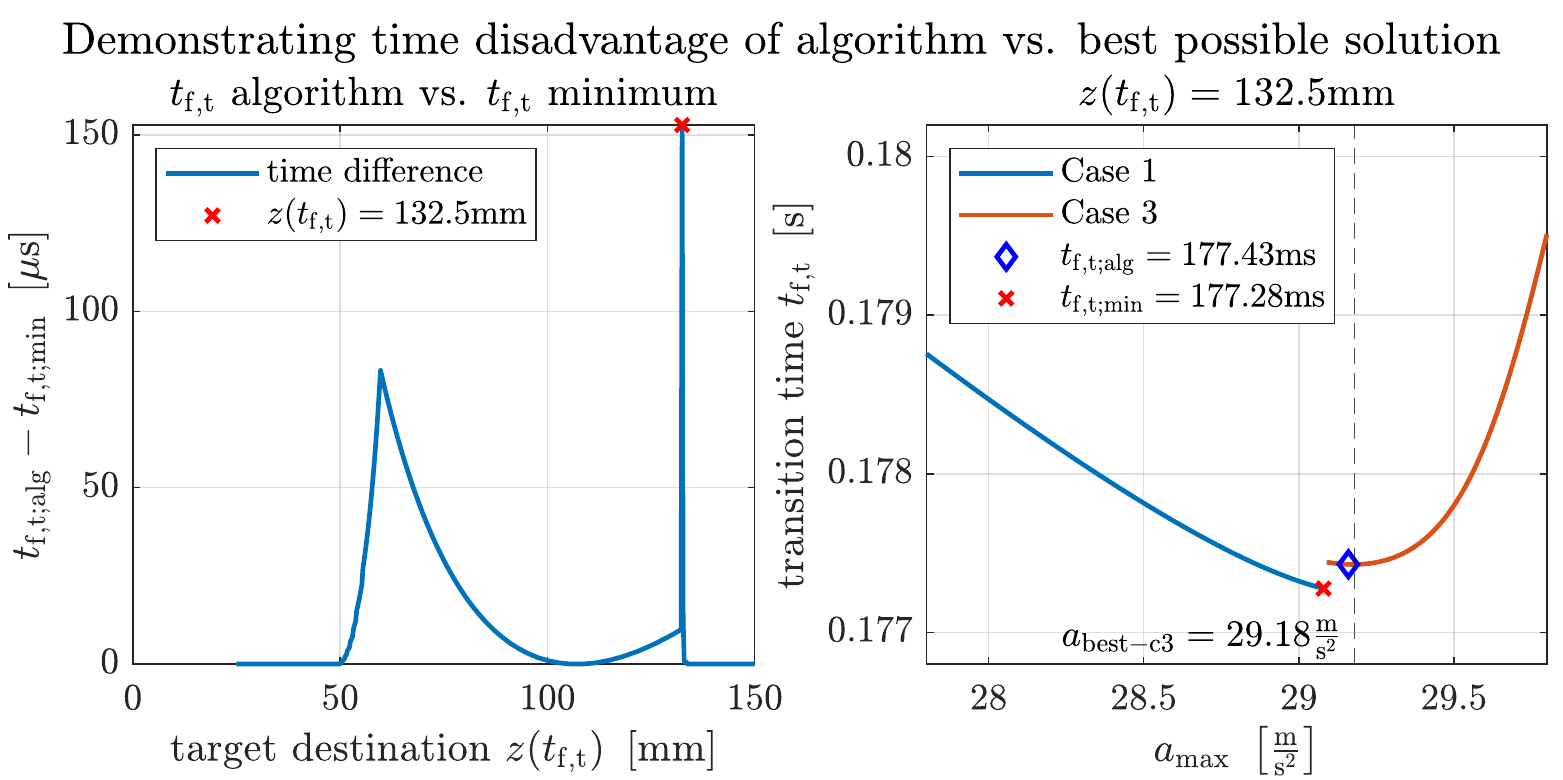 %
	\caption{Applying the algorithm to the trajectories and comparing the transition times with the best possible transition times to show the effectiveness of the algorithm (specific for distance).} %
	\label{fig:applying_algorithm_to_traj}
\end{figure}

In contrast to Case~2, the optimal $a_\text{max}$ for Cases~1~and~3 also depend on the overall transition distance, because of the terms $\nicefrac{\textcolor{\colorZtft}{z\tftt}}{a_\text{max}}$ in \eqref{eq:case1_calculate_t1} and \eqref{eq:case3_calculate_t_aMax} respectively. As \autoref{fig:showing_why_efficient_method_ocpJ_required_amax} suggests, the ideal acceleration for Case~1 and Case~3 can be slightly higher or lower than the ideal acceleration obtained for Case~2. This is underlined by \autoref{fig:applying_algorithm_to_traj}, which shows the results of a more detailed numerical analysis.
Here the globally optimal $a_\text{max}$ was evaluated and the transition time is compared to the transition time achieved with the algorithm described in \hyperref[App:alg_for_implementation]{Appendix~\ref*{App:alg_for_implementation}}. This result shows, that for a certain range of distances the result of the algorithms described below could be further improved. As the result in \autoref{fig:applying_algorithm_to_traj} shows, the potential for improvement for the parameter set considered here is below $\SI{200}{\micro\second}$, which is smaller, than a single PLC cycle. For this reason, it was postponed for future work.

\section{Comparison with existing methods}\label{sec:comparison_to_regular}
This section is devoted to a comparison with established and recently developed trajectory planning methods. Trajectories are compared \wrt transition times, parameter sensitivity and general limitations are listed. First, the trajectory planning approaches are briefly explained in \autoref{SubSec:Trajectory_planning_we_benchmark_against_explan}. Afterwards the parameters from \autoref{tab:table_machine_parameters_ex_Pap} and the kinematic constraints from \autoref{tab:table_listing_kincont_ex_Pap} are used to demonstrate the performance characteristics of the $\ocpJ$ approach in \autoref{SubSec:Comparison_ocpJ_others_ourMachineParameters}. These are the parameters that most closely resemble the production machines, and are therefore used to highlight the performance that can be expected in a real-world setting. Additional parameter sets (one from the laboratory system used for validation and one from \cite{Yalamanchili2024}) are taken and the performance of the $\ocpJ$ approach is shown for these parameter sets. This analysis and the parameters are given in \autoref{SubSec:Comp_ocpJ_others_otherParam}.

\subsection{Planning approaches considered in comparison}\label{SubSec:Trajectory_planning_we_benchmark_against_explan}
A regular $\SCurve$ \cite{ClaudioMelchiorri2008}, that does not consider any oscillation of the system, is used as a baseline. This trajectory represents the fastest way to reach an endpoint, taking into account all kinematic constraints of the actuator (velocity, acceleration and jerk), but not the system's internal dynamics. The second trajectory planning approach, is the $\ZV$-shaped $\SCurve$ \cite{Singhose1990,Singhose1994}, a well established method in the field and has been shown to work very well \cite{Kruk2023}. This approach was chosen for comparisons over more involved shapers \cite{Pao1998,Kasprowiak2022,Kang2019,Singhose1997_EI_shaper,Singer1999, Vaughan2008,Singhose2009} because it represents the fastest positive impulse shaper \cite{Singhose1990,Singhose1994}. Due to the nature of negative impulse shapers \cite{SinghoseMarch1996,Sorensen2008}, kinematic constraints may be violated (although there are strategies to avoid violation \cite{Sorensen2008}), they are not included in the comparison. The third method is the $\ocpS$ approach \cite{Papageorgiou2015,Hargraves1987,Betts1998,TSANG1975, Auer2021_7te_IftommDach}, outlined in \autoref{SubSec:OcpS_formulation}, which uses full discretisation to solve the trajectory planning problem in a time-optimal manner. The fourth approach is an implementation of an improved $\Fir$ filter as given in \cite{Yalamanchili2024}. It is denoted $\FirImp$ in the figures. It is an improved version of \cite{Biagiotti2015,Biagiotti2019,Biagiotti2021}, an implementation of which can be found in \cite{Biagiotti2020}. For the third order trajectories (limited jerk, continuous acceleration, velocity and position), \cite{Biagiotti2015,Besset2017,Biagiotti2019,Biagiotti2021,Biagiotti2020,Yalamanchili2024} is a variation of \cite{Meckl1998,Bearee2014,Kim2018} which allows a further reduction of the transition time in selected cases at the expense of parameter sensitivity. A direct comparison with \cite{Meckl1998,Bearee2014,Kim2018} would therefore be redundant. Approaches that also consider derivatives of the jerk \cite{Lee2020,Tatte2024,Han2024} are not included because they were developed for a different purpose and would therefore not provide the best comparison. Since \cite{Yalamanchili2024} is a further development of \cite{Biagiotti2021} that helps to remove violations of kinematic constraints, it was chosen for the comparisons instead of \cite{Biagiotti2021}.

\subsection{Comparison with parameters presented here}\label{SubSec:Comparison_ocpJ_others_ourMachineParameters}
\begin{figure}[]
	\centering
	\def\svgwidth{0.95\linewidth}
	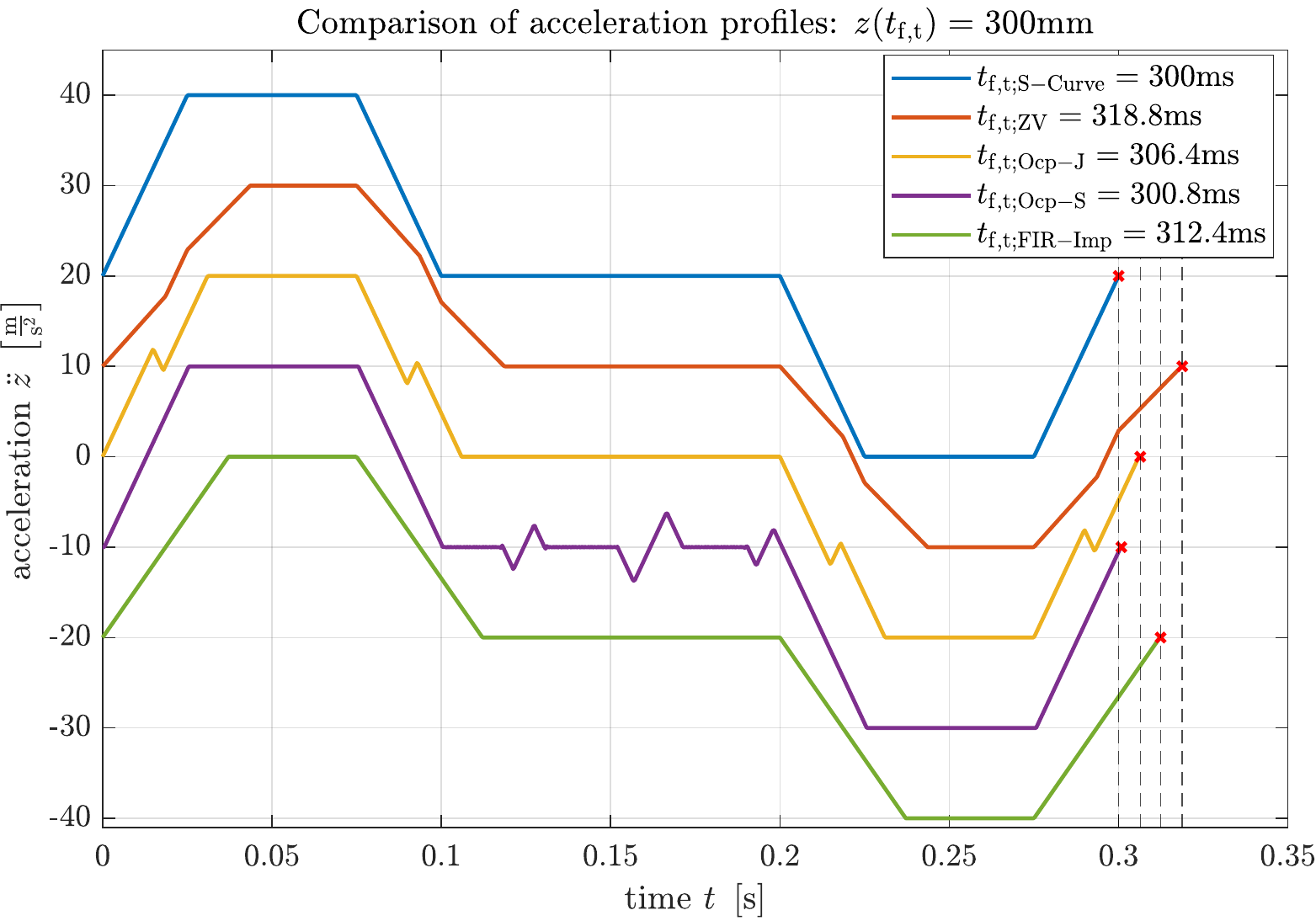 %
	\caption{Trajectories: $\SCurve$, $\ZV$, $\ocpS$, $\ocpJ$ and $\FirImp$ for a comparison of transition times for the motion distance $z\tftt = \SI{300}{\milli\meter}$.}
	\label{fig:exampleDist_SC_ZV_ocpS_ocpJ_Fir_exempl_pap_param_distLong} %
\end{figure}

\begin{figure}[]
	\centering
	\def\svgwidth{0.95\linewidth}
	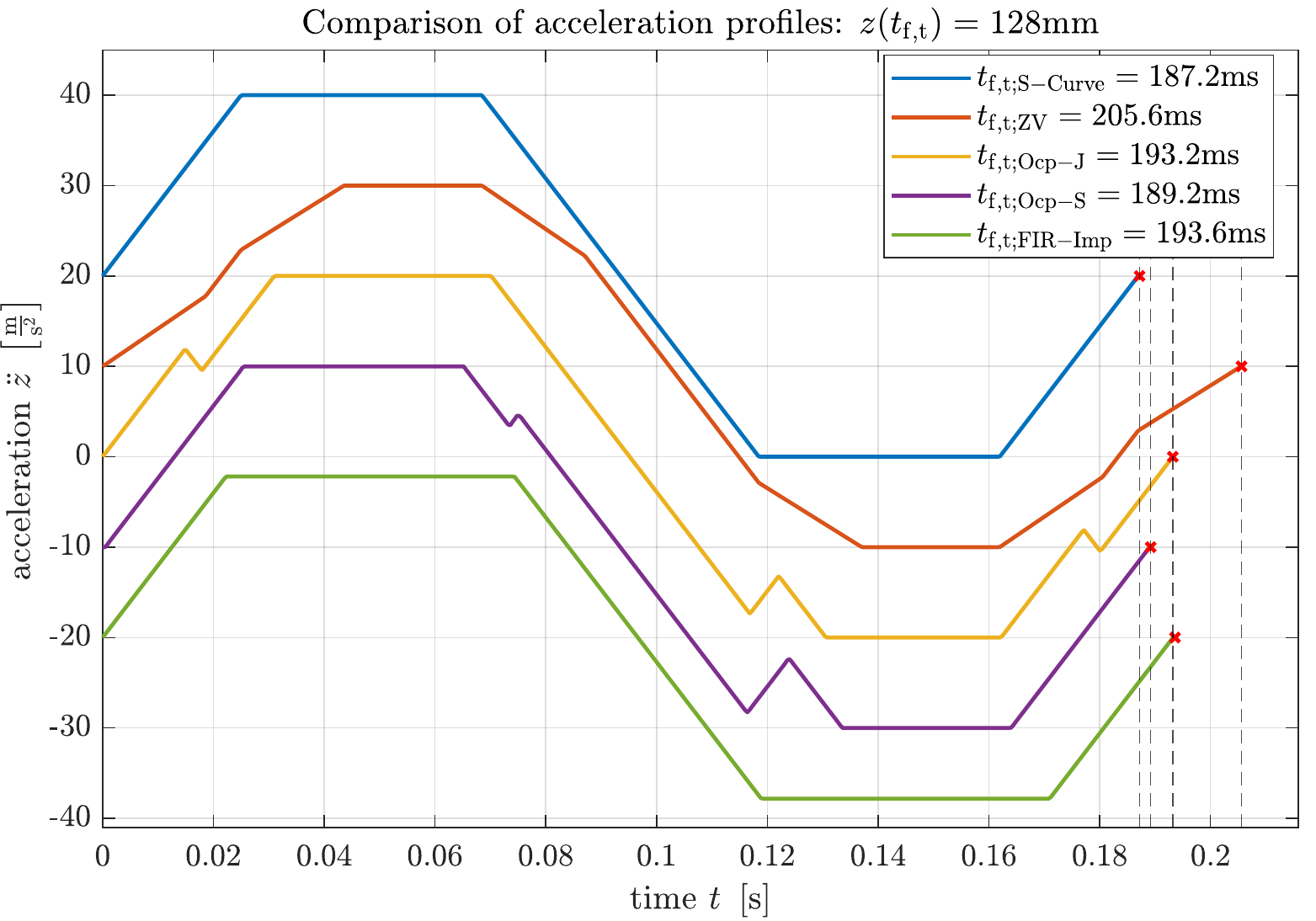 %
	\caption{Trajectories: $\SCurve$, $\ZV$, $\ocpS$, $\ocpJ$ and $\FirImp$ for a comparison of transition times for the motion distance $z\tftt = \SI{128}{\milli\meter}$.}
	\label{fig:exampleDist_SC_ZV_ocpS_ocpJ_Fir_exempl_pap_param_distMiddle} %
\end{figure}

\begin{figure}[!ht]
	\centering
	\def\svgwidth{0.95\linewidth}
	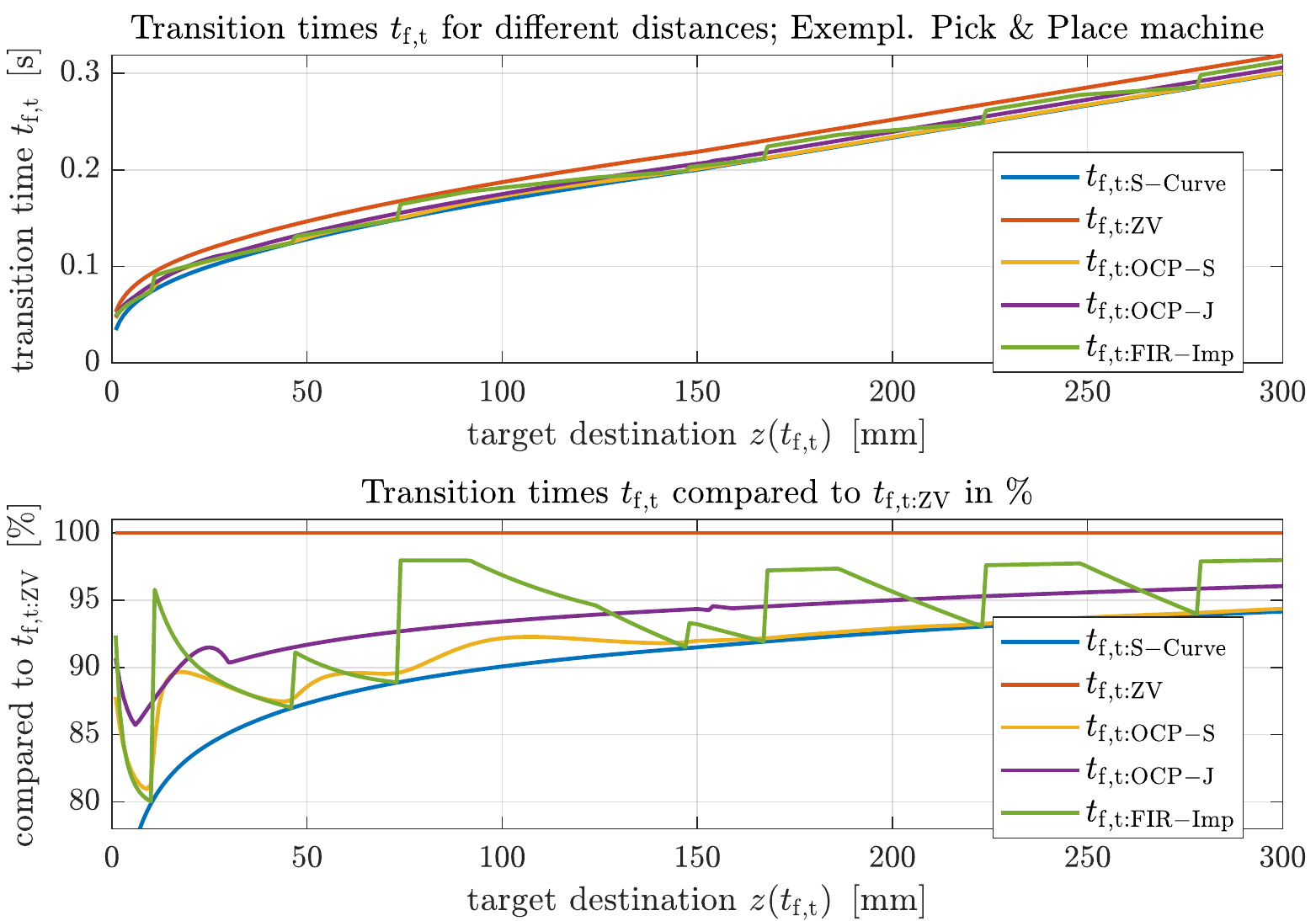 %
	\caption{Transition times for all methods. Shown are the absolute times and the times relative to the $\ZV$-shaper. The comparison is made to the $\ZV$-shaper, because it is the established method (plot not taken from \cite{Auer2023IFAC}).}
	\label{fig:distSweep_tfComp_SC_ZV_ocpS_ocpJ_Fir_exempl_pap_param}
\end{figure}

\begin{figure}[!ht]
	\centering
	\def\svgwidth{0.95\linewidth}
	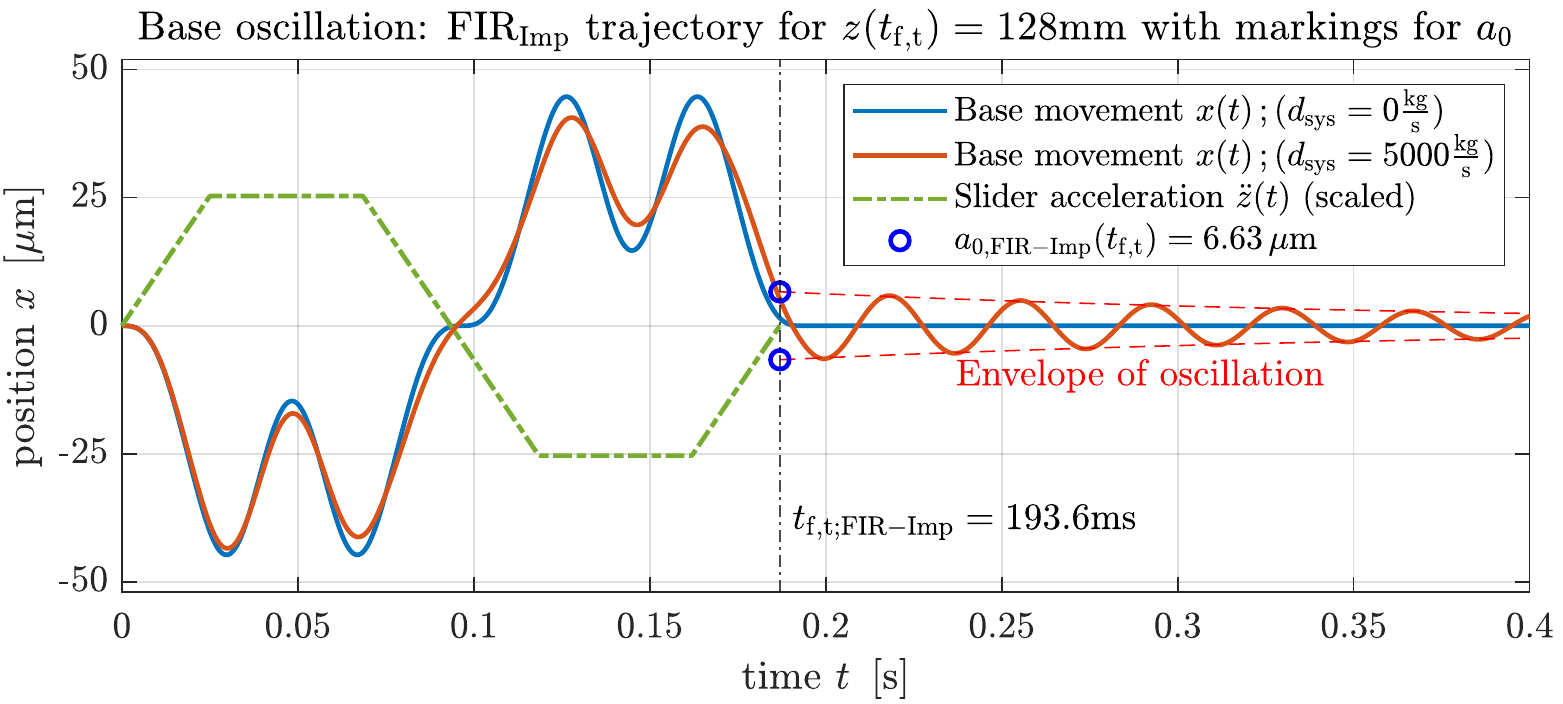 %
	\caption{Oscillation in base to mark, where and how the oscillation $a_0$ for a $\FirImp$ trajectory is determined.}
	\label{fig:exampleDist_oscillation_in_base_FirImp}
\end{figure}

\begin{figure}[!ht]
	\centering
	\def\svgwidth{0.95\linewidth}
	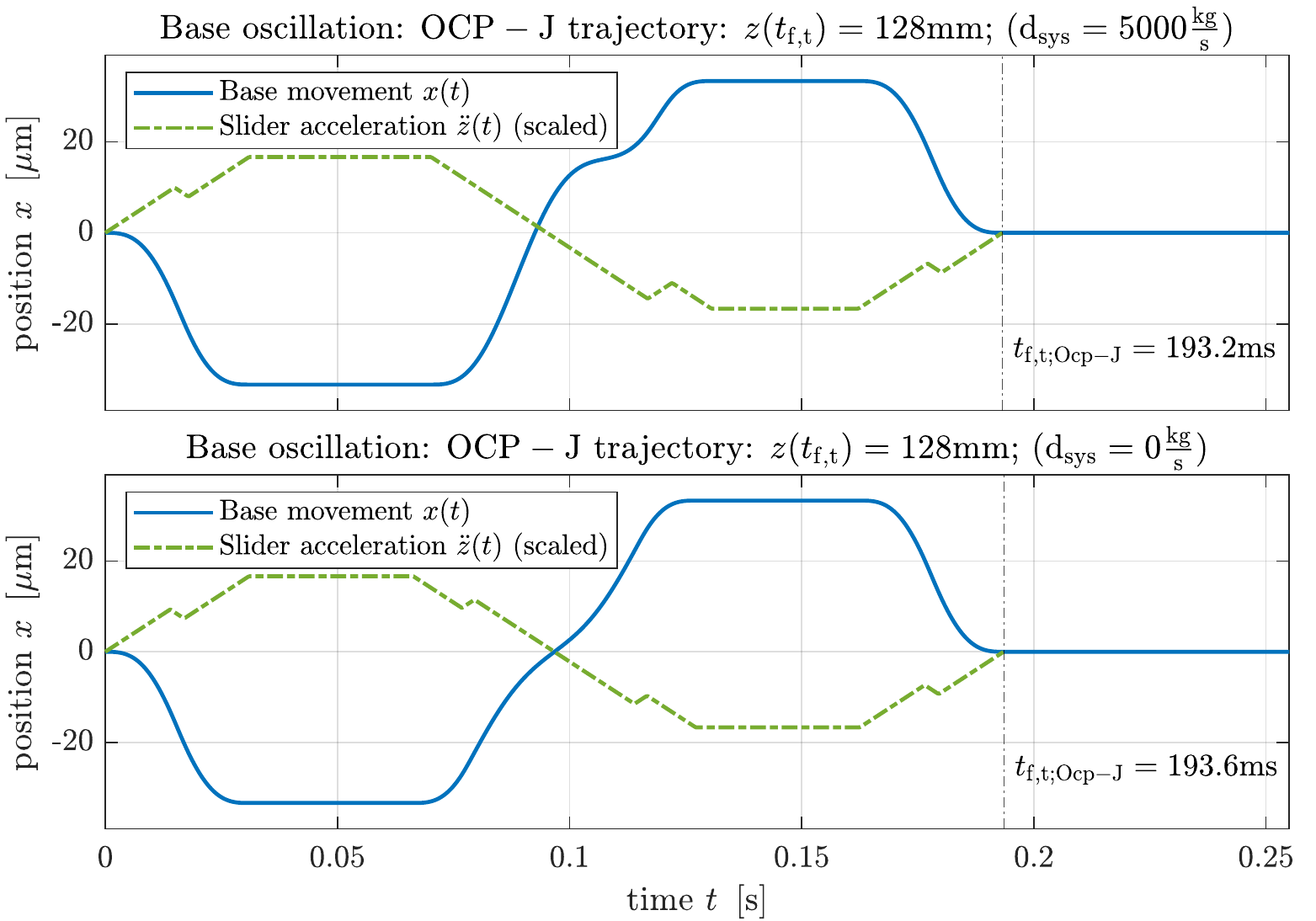 %
	\caption{Oscillation in base to mark, where and how the oscillation $a_0$ for a $\ocpJ$ trajectory is determined.}
	\label{fig:exampleDist_oscillation_in_base_OcpJ}
\end{figure}

For the first comparison, the parameters from \autoref{tab:table_machine_parameters_ex_Pap} and the kinematic constraints from \autoref{tab:table_listing_kincont_ex_Pap} are used to calculate and compare different trajectories. Trajectories for a single distance are shown in \autoref{fig:exampleDist_SC_ZV_ocpS_ocpJ_Fir_exempl_pap_param_distLong} (contains the trajectories shown in \autoref{fig:example_optimal_control_sCurve_ZV} and in addition the $\ocpJ$ and $\FirImp$ approaches). All of the acceleration profiles start at $a\tzt=\SI{0}{\meter\per\second\squared}$ respectively, but they have been separated by $\SI{10}{\meter\per\second\squared}$ in the plot for better visibility and comparison.
Two more specific distances are shown, that are of interest. The first in \autoref{fig:exampleDist_SC_ZV_ocpS_ocpJ_Fir_exempl_pap_param_distMiddle}, where the transition time of the $\ocpJ$ approach is within $\SI{1}{\milli\second}$ of the $\FirImp$ and a second, where the original $\SCurve$ has the same transition time, as the $\FirImp$.
\begin{figure}[]
	\centering
	\def\svgwidth{0.95\linewidth}
	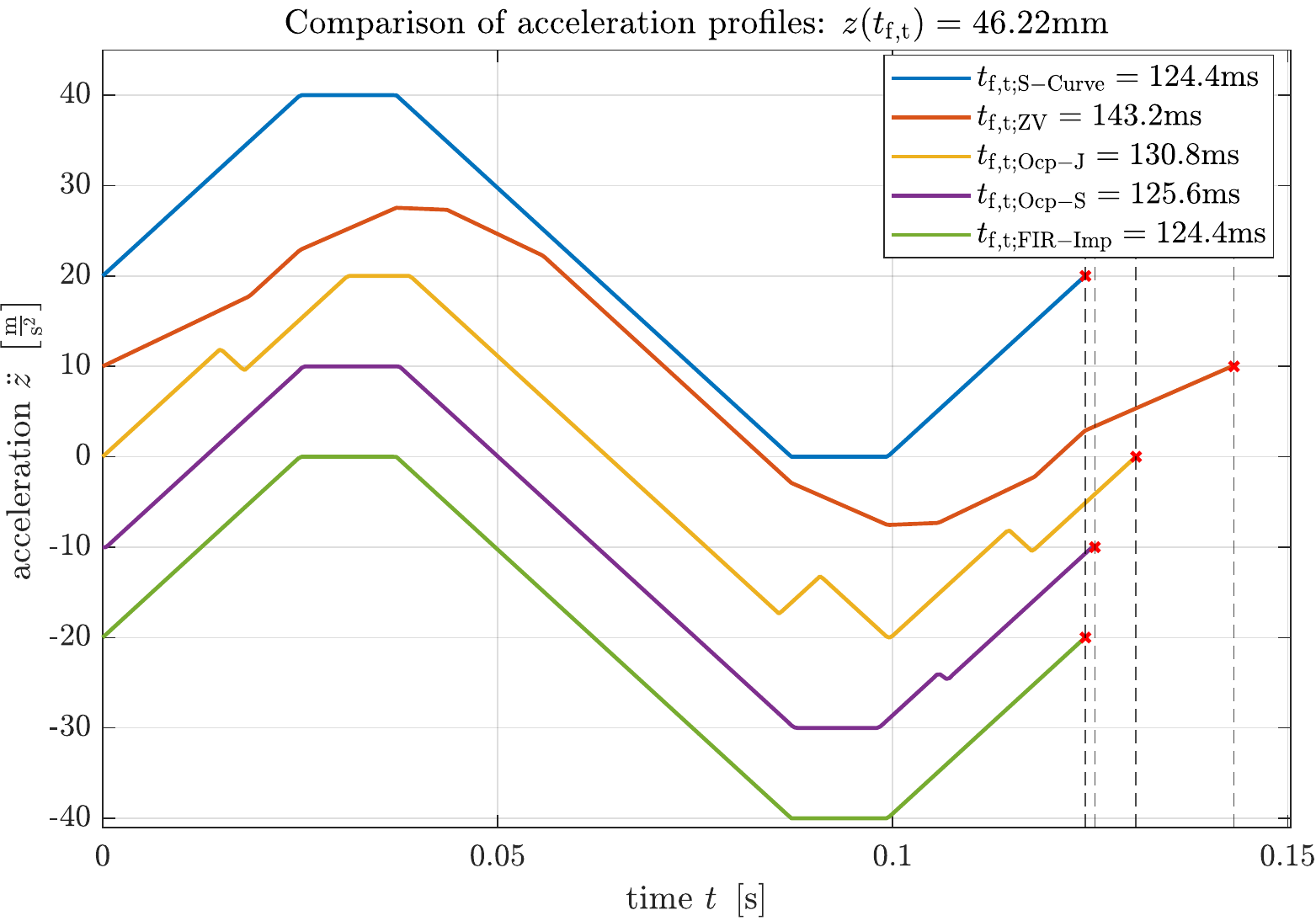 %
	\caption{Trajectories: $\SCurve$, $\ZV$, $\ocpS$, $\ocpJ$ and $\FirImp$ for a comparison of transition times for the motion distance $z\tftt = \SI{46.22}{\milli\meter}$.}
	\label{fig:exampleDist_SC_ZV_ocpS_ocpJ_Fir_exempl_pap_param_distShort} %
\end{figure}
The amount of time saved or gained depends on the distance and can additionally vary depending on the approach. The $\ZV$-shaped $\SCurve$ is always exactly half an oscillation period slower, than the underlying $\SCurve$. For the other approaches, the time saved varies relative to a $\ZV$-shaper. For this reason, distances between $\SI{1}{\milli\meter}$ and $\SI{300}{\milli\meter}$ were analysed. The result of this is shown in \autoref{fig:distSweep_tfComp_SC_ZV_ocpS_ocpJ_Fir_exempl_pap_param}, comprising the total transition times and times gained or lost compared to the $\ZV$-shaped $\SCurve$.
The comparison with the $\ZV$ shaped $\SCurve$ was chosen because it is a widely used and universally applicable technique. The approaches $\ZV$, $\ocpS$ and $\ocpJ$ approaches allow damping to be taken into account, while $\FirImp$ does not. Whenever the system response to a trajectory exhibits oscillation, the amplitude is determined using the method shown in \autoref{fig:exampleDist_oscillation_in_base_FirImp}. An envelope is fitted on the decaying oscillation and the resulting amplitude at $t_\text{f,t}$ (end of the slider movement) is evaluated and labelled with $a_0$, as shown in the plot. In \autoref{fig:exampleDist_oscillation_in_base_FirImp} the slider acceleration for a $\FirImp$ trajectory is shown together with the system response for two systems. First, the system response for a system without damping ($d=0$) is shown to demonstrate, that oscillation-free transitions can be achieved for those systems as mentioned in \cite{Yalamanchili2024}. Secondly, the system response with the actual damping from \autoref{tab:table_machine_parameters_ex_Pap} is shown to show the influence of the neglected system damping.
Another comparison for the same distance, but for the $\ocpJ$ trajectories developed in this contribution, is shown in \autoref{fig:exampleDist_oscillation_in_base_OcpJ}. As this approach can take damping into account, the trajectories calculated for the system with and without damping are different.
The possibility of taking system damping into account in the $\ocpJ$ approach opens the possibility to use it on system with damping.

\subsection{Comparison for different parameters}\label{SubSec:Comp_ocpJ_others_otherParam}
\begin{figure}[!ht]
	\centering
	\def\svgwidth{0.95\linewidth}
	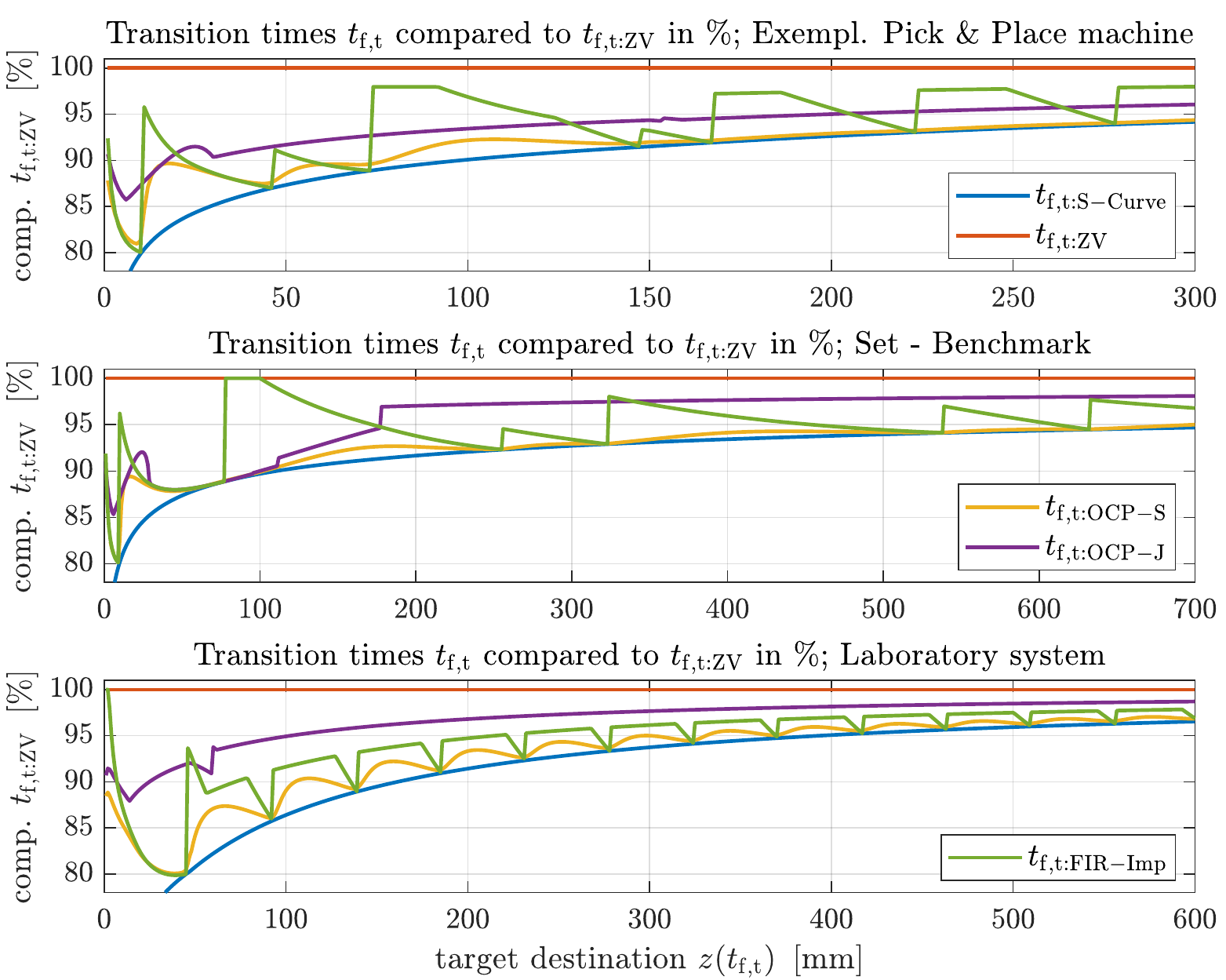 %
	\caption{Transition times relative to the $\ZV$-shaper for all of the parameter sets from \autoref{tab:table_all_parameters}.}
	\label{fig:comparison_transitionTimes_relative_ZV_allParamSets}
\end{figure}

To give a broader analysis and show the performance \wrt transition time (and also parameter uncertainty), comparisons with additional parameter sets are provided. All parameter sets are summarized in \autoref{tab:table_all_parameters}. The abbreviation \hbox{\textit{Ex.-PaP.}} stands for parameter set of exemplary pick and place machine which are given in \autoref{tab:table_machine_parameters_ex_Pap} and \autoref{tab:table_listing_kincont_ex_Pap}. The corresponding results are shown in \autoref{fig:distSweep_tfComp_SC_ZV_ocpS_ocpJ_Fir_exempl_pap_param}. The parameters \hbox{\textit{Set: \cite{Yalamanchili2024}}} are the parameters used in \cite{Yalamanchili2024} for the distance sweep. The last abbreviation \hbox{\textit{Lab.-Sys.}} marks the parameters of the laboratory system used for the measurements (more details for this system follow in \autoref{sec:measurements_laboratory_system}). In the table $\omega_0$ refers to the angular eigenfrequency and \hbox{$\delta = \nicefrac{d}{2 \cdot m_\text{g}}$} refers to the damping of the system.
\begin{table}[!ht]
	\captionsetup{width=\linewidth}
	\caption{All parameter sets listed for overview}
	\vspace{-0.5em}
	\renewcommand{\arraystretch}{1.25}
	\centering
	\begin{tabular}{|l||l|l|l|l|l|}
		\hline
		& & & & & \\ [-1em] 
		Param-Set  & $v_\text{lim} \! \left[\frac{\text{m}}{\text{s}}\right] $ & $a_\text{lim} \! \left[\frac{\text{m}}{\text{s}^2}\right]$ & $j_\text{lim} \! \left[\frac{\text{m}}{\text{s}^3}\right]$ & $\omega_0 \! \left[\frac{\text{rad}}{\text{s}}\right]$ & $\delta \! \left[\frac{1}{\text{s}}\right]$ \\[4pt] \hline\hline
		Ex.-PaP.                        & 1.5            & 20             & 800            & 169.03    & 4.762     \\ \hline
		Set from: \cite{Yalamanchili2024}  & 1              & 2              & 10             & 40         & 0        \\ \hline
		Lab.-Sys.                       & 0.45           & 6              & 200            & 61.02      & 0.799    \\ \hline
	\end{tabular}
	\label{tab:table_all_parameters}
\end{table}
The analysis shown in \autoref{fig:distSweep_tfComp_SC_ZV_ocpS_ocpJ_Fir_exempl_pap_param} has been repeated for the additional parameters listed in \autoref{tab:table_all_parameters}. The transition times relative to the $\ZV$-shaped trajectory are shown in \autoref{fig:comparison_transitionTimes_relative_ZV_allParamSets}.
In analysing these results, some observations can be made which are summarised in the discussion (\autoref{sec:summary_and_comparison}) together with the results of the rest of this section.

\subsection{Analysis of parameter sensitivity}
\begin{figure}[!ht]
	\centering
	\def\svgwidth{0.95\linewidth}
	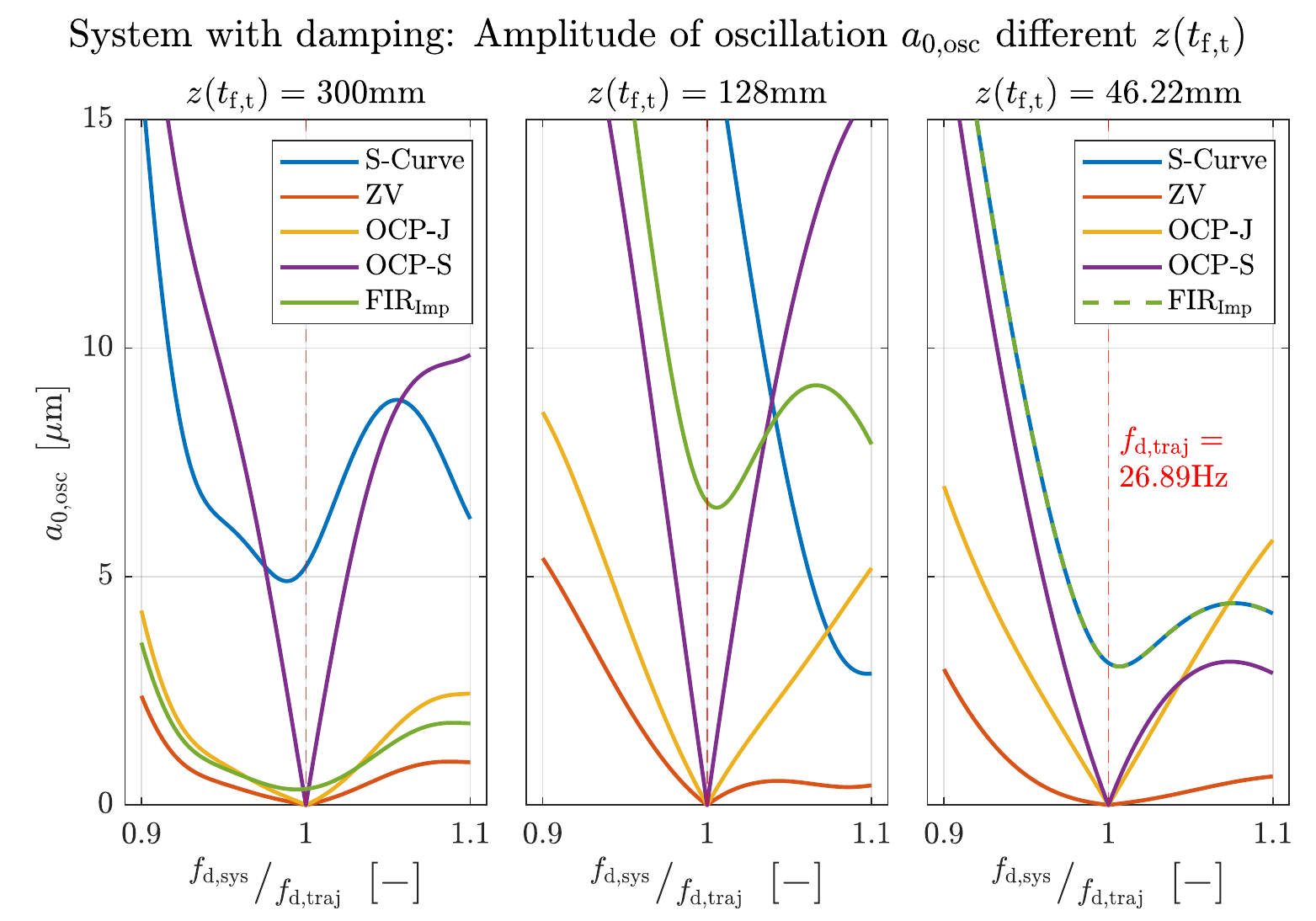 %
	\caption{Sensitivity to parameter uncertainty: Comparison of approaches for the distances shown in \autoref{fig:exampleDist_SC_ZV_ocpS_ocpJ_Fir_exempl_pap_param_distLong}, \autoref{fig:exampleDist_SC_ZV_ocpS_ocpJ_Fir_exempl_pap_param_distMiddle} and \autoref{fig:exampleDist_SC_ZV_ocpS_ocpJ_Fir_exempl_pap_param_distShort}.}
	\label{fig:sensibility_functions_for_two_distances_ocpJ_vs_ZV}
\end{figure}

The previous section showed the transition times for the different trajectory planning approaches. Another important issue to be considered is the sensitivity of the approaches \wrt parameter uncertainty, which is analysed in this subsection. To this end, the trajectories are all calculated for the nominal parameters shown in \autoref{tab:table_machine_parameters_ex_Pap} with the kinematic constraints from \autoref{tab:table_listing_kincont_ex_Pap}. In the analysis $\omega_{\text{d}}$ is adjusted to show the residual amplitude of oscillation under parameter uncertainty. The method used to determine this amplitude is illustrated in \autoref{fig:exampleDist_oscillation_in_base_FirImp}.
The sensitivity to parameter uncertainty for the trajectories from \autoref{fig:exampleDist_SC_ZV_ocpS_ocpJ_Fir_exempl_pap_param_distLong}-\ref{fig:exampleDist_SC_ZV_ocpS_ocpJ_Fir_exempl_pap_param_distShort} is shown in \autoref{fig:sensibility_functions_for_two_distances_ocpJ_vs_ZV}. Another analysis for the parameter set of the laboratory system is shown together with measurements in \autoref{sec:measurements_laboratory_system}. Note that an uncertainty in the oscillation frequency of $10\%$ corresponds to about a $20\%$ error in the measured system stiffness $k$. Note that the area under consideration is larger than it would be in practice just to visualize the behaviour under larger uncertainties. %
All of the results are summarily discussed in \autoref{sec:summary_and_comparison}.

\section{Experiments on Laboratory system}\label{sec:measurements_laboratory_system}
This section describes the experiments carried out on a laboratory system to validate the approach. The laboratory system and parameter combinations are given and an implementation aspect is mentioned. The planning methods presented in \autoref{sec:comparison_to_regular} are implemented (except for $\ocpS$; see: \autoref{SubSec:OcpS_formulation}) on the PLC. The trajectories shown in the measurement have been calculated directly on the PLC. Measurements have been performed for the different methods, for different distances and parameter combinations.

\subsection{Laboratory setup}
\begin{figure*}[!ht]
	\centering
	\def\svgwidth{0.95\linewidth}
	\footnotesize
	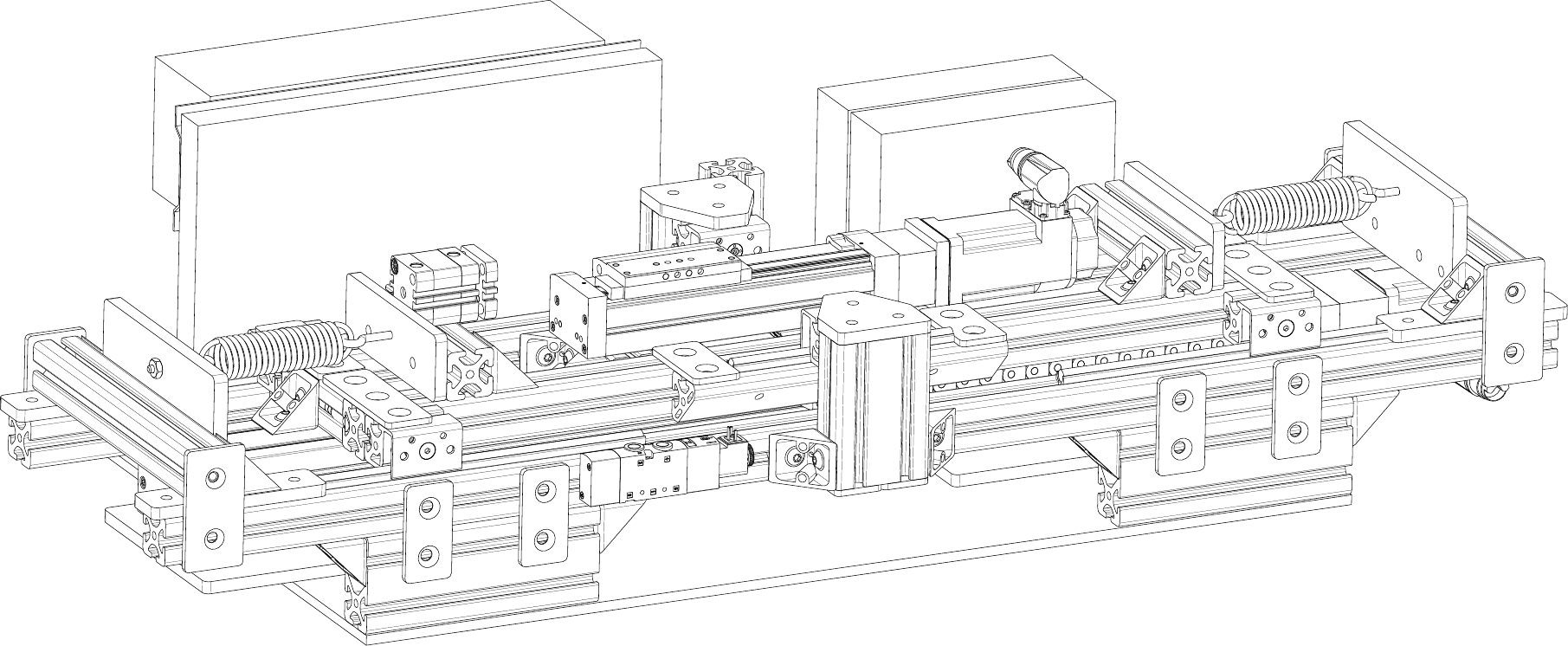
	\caption{CAD-picture of the laboratory system used for the measurements (picture also shown in \cite{Auer2024Case}).}
	\label{fig:CAD_of_laboratory_system_IACE}
\end{figure*}

A CAD-picture of the laboratory system is shown in \autoref{fig:CAD_of_laboratory_system_IACE}. The base movement $x\ttt$ from \autoref{fig:model_of_the_system} corresponds to the position of the main carriage mounted between the two springs and $z\ttt$ to the slider mounted on the bottom of the test rig (both the left and rightmost position of the slider are shown in the picture).
In order to study the influence of parameter uncertainty on trajectory planning approaches, the springs are easily interchangeable. Replacing the springs also alters the damping of the system and the five sets of parameters (cf. \cite{Auer2024Case}) are listed in \autoref{tab:table_adjusted_paramters}, where configuration $3_\text{nom}$ is the parameter set all for which all trajectories were planned.
\begin{table}[!ht]
	\captionsetup{width=\linewidth}
	\caption{Parameters of the system: Changes in stiffness to show effect of parameter variation (table taken from \cite{Auer2024Case}).}
	\vspace{-1em}
	\renewcommand{\arraystretch}{1.3}
	\centering
	\begin{tabular}{|l|l|l|l|}
		\hline
		$\text{N}_\text{sys}$ & $k_\text{sys}$ & $d_\text{sys}$ & $f_\text{d,sys}$ \\ \hline
		1 & $k_1 = \SI{94583}{\newton\per\meter}$ & $d_1 = \SI{68.1}{\kilo\gram\per\second}$ & $f_{\text{d,sys},1} = \SI{8.71}{\hertz}$ \\ \hline
		2 & $k_2 = \SI{105532}{\newton\per\meter}$ & $d_2 = \SI{67.2}{\kilo\gram\per\second}$ & $f_{\text{d,sys},2} = \SI{9.2}{\hertz}$ \\ \hline
		3$_\text{nom}$ & $k_\text{traj} = \SI{117499}{\newton\per\meter}$ & $d_\text{traj} = \SI{50.4}{\kilo\gram\per\second}$ & $f_\text{d,traj} =  \SI{9.71}{\hertz}$ \\ \hline
		4 & $k_4 = \SI{131367}{\newton\per\meter}$ & $d_4 = \SI{54.3}{\kilo\gram\per\second}$ & $f_{\text{d,sys},4} = \SI{10.27}{\hertz}$ \\ \hline
		5 & $k_5 = \SI{140042}{\newton\per\meter}$ & $d_5 = \SI{56.9}{\kilo\gram\per\second}$ & $f_{\text{d,sys},5} = \SI{10.601}{\hertz}$ \\ \hline
	\end{tabular}
	\label{tab:table_adjusted_paramters}
\end{table}
It is important to note, that the influence of parameter uncertainty is shown for comparison only. As mentioned previously, the trajectory planning approach has not been specifically optimized for behaviour under parameter uncertainty. In reality, some small uncertainty in the system is unavoidable so some level of robustness to parameter uncertainty is required.

\subsection{Implementation and computational complexity}\label{SubSec:Implementation_and_comp_complex}
This section outlines how the required code can be implemented on the PLC and explains why the presented method was chosen. The implementation for the full trajectories presented in \autoref{sec:calc_full_trajectory} and \autoref{sec:optim_full_trajectory} is best split into three separate functions when implementing to avoid unnecessary duplication of code. The three parts are given in \hyperref[App:alg_for_implementation]{Appendix~\ref*{App:alg_for_implementation}}. The first part given in \autoref{alg:get_ocp_ind_traject} implements the equations shown in \autoref{sec:calc_full_trajectory} and returns the $\ocpJ$ trajectory based on a given $a_\text{max}$. This $a_\text{max}$ is provided by the overlying function incorporating \autoref{alg:Get_ocpJ_total}. After a trajectory is returned by \autoref{sec:calc_full_trajectory}, the calculations from \autoref{alg:Get_ocpJ_total} check if any kinematic constraints are violated. This happens for very small distances due to the large negative overlap of the jerk segments mentioned before (see \autoref{fig:showing_overlapping_1p5mm} or \autoref{fig:showing_overlapping_200mm}). If a violation of the kinematic constraints occurs in \autoref{alg:Get_ocpJ_total}, the last method shown in \autoref{alg:Optimize_aMAx} is used to optimize $a_\text{max}$ to resolve these violations and to ensure minimal transition times (it is known that this can happen and has been taken into account). Performing the calculation in this way ensures that the trajectory respects the kinematic constraints and that the transition times are as small as possible with the $\ocpJ$ (cf. \autoref{sec:optim_full_trajectory}). Two additional details are provided in the following. First, the computational complexity is addressed, and subsequently, a post-processing step is outlined, necessary to provide the trajectory on the discrete controller intervals.

\subsubsection{Computational complexity and calculation times}
For the parameters and kinematic constraints given in \autoref{tab:table_machine_parameters_ex_Pap} and \autoref{tab:table_listing_kincont_ex_Pap}, the part of the algorithm to adjust $a_\text{max}$ (\autoref{alg:Optimize_aMAx}) only needs to be called for transition distances $z\tftt$ less than $\SI{30}{\milli\meter}$. For distances above this limit, the calculation of a complete trajectory is finished after calling \autoref{alg:get_ocp_ind_traject} once. Note that the calculation of the required jerk segments is completed after a fixed number of steps as well (the first two steps in \autoref{alg:get_ocp_ind_traject} and the method is as described in \cite{tau_ocpJ_assembly_part2}). For transition distances below this limit, the computational effort is higher because \autoref{alg:Optimize_aMAx} is executed. The total number of steps in \autoref{alg:Optimize_aMAx} can vary depending on which of the two terminal conditions (as explained in \hyperref[App:alg_for_implementation]{Appendix~\ref*{App:alg_for_implementation}}) is triggered first. In the worst case, the calculation is called $n_\text{max,iter}$ times. It turned out that using single precision on the PLC is sufficient, as the measurement results in \autoref{SubSec:Measurement_Results} show.

\begin{figure}[!ht]
	\centering
	\def\svgwidth{0.95\linewidth}
	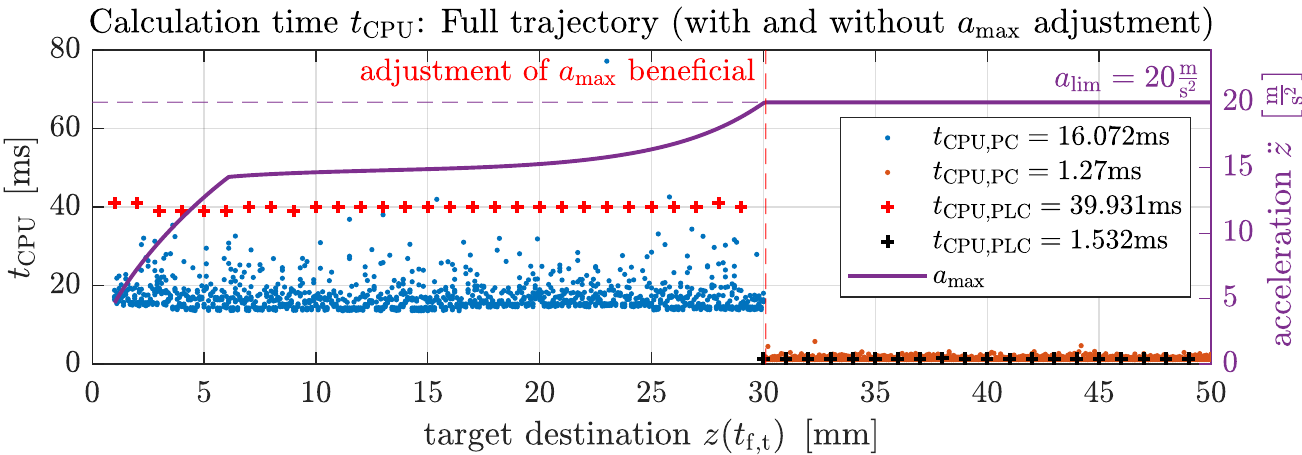 %
	\caption{Showing $a_\text{max}$ and $t_{\rm{CPU}}$ (average times over the points marked) used for different transition distances.}
	\label{fig:showing_aMax_for_diff_sTf}
\end{figure}

\bigskip
To demonstrate the frequency of adjustments of $a_\text{max}$ and to show real computation times, trajectories were planned for different distances. The resulting optimal values for $a_\text{max}$ and the total computation times are shown in \autoref{fig:showing_aMax_for_diff_sTf}. The times were measured on a regular PC (\hbox{i7-1185G7} processor and code running in MATLAB) and on a PLC (X20CP1585 and code programmed in structured text). Both times are shown in \autoref{fig:showing_aMax_for_diff_sTf}.
This shows, that adjustments of $a_\text{max}$ are required only for small overall distances (below $\SI{30}{\milli\meter}$ for the given parameters). 
Note that it is not always possible to work with pre-calculated jerk segments from previous runs, as the process parameters may change due to parameter drift. As \autoref{alg:get_ocp_ind_traject} may be called several times by \autoref{alg:Optimize_aMAx}, the calculation must be efficient and fast. For this reason, the algorithm presented in \cite{tau_ocpJ_assembly_part2}, which allows efficient calculation of the jerk segments, is advantageous. Listing the number of individual function evaluations are required is not useful, and instead the calculation time required on a PLC are given directly.

\subsubsection{Post-processing}
The calculation of the switching times corresponding to jerk changes from $j_\text{max}$ to $0$, $-j_\text{max}$ and vice versa is not bound to the specific controller cycle $t_\text{ctrl} = \SI{400}{\micro\second}$ of the PLC itself. To provide the position controller with the reference trajectory on the specific discrete interval, the equations given in \eqref{eq:calc_z_to_zppp_from_snap} have to be used directly (as the equations show, the computation time required for this step is basically negligible). This step results in rounding the final time of the trajectory $t_\text{f,t}$ to a full controller cycle. This approach is comparable to the supersampling approach outlined in \cite{ClaudioMelchiorri2008}. It enables the utilization of a continuously computed trajectory to derive a reference trajectory on the discrete controller interval.

\subsection{Measurements}\label{SubSec:Measurement_Results}
\begin{figure}[!ht]
	\centering
	\def\svgwidth{0.95\linewidth}
	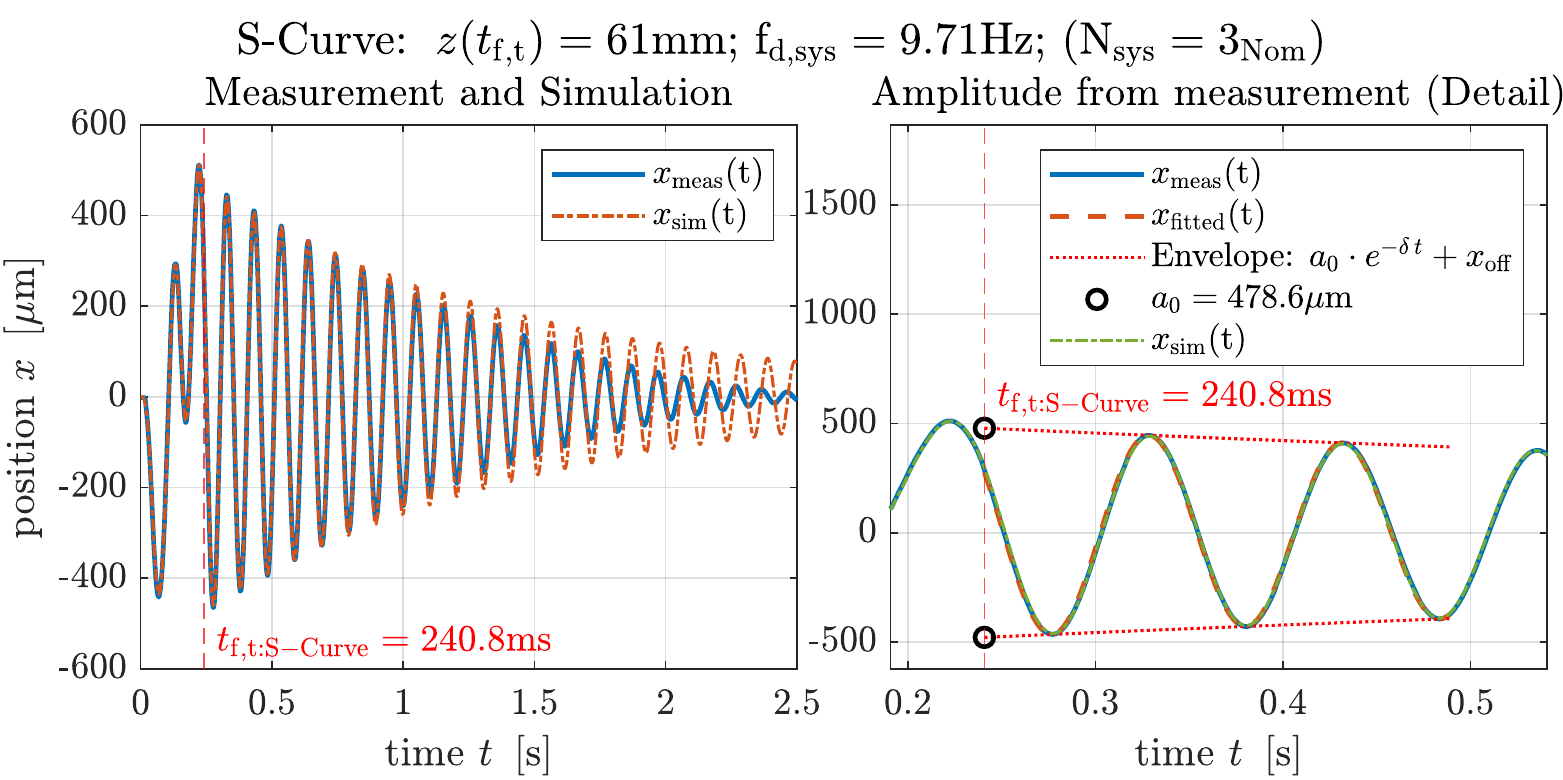 %
	\caption{Measurement of $\SCurve$ trajectory showing oscillation (oscillation of base).}
	\label{fig:showing_definition_of_a0_of_meas_1_SCurve}
\end{figure}

\begin{figure}[!ht]
	\centering
	\def\svgwidth{0.95\linewidth}
	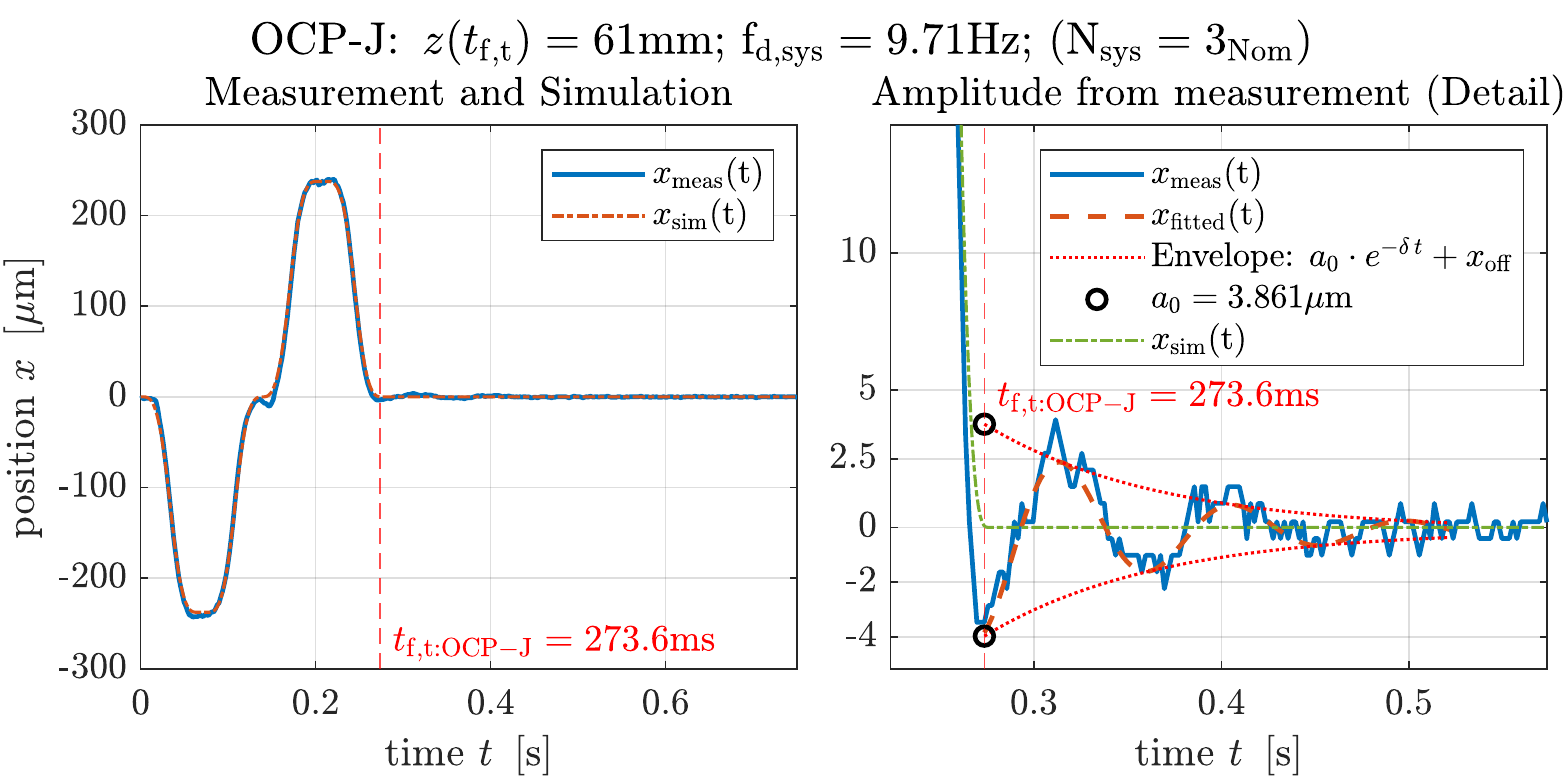 %
	\caption{Measurement of $\ocpJ$ trajectory with good system knowledge (oscillation of base).}
	\label{fig:showing_definition_of_a0_of_meas_2_OcpJ_paramknownExact}
\end{figure}

Since simulations of the trajectories under parameter uncertainty were shown, and the test rig specifically allows such measurements, measurements were also made for this reason. The depicted results show the amplitude of oscillation occurring after the slider reaches its final position. Here, an oscillation for a small timespan covering $\SI{0.25}{\second}$ or roughly 2 oscillation periods, corresponding to
\begin{equation}
	x\ttt = a_\text{0} \cdot e^{-\delta \, \left(t - t_\text{f,t}\right)} \sin \! \left(\omega_\text{d} \, t - \varphi_0\right)
\end{equation}
is fitted to the measurements.%
This is shown for three measurements in \autoref{fig:showing_definition_of_a0_of_meas_1_SCurve}-\ref{fig:showing_definition_of_a0_of_meas_2_OcpJ_paramknownExact}.
\begin{figure}[!ht]
	\centering
	\def\svgwidth{0.95\linewidth}
	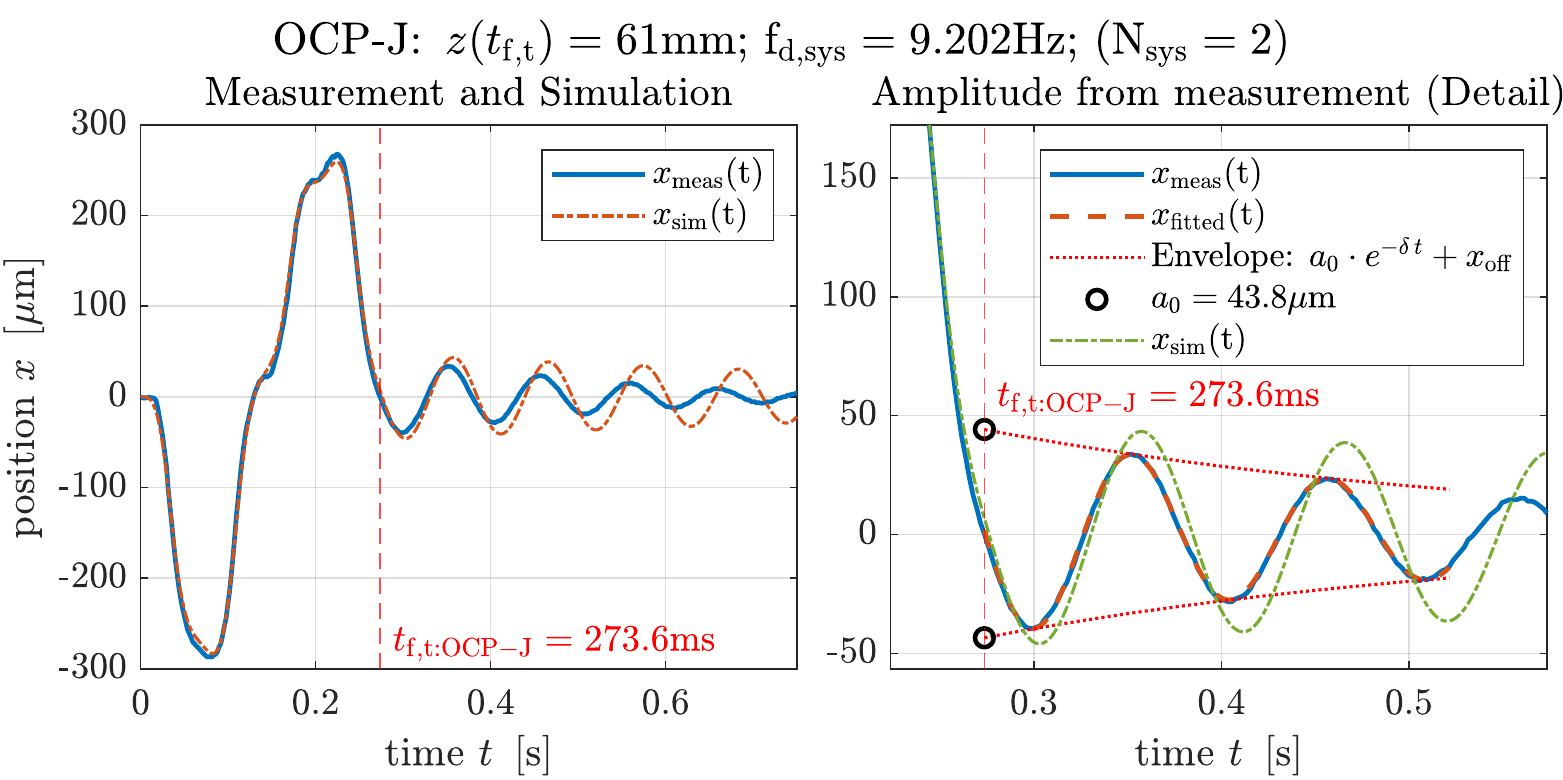 %
	\caption{Measurement of $\ocpJ$ trajectory under parameter uncertainty (oscillation of base).}
	\label{fig:showing_definition_of_a0_of_meas_2_OcpJ_paramUncert}
\end{figure}
The first measurement in \autoref{fig:showing_definition_of_a0_of_meas_1_SCurve} shows the system response to an $\SCurve$. The second and third trajectory show the $\ocpJ$ trajectory, planned for the frequency $f_\text{d,traj}$ as given in \autoref{tab:table_adjusted_paramters}. The response for a system, where $f_\text{d,sys} = f_\text{d,traj}$ is shown in \autoref{fig:showing_definition_of_a0_of_meas_2_OcpJ_paramknownExact}. The remaining oscillation is due to remaining model inaccuracies. In particular non-linear friction, which most likely occurs in the laboratory setup, is not covered in the model. This assumption is supported by the fact, that (as the plots show) the oscillation tends to decay quicker, as the amplitude decreases. The system response under larger parameter uncertainty for the same trajectory is then shown in \autoref{fig:showing_definition_of_a0_of_meas_2_OcpJ_paramUncert}. These measurements were then repeated for five different distances, the five different trajectory planning methods presented and for the system parameters given \autoref{tab:table_adjusted_paramters}. All of the distances, methods and the corresponding transition times are given in \autoref{tab:table_overview_transTimes}.
\begin{table}[!ht] 
 \captionsetup{width=\linewidth} 
 \caption{Transition times $t_\text{f,t}$ for measured distances and trajectory planning methods.} 
 \vspace{-0.75em} 
 \renewcommand{\arraystretch}{1.3} 
 \centering 
 \begin{tabular}{|l||l|l|l|l|} 
 \hline 
 Transition distance&$\SCurve$&$\ZV$&$\FirImp$&$\ocpJ$ \\ \hline 
$z\tftt = \SI{14.5}{\milli\meter}$&$\SI{132.8}{\milli\second}$&$\SI{184.4}{\milli\second}$&$\SI{156.4}{\milli\second}$&$\SI{162.4}{\milli\second}$ \\ \hline 
$z\tftt = \SI{61}{\milli\meter}$&$\SI{240.8}{\milli\second}$&$\SI{292.4}{\milli\second}$&$\SI{260.4}{\milli\second}$&$\SI{273.6}{\milli\second}$ \\ \hline 
$z\tftt = \SI{116}{\milli\meter}$&$\SI{362.8}{\milli\second}$&$\SI{414.4}{\milli\second}$&$\SI{382.8}{\milli\second}$&$\SI{395.2}{\milli\second}$ \\ \hline 
$z\tftt = \SI{139}{\milli\meter}$&$\SI{414}{\milli\second}$&$\SI{465.6}{\milli\second}$&$\SI{414}{\milli\second}$&$\SI{446.4}{\milli\second}$ \\ \hline 
$z\tftt = \SI{181}{\milli\meter}$&$\SI{507.6}{\milli\second}$&$\SI{558.8}{\milli\second}$&$\SI{515.2}{\milli\second}$&$\SI{539.6}{\milli\second}$ \\ \hline 
\end{tabular} 
 \label{tab:table_overview_transTimes} 
 \end{table}

\begin{figure}[!ht]
	\centering
	\def\svgwidth{0.95\linewidth}
	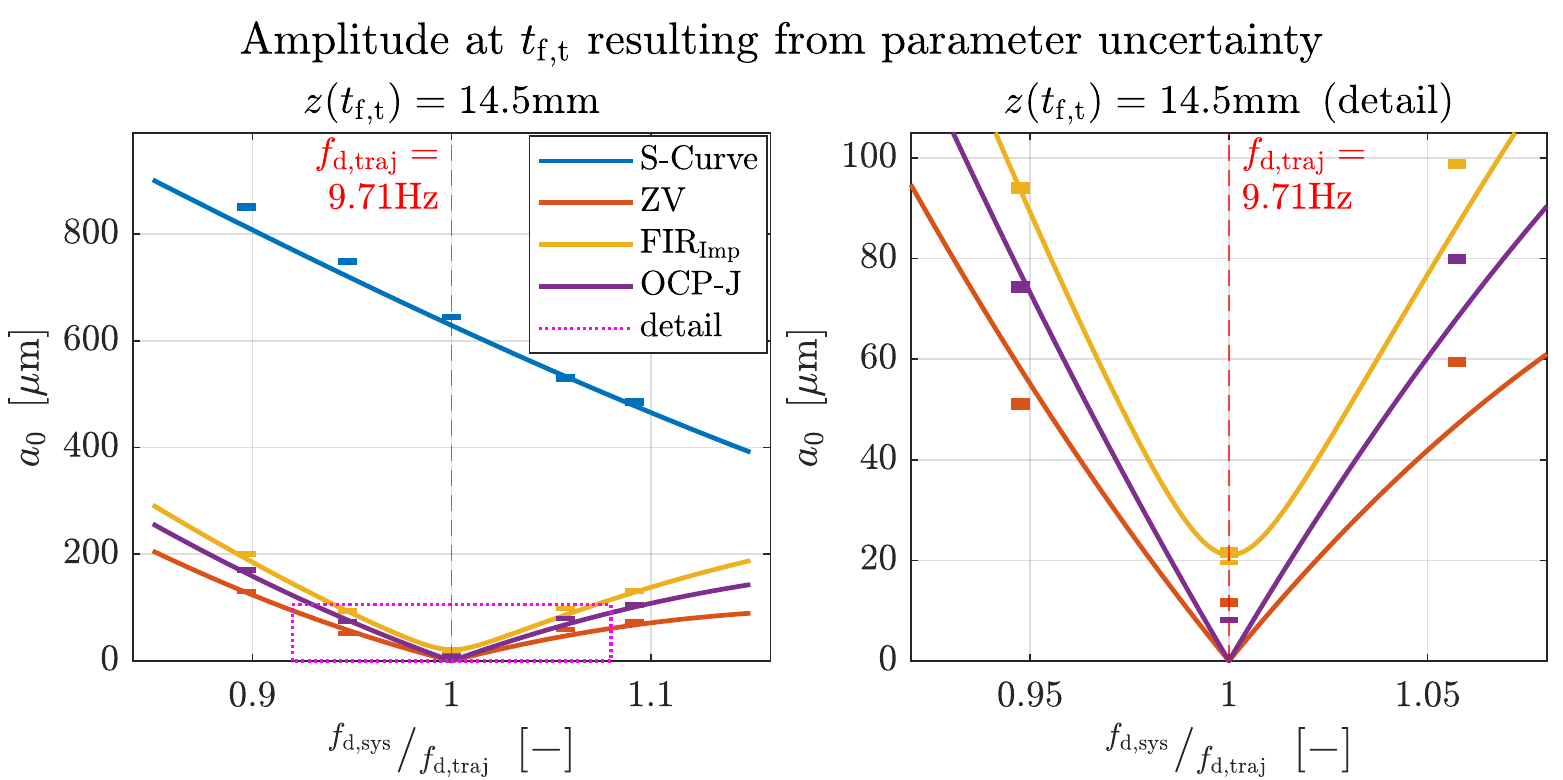 %
	\caption{Overview of all measurements and simulated amplitudes under parameter uncertainty (on distances with detail).}
	\label{fig:measurement_results_laboratory_system_first_dist}
\end{figure}

\begin{figure}[!ht]
	\centering
	\def\svgwidth{0.95\linewidth}
	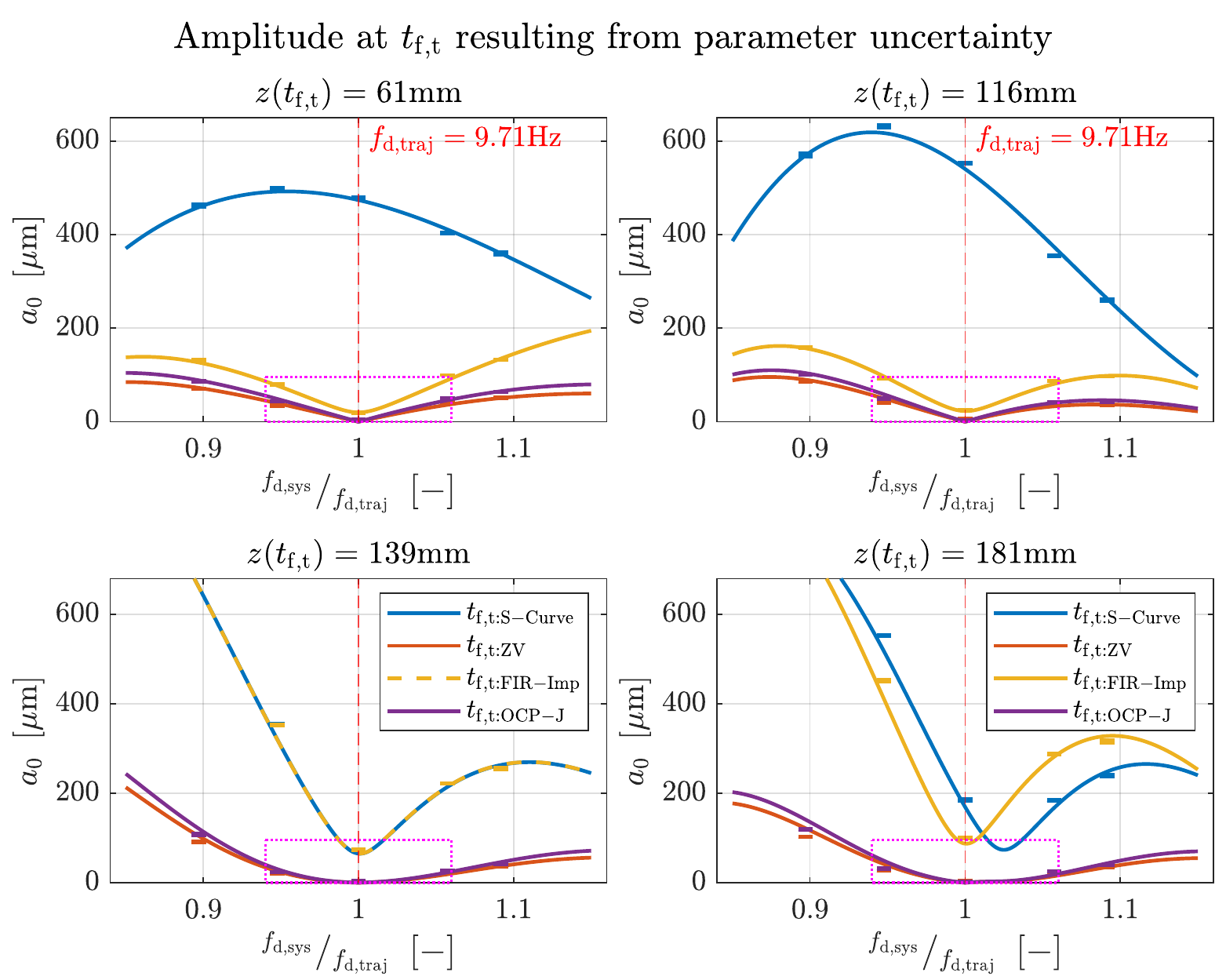 %
	\caption{Overview of all measurements (four distances) and simulated amplitudes under parameter uncertainty (marked details shown in \autoref{fig:measurement_results_laboratory_system_dist_2_to_5_details}).}
	\label{fig:measurement_results_laboratory_system_dist_2_to_5}
\end{figure}

\begin{figure}[!ht]
	\centering
	\def\svgwidth{0.95\linewidth}
	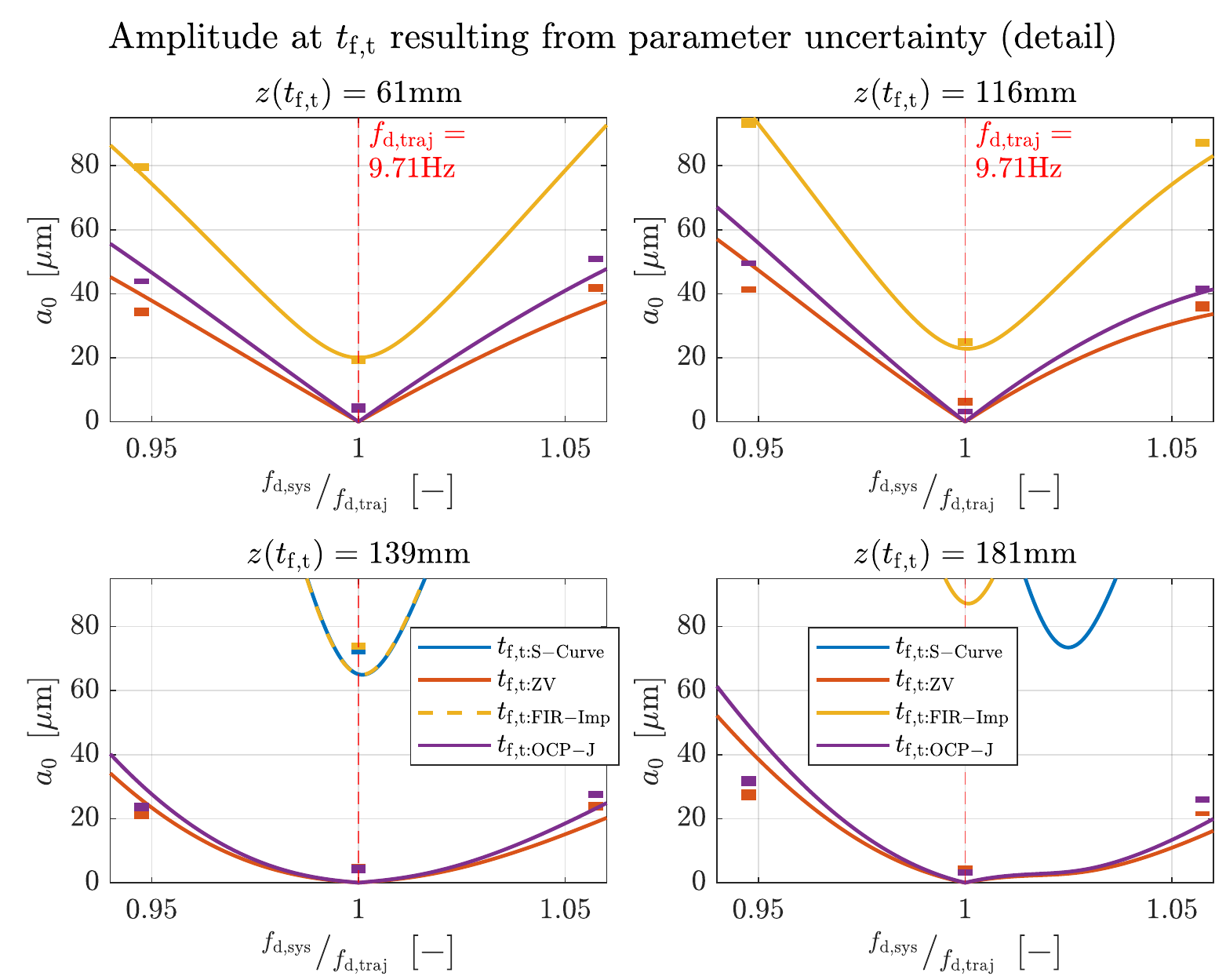 %
	\caption{Overview of all measurements (four distances) and simulated amplitudes under parameter uncertainty (details for measurement shown in \autoref{fig:measurement_results_laboratory_system_dist_2_to_5}).}
	\label{fig:measurement_results_laboratory_system_dist_2_to_5_details}
\end{figure}

Plots similar to \autoref{fig:sensibility_functions_for_two_distances_ocpJ_vs_ZV} show a comparison of simulated amplitudes and the measurements. These plots serve a dual purpose. First, they show the behaviour of the planned trajectories under parameter deviation (similar to \autoref{fig:sensibility_functions_for_two_distances_ocpJ_vs_ZV}). Secondly, a comparison of the measured and simulated amplitudes is shown in the plots, where all measurements and the deviation between simulated and measured amplitudes are visible. The simulated amplitudes are the solid lines in the plots. The measurements correspond to the small bars located at the different frequencies. Each measurement was repeated five times and the minimal and maximal amplitudes are given in \autoref{tab:table_measurement_results} located in \hyperref[App:Table_with_Measurement_Results]{Appendix~\ref*{App:Table_with_Measurement_Results}}. The measurements for the first distance of \autoref{tab:table_overview_transTimes} are shown in \autoref{fig:measurement_results_laboratory_system_first_dist}. The full picture of the measurements is shown on the left of the plot and a detail for a smaller range of amplitudes and system frequencies is shown on the right. The same is shown for the rest of the distances in \autoref{fig:measurement_results_laboratory_system_dist_2_to_5} and the details in \autoref{fig:measurement_results_laboratory_system_dist_2_to_5_details}.
The measurements, a comparison to the simulation results and the trajectory planning method are next discussed in \autoref{sec:summary_and_comparison}.

\section{Discussion of results}\label{sec:summary_and_comparison}
The selection of methods has already been discussed in \autoref{SubSec:Trajectory_planning_we_benchmark_against_explan}. As the results from \autoref{fig:comparison_transitionTimes_relative_ZV_allParamSets} show, the methods presented offer slightly faster transitions than $\ZV$ shaping at the expense of increased parameter sensitivity. Since small uncertainties in the model and parameters are unavoidable, the analysis of the approaches \wrt parameter uncertainty was included. It is important to note however, that none of the methods presented is specifically adapted to parameter uncertainty. For $\ocpS$, the mathematical formulation only ensures, that the internal dynamic is at rest at the end of the transition. For $\ocpJ$, the system is (with perfect system knowledge) oscillation free, whenever hitting $\pm a_\text{max}$ and $v_\text{max}$ respectively. In other words, the reduces sensitivity to parameter uncertainty the $\ocpJ$ results from the fact, that all oscillations are cancelled in the jerk-phase of the trajectory. For the $\ZV$-shaped trajectory, every oscillation induced in the system is always cancelled half of an oscillation period later, leading to the even better behaviour \wrt parameter uncertainty. For $\FirImp$ this is not the case, as (for the third-order trajectories shown here) oscillation can be removed by the spacing of three different times, depending which is fastest, as explained in \cite{Yalamanchili2024}. The $\ocpJ$ approach was developed specifically to be able to take system damping into account, as it is possible with $\ZV$-shaping.

\bigskip
The possibility of taking system damping into account is allows to plan trajectories with significantly smaller oscillation, as the measurement shown in \autoref{sec:measurements_laboratory_system} demonstrates. Note, that the method also applies to systems without damping. Considering the small stiffness of the laboratory system, when comparing to an actual pick-and-place machine the practicality of the $\ocpJ$ was demonstrated. Measurement results shown in \autoref{fig:showing_definition_of_a0_of_meas_1_SCurve} and \autoref{fig:showing_definition_of_a0_of_meas_2_OcpJ_paramknownExact} show the substantial reduction in oscillation amplitude, if the system parameters are known with a high degree of accuracy. The system response under parameter uncertainty of the $\ZV$-shaped $\SCurve$ and the $\ocpJ$ trajectories are further quite similar, as demonstrated with the measurements shown in \autoref{fig:measurement_results_laboratory_system_first_dist}, \autoref{fig:measurement_results_laboratory_system_dist_2_to_5} and \autoref{fig:measurement_results_laboratory_system_dist_2_to_5_details}.

\bigskip
A detailed analysis of time improvement for different sets of parameters is provided in \autoref{fig:comparison_transitionTimes_relative_ZV_allParamSets}. It is important to note, that the $\ocpJ$ approach aims at a specific problem, where very low oscillation in a short amount of time has to be achieved. The measurements and simulation results for the exemplary pick-and-place machine shown in \autoref{fig:sensibility_functions_for_two_distances_ocpJ_vs_ZV} underline the usefulness of the approach. In this publication, we emphasised and focused comparisons to $\ZV$-shaping. It is an established approach, that works exceptionally well \cite{Kruk2023}, most crucially allows to take system damping into account and is simple to implement. Recent developments, like the original $\Fir$ approach \cite{Biagiotti2015,Biagiotti2019,Biagiotti2021} and the newer $\FirImp$ \cite{Yalamanchili2024} allow for a reduction of transition times, at the expense of stability to parameter uncertainty, as shown in \autoref{fig:sensibility_functions_for_two_distances_ocpJ_vs_ZV}. For highly accurate pick-and-place systems we consider the $\ocpJ$  approach to be a the better alternative, because damping can be taken into account.

\bigskip
Although a model of a pick-and-place system has been used to develop the trajectory planning approach, in the end it is only influenced by the system oscillation frequency $\omega_{\text{d}}$, the system damping $\delta$ and the kinematic constraints of the actuator. The latter are the maximum velocity $v_\text{lim}$, the acceleration $a_\text{lim}$ and the jerk $j_\text{lim}$. In other words, if $\ZV$-shaping can be used(/is currently in use on a system), the $\ocpJ$ approach is also applicable. The system parameters have to be measured and the kinetic constraints of the actuator usually come from the drives themselves (especially $v_\text{lim}$ and $a_\text{lim}$). It is recognised that higher jerk $j_\text{lim}$ usually leads to higher vibration \cite{Hayati2020,Dai2020} and higher wear \cite{Esau2022,Bilal2023}. For this reason, the limit for jerk is usually based on previous experience on a system and can be set and adjusted by analysing the system response and the tracking behaviour of the slider on a trajectory. If tracking is poor (resulting in higher residual oscillation), the jerk needs to be reduced. If tracking is satisfactory, the jerk can be increased, resulting in a reduction of transition times. A reasonable limit for the jerk to start (in the absence of prior experience) can be given as follows
\begin{equation}
	j_\text{lim,init} = \frac{a_\text{max} \cdot \omega_{\text{d}}}{2 \, \pi} \, \text{.}
\end{equation}
By analysing the system's response to this jerk using the method described above, it is possible to make adjustments until a satisfactory result is reached (see also \autoref{fig:showing_definition_of_a0_of_meas_2_OcpJ_paramknownExact}).

\section{Summary and outlook}\label{sec:outlook_and_conclusion}
\subsection{Summary}
A method for planning trajectories called $\ocpJ$ based on motion primitives (called jerk segments) was introduced and its effectiveness in terms of transition time and parameter uncertainty was analysed on a laboratory system. It was found that recalculating the jerk segments with acceleration levels lower than the physical maximum has a positive effect on the transition time and helps to remove violations of kinematic constraints. This led to the development of an efficient algorithm to compute the segments published in \cite{tau_ocpJ_assembly_part2}. The presented calculation method allows implementation on a PLC (which was done) without relying on any optimization libraries. As the comparisons \autoref{fig:showing_aMax_for_diff_sTf} and \autoref{fig:distSweep_tfComp_SC_ZV_ocpS_ocpJ_Fir_exempl_pap_param} show, the calculation time of the trajectories is less than the transition time, even on a system with limited computing power (10 year old CPU on the tested PLC). The fact that the calculation always performs the same steps also leads to known and consistent calculation times. This allows the trajectories to be calculated as they are needed, without the need for pre-calculation. As shown in \autoref{fig:comparison_transitionTimes_relative_ZV_allParamSets}, the $\ocpJ$ approach offers a reduction in transition times of between $5\%$ and $10\%$ compared to the $\ZV$ shaping approach.

\subsection{Outlook}
In future work, it is planned to combine the $\ocpJ$ approach with a continuous update of the system parameters, which is made possible by the fast trajectory computation. A detailed analysis of the algorithm used to optimise $a_\text{max}$ might also allow to solve for the actual best $a_\text{best}$ for Case~1 and Case~3. In this context, a mathematical proof of what $a_\text{best}$ is might prove useful for further development of the algorithm. The use of jerk segments specifically optimised for performance \wrt parameter uncertainty is planned for the future. Alternatively, adaptation to multiple eigenfrequencies may be considered. Changes to the jerk segments themselves (contents of \cite{tau_ocpJ_assembly_part2}) will only affect the calculation required for the jerk segments, but the calculation of the full trajectories shown here will remain the same. This is the subject of planned future work.

\appendices
\clearpage
\section{Algorithms for implementation}\label{App:alg_for_implementation}%
This section contains the calculations necessary to compute a complete $\ocpJ$ trajectory without violating any kinematic constraints. As mentioned earlier, a trajectory for a given $a_\text{max}$ is computed using \autoref{alg:get_ocp_ind_traject}.
\begin{algorithm}
	\caption{Get $\ocpJ$ trajectory for $a_\text{max}$}\label{alg:get_ocp_ind_traject}
	\begin{algorithmic}[1]
		\State Get segment for $a_\text{max}$ \Comment{Use \cite{tau_ocpJ_assembly_part2}} %
		\State Get segment for $2 \, a_\text{max}$ \Comment{Use \cite{tau_ocpJ_assembly_part2}} %
		\State Calculate $s_\text{f1},..., \, v_\text{f3}$
		\State Calculate $\ocpJ$ trajectory for Case~1
		\If{$v\ttt \leq v_\text{lim} \ \forall t \in \left[0, t_\text{f,t}\right]$}
		\State \Return $\ocpJ$ trajectory for Case~1
		\Else
		\If{$s\tft \geq s_\text{f,min-Case~2}$}
		\State Calculate $\ocpJ$ trajectory for Case~2
		\State \Return $\ocpJ$ trajectory for Case~2
		\Else
		\State Calculate $\ocpJ$ trajectory for Case~3
		\State \Return $\ocpJ$ trajectory for Case~3
		\EndIf
		\EndIf
	\end{algorithmic}
\end{algorithm}
This calculation is called in \autoref{alg:Get_ocpJ_total} and the kinematic constraints are checked.
\begin{algorithm}
	\caption{Get $\ocpJ$ total}\label{alg:Get_ocpJ_total}
	\begin{algorithmic}[1]
		\State Calculate $a_\text{best}$ with \eqref{eq:calx_idal_aMax_Case2} \Comment{Only required once}
		\State Limit $a_\text{best}$ to $a_\text{lim}$
		\State Calculate $\ocpJ$ trajectory with \autoref{alg:get_ocp_ind_traject} for $a_\text{best}$
		\If{$a\ttt \leq a_\text{lim}$ and $j\ttt \leq j_\text{lim} \ \forall t \in \left[0, t_\text{f,t}\right]$ }
		\State \Return $\ocpJ$ trajectory
		\Else
		\State Run \autoref{alg:Optimize_aMAx}
		\State \Return $\ocpJ$ trajectory
		\EndIf
	\end{algorithmic}
\end{algorithm}
If the kinematic constraints are not violated, the computation is finished and the computed trajectory can be exported, otherwise \autoref{alg:Optimize_aMAx} must be called. A search algorithm iterates over possible $a_\text{max}$ values and searches for the highest possible $a_\text{max}$ where no kinematic constraint is violated. This can be achieved efficiently with a binary search as described in \autoref{alg:Optimize_aMAx}. The bounds for the terminal conditions $\Delta t_\text{boundary}$ and $n_\text{max-iter}$ can be set as follows. The time $\Delta t_\text{boundary}$ can be set to one controller cycle interval from the system. If the improvement in time $\Delta t_\text{f}$ is less than one controller cycle, the calculation can be stopped. When implementing the algorithm in single precision (the measurements in \autoref{SubSec:Measurement_Results} show, that this can already be sufficient), the upper limit for $n_\text{max-iter}$ can be set to $n_\text{max-iter} = 23$. Iterating more than the number of significant digits of the single precision variable does not provide any benefit. Setting these limits for $n_\text{max-iter}$ and $\Delta t_\text{boundary}$ ensures that the search algorithm will stop after a given number of iterations.
\begin{algorithm}
	\caption{Optimize $a_\text{max}$}\label{alg:Optimize_aMAx}
	\begin{algorithmic}[1]
		\State Initialize $\Delta a_\text{ges} = a_\text{best}$
		\State Set $n_\text{iter} = 1$
		\State Set $a_\text{iter} = \frac{1}{2} \Delta a_\text{ges}$
		\Repeat
		\State Get trajectory with \autoref{alg:get_ocp_ind_traject} for $a_\text{max} = a_\text{iter}$
		\State $n_\text{iter} = n_\text{iter} + 1 $
		\If{$a\ttt \leq a_\text{lim}$ and $j\ttt \leq j_\text{lim} \ \forall t \in \left[0, t_\text{f,t}\right]$}
		\State Save $t_\text{f,t}$ and calculate $\Delta t_\text{f} = t_{\text{f,t;}\left(k-1\right)} - t_{\text{f,t;}k}$
		\If{$t_{\text{f,t;}k} < t_{\text{f,best}}$}
		\State $t_{\text{f,best}} = t_{\text{f,t;}k}$
		\State Save trajectory for export\label{alg_step:best_traj_saved}
		\EndIf
		\State Set $a_\text{iter} = a_\text{iter} + \left(\frac{1}{2}\right)^{n_\text{iter}} \cdot \Delta a_\text{ges}$
		\Else
		\State Set $a_\text{iter} = a_\text{iter} - \left(\frac{1}{2}\right)^{n_\text{iter}} \cdot \Delta a_\text{ges}$
		\EndIf
		\Until{$\Delta t_\text{f} \leq \Delta t_\text{boundary} \vee n_\text{iter} > n_\text{max-iter}$}
		\State \Return Trajectory from step~\ref{alg_step:best_traj_saved} of Alg.
	\end{algorithmic}
\end{algorithm}%

\section{Measurement results}\label{App:Table_with_Measurement_Results}%

\begin{table*}[!ht] 
 \captionsetup{width=\linewidth} 
 \caption{Measurement results: Listing amplitudes at $t_\text{f,t}$ for measured distances, trajectory planning methods and parameter combinations according to \autoref{tab:table_adjusted_paramters}.} 
 \vspace{-1em} 
 \renewcommand{\arraystretch}{1.3} 
 \centering 
 \begin{tabular}{|l|l||l|l|l|l|l|} 
  \hline 
 \multicolumn{7}{|l|}{$z\tft = \SI{14.5}{\milli\meter}$} \\ \hline 
Method & $t_\text{f,t}$& $f_{\text{sys}}=\SI{8.71}{\hertz}$& $f_{\text{sys}}=\SI{9.20}{\hertz}$& $f_{\text{sys}}=f_{\text{nom}}=\SI{9.71}{\hertz}$& $f_{\text{sys}}=\SI{10.27}{\hertz}$& $f_{\text{sys}}=\SI{10.60}{\hertz}$ \\ \hline 
\multirow{2}{*}{$\SCurve$}& \multirow{2}{*}{$\SI{132.8}{\milli\second}$}& $a_\text{max}=\SI{853.67}{\micro\meter}$& $a_\text{max}=\SI{751.30}{\micro\meter}$& $a_\text{max}=\SI{645.37}{\micro\meter}$& $a_\text{max}=\SI{533.54}{\micro\meter}$& $a_\text{max}=\SI{487.78}{\micro\meter}$\\ 
 && $a_\text{min}=\SI{846.98}{\micro\meter}$& $a_\text{min}=\SI{746.07}{\micro\meter}$& $a_\text{min}=\SI{643.77}{\micro\meter}$& $a_\text{min}=\SI{527.82}{\micro\meter}$& $a_\text{min}=\SI{482.18}{\micro\meter}$\\ \hline 
\multirow{2}{*}{$\ZV$}& \multirow{2}{*}{$\SI{184.4}{\milli\second}$}& $a_\text{max}=\SI{130.35}{\micro\meter}$& $a_\text{max}=\SI{51.83}{\micro\meter}$& $a_\text{max}=\SI{12.09}{\micro\meter}$& $a_\text{max}=\SI{60.03}{\micro\meter}$& $a_\text{max}=\SI{75.05}{\micro\meter}$\\ 
 && $a_\text{min}=\SI{128.99}{\micro\meter}$& $a_\text{min}=\SI{50.40}{\micro\meter}$& $a_\text{min}=\SI{11.28}{\micro\meter}$& $a_\text{min}=\SI{58.99}{\micro\meter}$& $a_\text{min}=\SI{73.74}{\micro\meter}$\\ \hline 
\multirow{2}{*}{$\FirImp$}& \multirow{2}{*}{$\SI{156.4}{\milli\second}$}& $a_\text{max}=\SI{200.82}{\micro\meter}$& $a_\text{max}=\SI{94.83}{\micro\meter}$& $a_\text{max}=\SI{22.17}{\micro\meter}$& $a_\text{max}=\SI{99.30}{\micro\meter}$& $a_\text{max}=\SI{131.88}{\micro\meter}$\\ 
 && $a_\text{min}=\SI{199.66}{\micro\meter}$& $a_\text{min}=\SI{93.23}{\micro\meter}$& $a_\text{min}=\SI{19.63}{\micro\meter}$& $a_\text{min}=\SI{98.21}{\micro\meter}$& $a_\text{min}=\SI{130.17}{\micro\meter}$\\ \hline 
\multirow{2}{*}{$\ocpJ$}& \multirow{2}{*}{$\SI{162.4}{\milli\second}$}& $a_\text{max}=\SI{171.35}{\micro\meter}$& $a_\text{max}=\SI{75.03}{\micro\meter}$& $a_\text{max}=\SI{8.35}{\micro\meter}$& $a_\text{max}=\SI{80.40}{\micro\meter}$& $a_\text{max}=\SI{106.50}{\micro\meter}$\\ 
 && $a_\text{min}=\SI{168.55}{\micro\meter}$& $a_\text{min}=\SI{73.64}{\micro\meter}$& $a_\text{min}=\SI{7.94}{\micro\meter}$& $a_\text{min}=\SI{79.46}{\micro\meter}$& $a_\text{min}=\SI{104.49}{\micro\meter}$\\ \hline 
 \hline 
 \multicolumn{7}{|l|}{$z\tft = \SI{61}{\milli\meter}$} \\ \hline 
Method & $t_\text{f,t}$& $f_{\text{sys}}=\SI{8.71}{\hertz}$& $f_{\text{sys}}=\SI{9.20}{\hertz}$& $f_{\text{sys}}=f_{\text{nom}}=\SI{9.71}{\hertz}$& $f_{\text{sys}}=\SI{10.27}{\hertz}$& $f_{\text{sys}}=\SI{10.60}{\hertz}$ \\ \hline 
\multirow{2}{*}{$\SCurve$}& \multirow{2}{*}{$\SI{240.8}{\milli\second}$}& $a_\text{max}=\SI{464.30}{\micro\meter}$& $a_\text{max}=\SI{500.02}{\micro\meter}$& $a_\text{max}=\SI{479.34}{\micro\meter}$& $a_\text{max}=\SI{404.07}{\micro\meter}$& $a_\text{max}=\SI{363.29}{\micro\meter}$\\ 
 && $a_\text{min}=\SI{459.90}{\micro\meter}$& $a_\text{min}=\SI{496.75}{\micro\meter}$& $a_\text{min}=\SI{478.30}{\micro\meter}$& $a_\text{min}=\SI{402.72}{\micro\meter}$& $a_\text{min}=\SI{357.85}{\micro\meter}$\\ \hline 
\multirow{2}{*}{$\ZV$}& \multirow{2}{*}{$\SI{292.4}{\milli\second}$}& $a_\text{max}=\SI{71.14}{\micro\meter}$& $a_\text{max}=\SI{35.21}{\micro\meter}$& $a_\text{max}=\SI{4.74}{\micro\meter}$& $a_\text{max}=\SI{42.40}{\micro\meter}$& $a_\text{max}=\SI{51.98}{\micro\meter}$\\ 
 && $a_\text{min}=\SI{69.80}{\micro\meter}$& $a_\text{min}=\SI{33.57}{\micro\meter}$& $a_\text{min}=\SI{3.86}{\micro\meter}$& $a_\text{min}=\SI{41.12}{\micro\meter}$& $a_\text{min}=\SI{50.89}{\micro\meter}$\\ \hline 
\multirow{2}{*}{$\FirImp$}& \multirow{2}{*}{$\SI{260.4}{\milli\second}$}& $a_\text{max}=\SI{132.01}{\micro\meter}$& $a_\text{max}=\SI{80.20}{\micro\meter}$& $a_\text{max}=\SI{20.13}{\micro\meter}$& $a_\text{max}=\SI{100.01}{\micro\meter}$& $a_\text{max}=\SI{133.39}{\micro\meter}$\\ 
 && $a_\text{min}=\SI{130.73}{\micro\meter}$& $a_\text{min}=\SI{78.90}{\micro\meter}$& $a_\text{min}=\SI{18.63}{\micro\meter}$& $a_\text{min}=\SI{99.57}{\micro\meter}$& $a_\text{min}=\SI{132.33}{\micro\meter}$\\ \hline 
\multirow{2}{*}{$\ocpJ$}& \multirow{2}{*}{$\SI{273.6}{\milli\second}$}& $a_\text{max}=\SI{87.32}{\micro\meter}$& $a_\text{max}=\SI{44.09}{\micro\meter}$& $a_\text{max}=\SI{5.24}{\micro\meter}$& $a_\text{max}=\SI{51.34}{\micro\meter}$& $a_\text{max}=\SI{64.60}{\micro\meter}$\\ 
 && $a_\text{min}=\SI{85.28}{\micro\meter}$& $a_\text{min}=\SI{43.72}{\micro\meter}$& $a_\text{min}=\SI{3.34}{\micro\meter}$& $a_\text{min}=\SI{50.49}{\micro\meter}$& $a_\text{min}=\SI{64.22}{\micro\meter}$\\ \hline 
 \hline 
 \multicolumn{7}{|l|}{$z\tft = \SI{116}{\milli\meter}$} \\ \hline 
Method & $t_\text{f,t}$& $f_{\text{sys}}=\SI{8.71}{\hertz}$& $f_{\text{sys}}=\SI{9.20}{\hertz}$& $f_{\text{sys}}=f_{\text{nom}}=\SI{9.71}{\hertz}$& $f_{\text{sys}}=\SI{10.27}{\hertz}$& $f_{\text{sys}}=\SI{10.60}{\hertz}$ \\ \hline 
\multirow{2}{*}{$\SCurve$}& \multirow{2}{*}{$\SI{362.8}{\milli\second}$}& $a_\text{max}=\SI{574.02}{\micro\meter}$& $a_\text{max}=\SI{633.81}{\micro\meter}$& $a_\text{max}=\SI{554.17}{\micro\meter}$& $a_\text{max}=\SI{356.40}{\micro\meter}$& $a_\text{max}=\SI{262.14}{\micro\meter}$\\ 
 && $a_\text{min}=\SI{567.43}{\micro\meter}$& $a_\text{min}=\SI{629.54}{\micro\meter}$& $a_\text{min}=\SI{552.22}{\micro\meter}$& $a_\text{min}=\SI{353.13}{\micro\meter}$& $a_\text{min}=\SI{257.48}{\micro\meter}$\\ \hline 
\multirow{2}{*}{$\ZV$}& \multirow{2}{*}{$\SI{414.4}{\milli\second}$}& $a_\text{max}=\SI{86.33}{\micro\meter}$& $a_\text{max}=\SI{41.83}{\micro\meter}$& $a_\text{max}=\SI{6.85}{\micro\meter}$& $a_\text{max}=\SI{37.07}{\micro\meter}$& $a_\text{max}=\SI{36.17}{\micro\meter}$\\ 
 && $a_\text{min}=\SI{86.12}{\micro\meter}$& $a_\text{min}=\SI{40.83}{\micro\meter}$& $a_\text{min}=\SI{5.51}{\micro\meter}$& $a_\text{min}=\SI{35.10}{\micro\meter}$& $a_\text{min}=\SI{35.48}{\micro\meter}$\\ \hline 
\multirow{2}{*}{$\FirImp$}& \multirow{2}{*}{$\SI{382.8}{\milli\second}$}& $a_\text{max}=\SI{159.30}{\micro\meter}$& $a_\text{max}=\SI{94.53}{\micro\meter}$& $a_\text{max}=\SI{25.64}{\micro\meter}$& $a_\text{max}=\SI{87.92}{\micro\meter}$& $a_\text{max}=\SI{97.86}{\micro\meter}$\\ 
 && $a_\text{min}=\SI{157.40}{\micro\meter}$& $a_\text{min}=\SI{92.42}{\micro\meter}$& $a_\text{min}=\SI{24.15}{\micro\meter}$& $a_\text{min}=\SI{86.49}{\micro\meter}$& $a_\text{min}=\SI{96.66}{\micro\meter}$\\ \hline 
\multirow{2}{*}{$\ocpJ$}& \multirow{2}{*}{$\SI{395.2}{\milli\second}$}& $a_\text{max}=\SI{101.24}{\micro\meter}$& $a_\text{max}=\SI{49.87}{\micro\meter}$& $a_\text{max}=\SI{3.56}{\micro\meter}$& $a_\text{max}=\SI{42.07}{\micro\meter}$& $a_\text{max}=\SI{43.42}{\micro\meter}$\\ 
 && $a_\text{min}=\SI{99.84}{\micro\meter}$& $a_\text{min}=\SI{49.16}{\micro\meter}$& $a_\text{min}=\SI{2.93}{\micro\meter}$& $a_\text{min}=\SI{41.25}{\micro\meter}$& $a_\text{min}=\SI{41.93}{\micro\meter}$\\ \hline 
 \hline 
 \multicolumn{7}{|l|}{$z\tft = \SI{139}{\milli\meter}$} \\ \hline 
Method & $t_\text{f,t}$& $f_{\text{sys}}=\SI{8.71}{\hertz}$& $f_{\text{sys}}=\SI{9.20}{\hertz}$& $f_{\text{sys}}=f_{\text{nom}}=\SI{9.71}{\hertz}$& $f_{\text{sys}}=\SI{10.27}{\hertz}$& $f_{\text{sys}}=\SI{10.60}{\hertz}$ \\ \hline 
\multirow{2}{*}{$\SCurve$}& \multirow{2}{*}{$\SI{414}{\milli\second}$}& $a_\text{max}=\SI{724.58}{\micro\meter}$& $a_\text{max}=\SI{354.98}{\micro\meter}$& $a_\text{max}=\SI{74.41}{\micro\meter}$& $a_\text{max}=\SI{222.30}{\micro\meter}$& $a_\text{max}=\SI{258.53}{\micro\meter}$\\ 
 && $a_\text{min}=\SI{715.57}{\micro\meter}$& $a_\text{min}=\SI{353.54}{\micro\meter}$& $a_\text{min}=\SI{72.00}{\micro\meter}$& $a_\text{min}=\SI{220.94}{\micro\meter}$& $a_\text{min}=\SI{253.75}{\micro\meter}$\\ \hline 
\multirow{2}{*}{$\ZV$}& \multirow{2}{*}{$\SI{465.6}{\milli\second}$}& $a_\text{max}=\SI{92.36}{\micro\meter}$& $a_\text{max}=\SI{21.79}{\micro\meter}$& $a_\text{max}=\SI{5.25}{\micro\meter}$& $a_\text{max}=\SI{24.68}{\micro\meter}$& $a_\text{max}=\SI{35.99}{\micro\meter}$\\ 
 && $a_\text{min}=\SI{89.93}{\micro\meter}$& $a_\text{min}=\SI{20.34}{\micro\meter}$& $a_\text{min}=\SI{4.82}{\micro\meter}$& $a_\text{min}=\SI{23.12}{\micro\meter}$& $a_\text{min}=\SI{34.95}{\micro\meter}$\\ \hline 
\multirow{2}{*}{$\FirImp$}& \multirow{2}{*}{$\SI{414}{\milli\second}$}& $a_\text{max}=\SI{723.64}{\micro\meter}$& $a_\text{max}=\SI{353.73}{\micro\meter}$& $a_\text{max}=\SI{74.41}{\micro\meter}$& $a_\text{max}=\SI{222.23}{\micro\meter}$& $a_\text{max}=\SI{257.77}{\micro\meter}$\\ 
 && $a_\text{min}=\SI{718.85}{\micro\meter}$& $a_\text{min}=\SI{351.89}{\micro\meter}$& $a_\text{min}=\SI{73.36}{\micro\meter}$& $a_\text{min}=\SI{221.46}{\micro\meter}$& $a_\text{min}=\SI{252.73}{\micro\meter}$\\ \hline 
\multirow{2}{*}{$\ocpJ$}& \multirow{2}{*}{$\SI{446.4}{\milli\second}$}& $a_\text{max}=\SI{109.91}{\micro\meter}$& $a_\text{max}=\SI{24.51}{\micro\meter}$& $a_\text{max}=\SI{5.01}{\micro\meter}$& $a_\text{max}=\SI{28.14}{\micro\meter}$& $a_\text{max}=\SI{42.80}{\micro\meter}$\\ 
 && $a_\text{min}=\SI{106.53}{\micro\meter}$& $a_\text{min}=\SI{22.91}{\micro\meter}$& $a_\text{min}=\SI{3.45}{\micro\meter}$& $a_\text{min}=\SI{26.98}{\micro\meter}$& $a_\text{min}=\SI{41.74}{\micro\meter}$\\ \hline 
 \hline 
 \multicolumn{7}{|l|}{$z\tft = \SI{181}{\milli\meter}$} \\ \hline 
Method & $t_\text{f,t}$& $f_{\text{sys}}=\SI{8.71}{\hertz}$& $f_{\text{sys}}=\SI{9.20}{\hertz}$& $f_{\text{sys}}=f_{\text{nom}}=\SI{9.71}{\hertz}$& $f_{\text{sys}}=\SI{10.27}{\hertz}$& $f_{\text{sys}}=\SI{10.60}{\hertz}$ \\ \hline 
\multirow{2}{*}{$\SCurve$}& \multirow{2}{*}{$\SI{507.6}{\milli\second}$}& $a_\text{max}=\SI{817.42}{\micro\meter}$& $a_\text{max}=\SI{554.20}{\micro\meter}$& $a_\text{max}=\SI{186.82}{\micro\meter}$& $a_\text{max}=\SI{184.74}{\micro\meter}$& $a_\text{max}=\SI{241.01}{\micro\meter}$\\ 
 && $a_\text{min}=\SI{806.26}{\micro\meter}$& $a_\text{min}=\SI{551.92}{\micro\meter}$& $a_\text{min}=\SI{183.39}{\micro\meter}$& $a_\text{min}=\SI{183.51}{\micro\meter}$& $a_\text{min}=\SI{238.01}{\micro\meter}$\\ \hline 
\multirow{2}{*}{$\ZV$}& \multirow{2}{*}{$\SI{558.8}{\milli\second}$}& $a_\text{max}=\SI{103.51}{\micro\meter}$& $a_\text{max}=\SI{28.53}{\micro\meter}$& $a_\text{max}=\SI{4.87}{\micro\meter}$& $a_\text{max}=\SI{21.84}{\micro\meter}$& $a_\text{max}=\SI{33.83}{\micro\meter}$\\ 
 && $a_\text{min}=\SI{102.12}{\micro\meter}$& $a_\text{min}=\SI{26.37}{\micro\meter}$& $a_\text{min}=\SI{3.48}{\micro\meter}$& $a_\text{min}=\SI{21.37}{\micro\meter}$& $a_\text{min}=\SI{33.19}{\micro\meter}$\\ \hline 
\multirow{2}{*}{$\FirImp$}& \multirow{2}{*}{$\SI{515.2}{\milli\second}$}& $a_\text{max}=\SI{818.45}{\micro\meter}$& $a_\text{max}=\SI{453.56}{\micro\meter}$& $a_\text{max}=\SI{101.22}{\micro\meter}$& $a_\text{max}=\SI{289.66}{\micro\meter}$& $a_\text{max}=\SI{318.30}{\micro\meter}$\\ 
 && $a_\text{min}=\SI{817.10}{\micro\meter}$& $a_\text{min}=\SI{450.09}{\micro\meter}$& $a_\text{min}=\SI{98.77}{\micro\meter}$& $a_\text{min}=\SI{286.55}{\micro\meter}$& $a_\text{min}=\SI{313.76}{\micro\meter}$\\ \hline 
\multirow{2}{*}{$\ocpJ$}& \multirow{2}{*}{$\SI{539.6}{\milli\second}$}& $a_\text{max}=\SI{120.42}{\micro\meter}$& $a_\text{max}=\SI{32.65}{\micro\meter}$& $a_\text{max}=\SI{3.56}{\micro\meter}$& $a_\text{max}=\SI{26.40}{\micro\meter}$& $a_\text{max}=\SI{40.41}{\micro\meter}$\\ 
 && $a_\text{min}=\SI{118.61}{\micro\meter}$& $a_\text{min}=\SI{30.77}{\micro\meter}$& $a_\text{min}=\SI{2.75}{\micro\meter}$& $a_\text{min}=\SI{25.65}{\micro\meter}$& $a_\text{min}=\SI{39.22}{\micro\meter}$\\ \hline 
\end{tabular} 
 \label{tab:table_measurement_results} 
 \end{table*}

\end{document}